\documentclass[sn-basic, linktocpage]{sn-jnl}
\let\orcidlogo\relax
\usepackage{ulem}
\usepackage{orcidlink}
\usepackage{tablefootnote}
\usepackage[symbol]{footmisc}
\usepackage{graphicx}%
\usepackage{multirow}%
\usepackage{amsmath,amssymb,amsfonts}%
\usepackage{amsthm}%
\usepackage{mathrsfs}%
\usepackage[title]{appendix}%
\usepackage{xcolor}
\usepackage{tabularx}
\usepackage{aas_macros}
\def\kms{\hbox{km$\;$s$^{-1}$}} 
\def\arcsec{\hbox{$^{\prime\prime}$}}
 
\newcommand{\rs}{$\mathrm{R_{\odot}}$}

\begin{document}
\title{Potential Solar Precursors to Magnetic Switchbacks}
\author[1]{\fnm{Durgesh} \sur{Tripathi}
\orcidlink{0000-0003-1689-6254}}
\email{durgesh@iucaa.in}
\equalcont{These authors contributed equally to this work.}
\author*[2,3,4]{\fnm{Maria S.} \sur{Madjarska}
\orcidlink{0000-0001-9806-2485}}
\email{madjarska@mps.mpg.de}
\equalcont{These authors contributed equally to this work.}
\author[5]{\fnm{Judy}\sur{Karpen}
\orcidlink{0000-0002-6975-5642}}
\email{judy.karpen@nasa.gov}
\equalcont{These authors contributed equally to this work.}
\author[6]{\fnm{Marco}\sur{Velli}
\orcidlink{0000-0002-2381-3106}}
\email{mvelli@ucla.edu}
\author[7]{\fnm{Clara} \sur{Froment}
\orcidlink{0000-0001-5315-2890}}
\email{clara.froment@cnrs-orleans.fr}
\author[8,9]{\fnm{Etienne} \sur{Pariat}
\orcidlink{0000-0002-2900-0608}}
\email{etienne.pariat@lpp.polytechnique.fr}
\author[10]{\fnm{Spiros} \sur{Patsourakos}
\orcidlink{0000-0003-3345-9697}}
\email{spatsour@uoi.gr}
\author[11]{\fnm{Nour E.} \sur{Raouafi}
\orcidlink{0000-0003-2409-3742}}
\email{nour.raouafi@jhuapl.edu}
\author[12,13]{\fnm{Alexis P.}
\sur{Rouillard}
\orcidlink{0000-0003-4039-5767}}
\email{arouillard@irap.omp.eu}
\author[14]{\fnm{Alphonse C.} \sur{Sterling}
\orcidlink{0000-0003-1281-897X}}
\email{alphonse.sterling@nasa.gov}
\author[15]{\fnm{Kostas} \sur{Tziotziou}
\orcidlink{0000-0002-2542-7073}}
\email{kostas@noa.gr}
\author[16]{\fnm{Peter F.} \sur{Wyper}
\orcidlink{0000-0002-6442-7818}}
\email{peter.f.wyper@durham.ac.uk}

\affil[1]{\orgdiv{Inter-University Centre for Astronomy and Astrophysics},  \orgaddress{\street{Post Bag 4}, \city{Ganeshkhind}, \state{State}, \postcode{411007},  \country{India}}}
\affil*[2]{Max Planck Institute for Solar System Research, Justus-von-Liebig-Weg 3, 37077, G\"ottingen, Germany}
\affil[3]{Korea Astronomy and Space Science Institute, 34055, Daejeon, Republic of Korea}
\affil[4]{Space Research and Technology Institute, Bulgarian Academy of Sciences, Acad. G. Bonchev Str., Bl. 1, 1113, Sofia, Bulgaria}
\affil[5]{\orgdiv{Heliophysics Science Division}, \orgname{NASA Goddard Space Flight Center}, \orgaddress{\street{8800 Greenbelt Rd.}, \city{Greenbelt}, \postcode{20771}, \state{MD}, \country{USA}}}
\affil[6]{\orgname{Earth, Planetary and Space Sciences, University of California, Los Angeles,} \postcode{91109},\country{USA}}
\affil[7]{LPC2E, OSUC, Univ Orleans, CNRS, CNES, F-45071 Orleans, France}
\affil[8]{\orgdiv{French - Spanish Laboratory for Astrophysics in Canarias (FSLAC), CNRS IRL2009, Instituto de AstrofÃ­sica de Canarias},\orgaddress{ \postcode{38205}, \city{La Laguna},  \state{Tenerife}, \country{Spain}}}
\affil[9]{\orgdiv{Sorbonne Universit\'e, \'Ecole polytechnique, Institut Polytechnique de Paris, Universit\'e Paris Saclay, Observatoire de Paris-PSL, CNRS, Laboratoire de Physique des Plasmas (LPP)}, \orgaddress{\street{4 place Jussieu}, \postcode{75005}, \city{Paris}, \country{France}}}
\affil[10]{\orgname{University of Ioannina, Department of Physics, Section of AstroGeophysics}, \city{Ioannina}, \postcode{45110},\country{Greece}}
\affil[11]{\orgdiv{Johns Hopkins Applied Physics Laboratory}, \orgaddress{\street{11100 Johns Hopkins Road}, \city{Laurel}, \postcode{20723}, \state{Maryland}, \country{USA}}}
\affil[12]{\orgname{Institute of Research in Astrophysics and Planetology, UPS, CNRS, CNES}, \postcode{31000}, \country{France}}
\affil[13]{\orgname{Leibniz-Institut f\:{u}r Astrophysik Potsdam (AIP), An der Sternwarte 16}, \postcode{14482}, \country{Germany}}
\affil[14]{\orgdiv{Heliophysics and Planetary Science Branch}, \orgname{NASA Marshall Space Flight Center}, \orgaddress{\street{320 Sparkman Dr.}, \city{Huntsville}, \postcode{35812}, \state{AL}, \country{USA}}}
\affil[15]{\orgdiv{Institute for Astronomy, Astrophysics, Space Applications and Remote Sensing}, \orgname{National Observatory of Athens}, \orgaddress{ \city{Penteli}, \postcode{15236},  \country{Greece}}}
\affil[16]{\orgdiv{Department of Mathematical Sciences}, \orgname{Durham University}, \orgaddress{\street{Stockton Road}, \city{Durham}, \postcode{DH1 3LE}, \country{UK}}}
\abstract{The origin of magnetic switchbacks{---}large-amplitude, spherically polarized magnetic-field fluctuations with local polarity inversions shown to be nearly ubiquitous in the inner heliosphere by Parker Solar Probe{---}is presently one of the most outstanding open questions in solar and heliospheric science. The occurrence of these structures in the young solar wind is a topic of active research, focusing not only on their characteristics and evolution but also on unraveling the puzzle of their origin. We first discuss the potential influence on switchback (SB) formation of large-scale coronal dynamics and restructuring processes that shape and regulate the extended solar corona.
Following that, we review the dynamics, physical properties, occurrence rate, and energetics of numerous possible precursor small-scale dynamic events that occur throughout the solar atmosphere, emphasizing their possible roles in creating SBs. Finally, we discuss some recent studies attempting to connect in situ observations of SBs and SB patches with remote-sensing observations, and the related challenges in identifying solar wind source regions precisely and linking in situ observations to transient solar phenomena. We also clarify the terminology used by the solar community to describe small-scale solar phenomena.}
\maketitle

\keywords{Sun, atmosphere, heliosphere, solar activity, Solar wind, jets, chromosphere, corona}
\maketitle
\section{Introduction}
\label{sec1}
The prevalence of switchbacks (SBs){---}large-amplitude, spherically polarized magnetic-field fluctuations with local polarity inversions in the inner heliosphere solar wind{---}was one of the first surprising observational discoveries of the Parker Solar Probe \citep[][]{2016SSRv..204....7F} mission. The near constancy of the total magnetic field implied by the observed spherical polarization, and the ubiquitous presence of radial velocity ``jets" in correspondence with the magnetic polarity inversions, imply that SBs may be interpreted as nonlinear Alfv\'enic fluctuations propagating away from the Sun \citep[for details, see the reviews by][]{chapter3,chapter6}. Radial magnetic field variations in SBs are highly correlated with proton velocity variations within their structure. As SBs represent steepened, large-amplitude Alfv\'en waves in which the magnetic field and velocity vector vary together, most of their plasma density changes little relative to those of the surrounding solar wind. The sum of the SBs' kinetic (proton, alpha, and electron) and magnetic pressure remains constant. The proton temperature inside SBs appears to be higher than in the surrounding solar wind.

Furthermore, SBs tend to appear in patches with spatial scales comparable to those of the solar supergranulation. They can last up to several hours, and their occurrence rate shows little dependence on solar cycle activity \citep{2026A&A...708A.273D}. For details and references, see \citet{chapter3}. Although SBs are detected far from the Sun, several theoretical and observational clues suggest that impulsive phenomena in the solar atmosphere may be fundamental to their generation. 

The fact that the interplanetary magnetic field, generated within the Sun, permeates the entire heliosphere (outside of planetary magnetospheres) suggests a causal link between various dynamical phenomena occurring on the Sun and their subsequent detection at remote (heliospheric) locations \citep[see][]{chapter3,chapter6}. Ubiquitous energetic transient events across various spatio-temporal scales that eject mass, momentum, and/or energy into the corona and solar wind, have been recorded over the past few decades at various spatio-temporal resolutions \citep[e.g.,][]{Beckers1968, 2012SSRv..169..181T, Sch_2006, Pick_2006, 2016SSRv..201....1R, 2021RSPSA.47700217S, Temmer_2021}. Those could provide precursors to an SB structure observed in the young solar wind.

\citet{Bale2021} and \citet{Fargette2021} suggested a possible magnetic connection between SB patches and the chromospheric network (boundaries of supergranules), where concentrations of mixed magnetic polarities are abundant. \cite{Fargette2021} established the link by converting the temporal scales of PSP observations into spatial scales by directly projecting PSP’s orbit. However, \cite{Bale2021} used a ballistic projection of the orbit into the corona, followed by a potential field magnetic map to project the spacecraft's path onto the solar surface, and overplotting it onto images and magnetograms from the Atmospheric Imaging Assembly (AIA) and the Helioseismic and Magnetic Imager \citep[HMI;][]{2012SoPh..275..207S}, respectively, on board the Solar Dynamics Observatory (SDO). The open magnetic field rooted in these concentrations broadens into space-filling coronal ``funnels'', most readily detectable in open-flux-dominated regions such as coronal holes (CHs). Plasma and waves can travel outward along these funnel-like open-field structures. The persistence of SBs during the PSP fast radial scans suggested that a dynamical connection with emerging flux within the supergranular cells might also be present \citep{shi_patches_2022}.

 Several studies have been dedicated to determining the possible origin of SBs and examining their properties \citep[e.g.,][]{DudokdeWit2020, Woodham2021,2025A&A...694A.181B}. The various mechanisms that could explain the formation of SBs are discussed in detail in \cite{chapter5}. Here, we focus on the well-established connection between the underlying magnetic organization of the solar atmosphere and its inferred temporal and spatial imprints on the inner heliosphere. In particular, given the inherent intermittency of SBs and their patches, we primarily focus on both structures and transient activity that might form or enable the propagation of SB progenitors.  

Eruptive behavior occurs on many spatial and temporal scales on the Sun. In particular, jet-like solar-atmospheric dynamic phenomena can be large and energetic enough to appear as narrow-angle coronal mass ejections (CMEs) in coronagraph images \citep{1998ApJ...508..899W} and small enough to match or fall below the UV and EUV instrumental resolution limits  \citep[e.g.,][]{2014Sci...346A.315T,2021ApJ...918L..20H, 2023Sci...381..867C}.  These eruptive events increase in number as the spatial scale decreases, perhaps even following a power-law distribution in size and energy \citep[e.g.,][]{1999Ap&SS.264..129S,2023ApJ...955L..38U,2024ApJ...963....4S}. Since these events expel material, energy, and/or magnetic field outward into the corona and perhaps into the heliosphere, they are considered to be one of the prime candidates for generating SBs \citep{2020ApJS..246...45H,2020ApJ...896L..18S,2021ApJ...920L..31N}, and may contribute more generally to the solar wind \citep[e.g.,][]{2012ApJ...750...50N,2023ApJ...945...28R}. 
Another class of candidate mechanisms for SB formation involves the restructuring of the coronal magnetic field along topological features such as magnetic separatrices and quasi-separatrix layers, which organize the low-plasma-beta solar corona.
Numerous localized events driving these reconfigurations, likely impulsive but smaller in scale than the previously discussed jet-like events, can generate  
Alfv\'enic fluctuations, velocity shear, and compressive disturbances in and around these topological regions. We caution the reader, however, that {--} as of now {--} whether such eruptive phenomena at diverse spatiotemporal scales are formed by the same physical process, and whether and how much they contribute to SBs and the solar wind, are the subject of active study and debate in the heliophysics community.

This paper reviews the large-scale structure of the Sun's magnetized atmosphere and solar active events that are most likely to precede or be potential progenitors of SBs. We provide a comprehensive overview aimed at non-specialist readers, with references to guide deeper exploration of specific coronal structures and 
phenomena.

\section{Data Sources for Studying Solar SB Precursors}
\label{sec:Data}
The observations described in this paper were obtained using remote-sensing instruments with imaging and spectroscopic capabilities.  Many jet-like phenomena were first detected using ground-based observations \citep[e.g.,][]{1972ARA&A..10...73B, Roy_1973}. Further studies have been performed since the era of {\sl Yohkoh} when the Soft X-ray Telescope \citep[SXT;][]{sxt} revealed a variety of coronal jet-like features. SXT was one of the four instruments on board {\sl Yohkoh}, which obtained full-disk as well as partial-disk observations of the Sun in the wavelength range 0.5{--}4~nm with spatial and temporal resolutions of 2.5{\arcsec} and 4~s, respectively. Since then, jets and other transient features in the upper solar atmosphere have been regularly observed by various instruments on space missions, including the Solar and Heliospheric Observatory \citep[SOHO;][]{soho}, the Transition Region And Corona Explorer \citep[TRACE;][]{trace}, Hinode \citep{hinode}, the Solar TErrestrial RElations Observatory \cite[STEREO;][]{stereo} (composed of two spacecraft, A and B; since 2014 only STEREO-A is operational), SDO \citep{2012SoPh..275....3P}, the Interface Region Imaging Spectrograph \citep[IRIS;][]{iris}, and most recently with Solar Orbiter \citep[SO;][]{2020A&A...642A...1M}. 

SOHO provided multiwavelength imaging and spectroscopic observations of different types of transients in observations recorded by the Extreme Ultraviolet Imaging Telescope \citep[EIT;][]{eit}, the Coronal Diagnostic Spectrometer \citep[CDS;][]{cds}, the Solar Ultraviolet Measurements of Emitted Radiation \citep[SUMER;][]{sumer} spectrometer, and the Large Angle and Spectrometric Coronagraph \citep[LASCO;][]{lasco}. EIT observed the solar disk using four different filters, at 171~{\AA}, 195~{\AA}, 284~{\AA}, and 304~{\AA},  with a spatial resolution of about 5.0{\arcsec}. It covered a temperature range of 0.5~MK to 3~MK in quiet conditions; during flares, it also captured plasma at the formation temperature of the Fe~{\sc xxiv}~192~\AA\ line \citep[][]{Trip_EIT}. In the regular mode of observations, EIT captured full-disk images of the Sun with a cadence of 12~min in 195~{\AA}. The three instruments on board LASCO cover different fields of view: LASCO/C1 (1.1 to 3.6~R$_{\odot}$), LASCO/C2 (2.5 to 4.5~R$_\odot$), and LASCO/C3 (3.6 to 32~R$_\odot$). While LASCO/C1 became non-operational relatively early in the mission, C2 and C3 continue to provide continuous white-light solar corona observations. SOHO also carried two spectrometers: CDS covered the wavelength range of 150 to 800~{\AA}, and SUMER covered the wavelength range from 500 to 1610~{\AA}, sampling plasmas from chromospheric to coronal (including flare) temperatures. 

The Transition Region And Coronal Explorer (TRACE) was launched in 1998 to perform high-resolution imaging of the solar transition region and corona using different filters:  171~{\AA}, 195~{\AA}, 284~{\AA}, 1216~{\AA}, 1550~{\AA}, 1600~{\AA}, 1700~{\AA} and 5000~{\AA}. TRACE observed a limited field-of-view (FOV) with a spatial resolution of 1{\arcsec} and varying cadence depending on the mode of observations. Dynamic features, such as jets at various spatiotemporal scales, were primarily studied in images taken with the 171~{\AA}, 195~{\AA}, and 284~{\AA} filters. The temperature coverage of TRACE was similar to that of EIT, including the high-temperature coverage during flares \citep[e.g.,][]{Trace_1, Trip_EIT}.

Hinode and STEREO-A (STEREO-B was decommissioned in 2018) were launched in 2006 and remain largely operational to date. Hinode carries three instruments: the X-ray Telescope \cite[XRT;][]{xrt} and the Extreme-ultraviolet Imaging Spectrometer \cite[EIS;][]{eis}, which observe the transition region and corona with high spatial and temporal cadence in a broad range of temperatures; and the Solar Optical telescope \citep{2008SoPh..249..167T}, which included the Narrowband Filter Imager (NFI) and  Broadband Filter Imager (BFI)(no longer operational since February 2016), and the Spectro-polarimeter (SP) that records all four Stokes I, Q, U, V polarimetric spectra. While XRT provides images of solar phenomena in the soft X-ray band from 6 to 60~\AA\ (T $\sim$ 1--10 MK) with spatial resolution of 2{\arcsec}, EIS samples the transition region and corona in spectral lines from two wavelength ranges,  170{--}210~{\AA} and 250{--}290~{\AA}, that cover a broad range of temperatures (Log(T) $\approx$ 5.8{--}6.7~MK) allowing measurement of plasma properties such as electron temperature, electron density, and Doppler velocity in the observed events.  Hinode/SOT has been used to study small-scale transient events in the chromosphere. Transients have also been reported using the observations recorded by the Sun-Earth Connection Coronal and Heliospheric Investigation \citep[SECCHI;][]{secchi} on board the two STEREO spacecraft, mainly with the Extreme Ultraviolet Imager (EUVI) that observes the Sun in 171 and 195~{\AA} with a spatial resolution of 1.2\arcsec\ and at a cadence as high as 2.5~min.  

Most studies of transients thus far have been conducted using observations recorded by SDO/AIA. The AIA observes the transition region and corona of the Sun using seven EUV filters (centered around the 94, 131, 171, 193, 211, 304, and 335~{\AA} wavelengths) and two UV filters (1600 and 1700~{\AA}), with a spatial resolution of about 1.2{\arcsec}. In EUV channels, AIA provides a cadence of 12~s, whereas in the UV channels, the cadence is approximately 24~s. The near-simultaneous observations in six EUV filters help diagnose the plasma properties and derive the detailed thermal structure of transients at many spatiotemporal scales, which were quite sporadic with earlier observations. AIA covers a broad temperature range from 50\,000~K to $\approx$5~MK using eight filters. During flares, it can register emission from plasma as hot as 20~MK in the Fe~{\sc xxi} line of the 131~{\AA} filter \citep{aia_temp}.

While AIA observations have covered jets in the upper transition region and the corona, IRIS, launched in 2013, has opened a window to unprecedented high-resolution observations of transients in the chromosphere and lower transition region. IRIS observes in three wavelength ranges: 1332{--}1359~{\AA} and 1390{--}1407~{\AA} in far UV and 2782.56{--}2833.89~{\AA} in the near UV. It provides both images and UV spectra with very high spatial (0.33{--}0.4{\arcsec}) and temporal (1{--}2~s) resolution. The IRIS observations, combined with AIA and EIS observations, have provided a wealth of information on the thermodynamic properties of the transients. Small-scale transients at coronal temperatures are also observed with very high spatial (0.16{\arcsec}) and temporal ($\approx$~1~s) resolution using the 174~{\AA} filter of the Extreme-ultraviolet Imager (EUI) \citep{2020A&A...642A...8R} on board SO when the spacecraft is at perihelion. 

Aditya-L1 \citep{aditya} provides full-disk imaging of the Sun's chromosphere from the Solar Ultraviolet Imaging Telescope \cite[SUIT; ][]{suit1,suit2} in the wavelength range 2000{--}4000~{\AA} using 11 filters, including Mg~{\sc ii}~h\&k and Ca~{\sc ii}~K. Aditya-L1/SUIT enables the investigation of the plasma properties of the cooler counterparts of jets. SUIT data are now publicly available on https://pradan1.issdc.gov.in/al1/. In the region-of-interest mode, it provides images with a pixel size of  0.7\arcsec\ and a time cadence of 30~s, depending on the observation modes.

In addition to space-based telescopes, several ground-based observatories observe transients in photospheric and chromospheric lines. These include the Big Bear Solar Observatory (BBSO, USA), Udaipur Solar Observatory (India), the Swedish Solar Telescope (SST, Sweden), the Gregor Telescope (Germany), THEMIS (France and Spain), the New Vacuum Solar Telescope (NVST, China), and the Daniel K. Inouye Solar Telescope (DKIST, USA).

Complementary observations of the photospheric magnetic field underlying transient sources are crucial for understanding the physical mechanisms responsible for these events, as well as the extrapolated magnetic structure connecting these sources to the heliosphere. The Michelson Doppler Imager \citep[MDI;][]{1995SoPh..162..129S} on board SOHO, HMI on board SDO, and the Polarimetric and Helioseismic Imager \citep[PHI;][]{2020A&A...642A..11S} on board SO have provided magnetograms with increasing resolution and sensitivity, while DKIST ultimately will yield long-awaited coronal magnetic-field measurements. BBSO's Goode Solar Telescope (GST), SST (La Palma, Spain), and GREGOR (Tenerife, Spain), complemented by THEMIS  (Tenerife, Spain), currently measure photospheric fields in a small FOV with high spatial and temporal resolution.

\section{Large-scale Structures and Phenomena of the Solar Corona and Their Dynamics}
\label{sec:large-Bfield}
SBs evolve, and perhaps form, in a medium (the solar wind), that is inhomogeneous and time-dependent. Beyond the overall radial gradients due to solar wind expansion,  the inhomogeneous nature of the solar wind is rooted in the non-uniformity of the solar coronal magnetic field. Therefore, understanding SB properties and their possible formation mechanisms requires a thorough understanding of the large-scale structures that modulate the corona and its underlying magnetic field. Table~\ref{tab:table_largescale} summarizes some key properties of the main features of the large-scale magnetic field.

\begin{figure*}[!ht]
\includegraphics[width=1.0\textwidth]{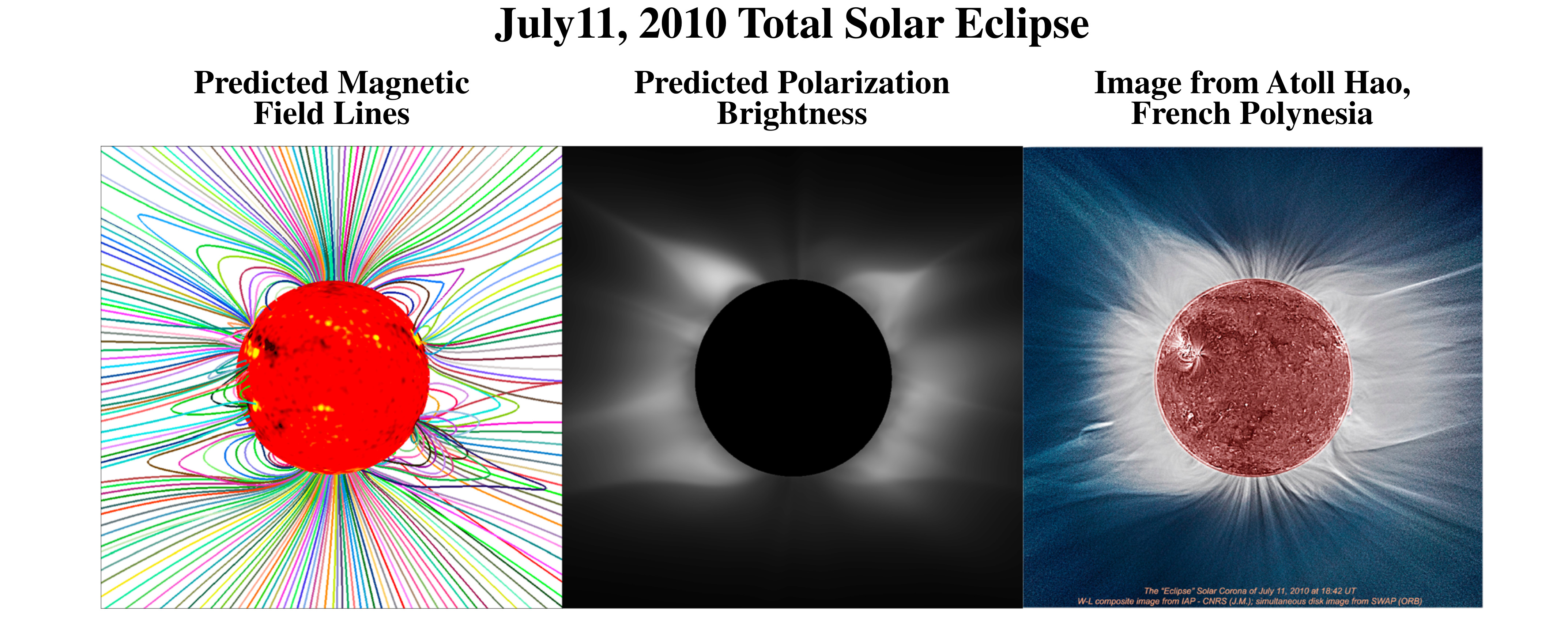}
\caption{Global structure of the solar corona. The left panel displays a global MHD model of the corona on 11 July 2010, along with its predicted polarized brightness (middle panel). The right panel displays a white-light observation of the corona taken during the total solar eclipse on 11 July 2010, revealing large-scale open and closed magnetic structures together with a solar disk from the Sun Watcher using the Active Pixel System detector and Image Processing (SWAP) instrument on board the Proba-2 mission. Images reproduced with permission from \citet{MacNeice18}, copyright by AGU}
\label{fig:LargescaleB}
\end{figure*}

\begin{sidewaystable}
\caption{Topological structure of the large-scale coronal magnetic field.}
\label{tab:table_largescale}
\begin{tabular}{lllll}
\hline 
Structure & Timescale & Size (in Lat/Lon$^\circ$)  & Topology & Relevant dynamics \\
\hline
Polar CHs & Almost\footnotemark[1] permanent & A few tens & Open magnetic field; & Source of fast solar wind; \\
& & & & \\
Helmet streamer & Solar-cycle dependent   & Several tens& Separate open  & Base of heliospheric \\
 &  &   &  and closed field; & current sheet;\\
 &  & & open field of opposite  & Interchange reconnection  \\
  &  &  & sign on each side; &  at its boundary; \\
 &  & & & Helmet streamer blobs; \\ 
  &  & & & Dynamic structure; \\ 
 &  & & & Potential source  \\ 
  &  & & &  of slow solar wind; \\ 
& & & & \\
Low-latitude CHs; & Solar cycle dependent& Several to a few tens  & Open magnetic field  & Source of fast solar wind; \\
 & Days to month &  & &  \\
& & & & \\
Pseudostreamers & Week to a year & Several to a few tens & Separate open and closed field & Interchange reconnection; \\
&&& of the same sign; & Potential source\\
& & & Separate open field regions & of slow solar wind ; \\
& & &   of same sign;& (Pseudo-)Streamer blobs;  \\
& & & One 3D null point  & \\
& & & + at least two other patches; &  \\
& & &  3D nulls/bald patches\footnotemark[2] & \\
& & & Central 3D null has   &   \\
& & & a partly open fan;  & \\
& & & & \\
Fan-spine (anemone) configurations & Hours to days & Fraction to several & One 3D null point & Jet-like events  \\
& & & Fully closed fan surface &  \\
\hline
\end{tabular}
\footnotemark[1]{During solar maximum, polarity reversal takes place,  which can cause the polar CHs to shrink, become less distinct, or even disappear temporarily \citep{2009LRSP....6....3C,2015LRSP...12....5P}.}
\footnotemark[2]{Locations where field lines with U-shaped topology are tangent to the photosphere}
\end{sidewaystable}

Open and closed magnetic fields provide the structure of the solar corona as demonstrated by coronagraph and eclipse observations (right panel in Fig.~\ref{fig:LargescaleB}). Closed structures appear brighter, while open structures tend to be dimmer because of their lower density and are more radially striated. The open magnetic-field regions make up CHs, identified most easily in EUV and X-ray images as areas with reduced emission. During solar minimum, CHs are primarily observed in the polar regions. However, low-latitude and equatorial CHs appear frequently during solar maximum, especially during the declining phase of the solar cycle. The brighter closed-field regions with cusp-shaped structures make streamers. These streamers are of two types: helmet streamers and unipolar streamers, or pseudostreamers.  Helmet streamers mark the large-scale boundaries separating the magnetically open and closed solar atmosphere, while pseudostreamers, usually presenting better-defined stalks extending lower down in the corona in eclipses, are surrounded by an unipolar field.

\subsection{Helmet Streamers} \label{subsec:streamers}
Helmet streamers are structures composed of magnetic loops bounded by open fields of opposing polarities in each hemisphere. ``Rays'' or ``stalks'' extend from the top of the streamers, delineating the heliospheric plasma sheet surrounding the heliospheric current sheet \citep[HCS; e.g.,][]{2022ApJ...933...95K}. These streamers are also known as bipolar streamers \citep{Pneuman71}. The cusps at the top of the helmet streamers mark the base of the HCS. Helmet streamers expand continuously due to photospheric flux emergence and added magnetic stress resulting from footpoint motions driven by subsurface convection. The interplay between gas dynamics and the magnetic field at the inverted Y-type null line at the cusp causes instability, producing ``streamer blobs", which are characterized as density variations with large azimuthal (longitudinal) extents and radial scales of up to 12~{\rs} at 30~{\rs} \citep[e.g.,][]{1997ApJ...484..472S,SanchezDiaz2017a}. The largest blobs are released in the HCS quasi-periodically at  8{--}16-hour intervals \citep[e.g.,][]{1997ApJ...484..472S, 2009ApJ...694.1471S, SanchezDiaz2017a, SanchezDiaz2017b}. Recent observations from the  Wide-Field Imager for Solar Probe Plus \citep[WISPR;][]{2016SSRv..204...83V} on board PSP
show a frequency of $\approx$3~blobs a day beyond 15~{\rs} \citep{Liewer2024}, consistent with earlier estimates from LASCO data \citep{Wang2000}. These blobs are generally interpreted as magnetic flux ropes that form dynamically at the tip of helmet streamers via pinching-off magnetic reconnection \citep{Higginson18, Lynch2020, Reville2020, Reville2022}. However, the frequency of these large-scale pinched-off blobs is well below what is necessary to credibly account for most SBs. 

In addition to the large blobs, helmet streamers also produce smaller structures with a quasi-periodic nature. Solar-wind density fluctuations identified in in situ data at 1~AU \citep{Kepko2003,Viall2010} are omnipresent in STEREO-A SECCHI/HI1  white-light images of streamer outflows \citep{2018ApJ...862...18D}. About 30\% of the solar wind measured in the ecliptic comprises Periodic Density Structures (PDSs) with periods ranging from tens of minutes to 2{--}3~hours \citep[][]{Viall2009}. PDSs with length scales greater than 600~Mm have been measured inside the HCS \citep{Kepko2016, DiMatteo2019}. It has been suggested that PDSs are produced at the base of the corona \citep{Ofman2000} or near the tips of helmet streamers \citep{Kepko2016, SanchezDiaz2017b, Kepko2024}. PDSs at an 11\,000~Mm scale have also been observed in time-dependent MHD simulations involving tearing mode instabilities \citep{Reville2020, Poirier2023}. 

One possible source of PDSs is interchange reconnection near the top of helmet streamers, driven by supergranular evolution, which continuously changes the coronal field connectivity, modifies the large-scale boundary between closed and open flux, and intermittently releases slow-wind plasma along corridors far from the HCS \citep{Higginson17}. Torsional Alfv\'en waves with compressive components are other possible sources of these density fluctuations \citep{Higginson18}. Solar wind composition measurements suggest that the smaller PDSs (sizes from 100~Mm to 600~Mm) are most likely produced in the solar corona, possibly due to periodic reconnection driven by velocity fluctuations induced by the conversion of p-mode waves into transverse Alfv\'enic fluctuations \citep{Kepko2024}. It is speculated that such density fluctuations may evolve into SBs further out in the solar wind  \citep[see][]{chapter5}. Note, however, that SBs are defined by their magnetic properties and velocity--magnetic field correlations, most often with small or negligible density fluctuations, so that some or even most observed density structures may have little to do with the magnetic perturbations that could evolve into SBs.

Linking observed blobs (detected in remote-sensing observations) with PDSs (detected in situ) has been difficult because of the lack of observations in the corona from the outer limits of UV/EUV/X-ray imagers to the inner limits of coronagraphs. Presently, Metis \citep[][]{metis} on board SO provides coverage from 0.7 to 4~\rs\ at perihelion, while the Association of Spacecraft
for Polarimetric and Imaging Investigation of the Corona of
the Sun (ASPIICS) on board Proba-3 covers the region from 0.97 to 2~\rs\ (measured from the photosphere). However, some estimates can be made by combining solar wind observations and modelling.
\citet{2023A&A...676A.125P} performed idealized simulations of transient solar wind density structures launched from the streamer belt during the fourth solar encounter of PSP. The modeled density structures were released at a heliocentric distance of 5~{\rs}, consistent with observations of streamer blobs \citep[][]{1999JGR...10424739S}. Comparison between the simulation results and the actual PSP/WISPR observations during the same interval \citep{2021A&A...650A..30N} showed that 35--127 transient density structures were launched from the entire solar equatorial plane per day. Of these, only 4{--}5 were related to streamer blobs. These inferences, derived from the analysis of remote sensing PSP/WISPR observations, predict the in situ detection of density structures spanning sizes 2{--}8~{\rs} and covering 1{--}20~\% of the fourth PSP encounter, a prediction that awaits scrutiny against actual PSP in situ observations. Note that smaller-scale density fluctuations that are undetected by WISPR due to the imaging cadence are also always present in the in situ measured turbulent density spectra.

While the helmet streamers are always present, as mentioned above, more transient structures also contribute to the large-scale coronal magnetic field. At the smallest scales, the photospheric field is comprised of a myriad of opposite polarities, the global dipole emerging only at much larger heights. With the rise of the solar cycle, active regions (ARs)
and other magnetic concentrations emerge, creating a larger-scale multi-polar global magnetic field on the Sun. The presence of ARs distorts the equatorial polarity inversion line (PIL) and results in the formation of equatorial extensions of polar CHs. Complex magnetic-field concentrations contain multiple regions with large-scale gradients, and the mapping of fields from the corona into the solar wind is far from smooth. 3D magnetic null points, fan surfaces, spines, separatrices, and quasi-separatrix layers define the complex magnetic structure: a porous magnetic cage gradually opening up at larger heights. The complexity is often imaged in connectivity maps starting a few radii away in the solar wind, the separatrix web \citep[S-web;][]{Antiochos11,Antiochos12,Titov11} as measured by the squashing factor, further defined below. Null points and regions of separatrix-type layers have important consequences for coronal dynamics, as they are regions where even small stresses can form large currents, magnetic reconnection, and energy release. 

To summarize, while helmet streamers are one of the potential sources of SBs resulting from continuous interchange reconnection that produces density variations in the helmet streamer plasma, they will produce only a very small number of SBs. It should also be noted that helmet streamers do not trace the magnetic network where SBs are believed to emanate.

\subsection{Pseudostreamers}\label{subsec:psfstne}

Pseudostreamers, or unipolar streamers, are another large-scale feature of the corona appearing in globally unipolar regions above multiple polarity reversal boundaries and can be observed at the solar limb as thin bright rays  \citep{Wang07}. 
Like streamers, pseudostreamers are elongated structures surmounted by cusps, with pseudostreamers usually being smaller in height and width \citep{Wang07}. However, a fundamental difference between helmet streamers and pseudostreamers is their magnetic topology and longitudinal extent \citep{Wang07,Wang12,Titov12,Rachmeler14,Scott18}. Pseudostreamers are rooted in relatively large minority-polarity magnetic concentrations surrounded by an open field of a single majority polarity. They occur in various forms, with internal structures shaped by their formation history, the number of magnetic polarities in the underlying closed regions, and their stability against disruption by solar differential rotation or nearby flux emergence. They may also contain filament channels, often appearing as twin pairs. When present, these channels with their strongly sheared fields and skewed overlying arcades can strongly influence how the surrounding coronal field expands near pseudostreamers \citep{Titov12}. The filaments in pseudostreamers can erupt, sometimes producing narrow CMEs  \citep{Wang_2023}, but also triggering, when twin filaments exist, global eruptions \citep{Torok11}.
At solar minimum, pseudostreamers are primarily located at high latitudes. However, during solar maximum, they can be found at all latitudes where CHs are present. 

Similar to helmet streamers, high-cadence coronagraph images reveal that pseudostreamers show sporadic emissions of blobs and weak outflows contributing to the slow wind intermittency \citep{Crooker2012} and sometimes even CMEs \citep[e.g.,][]{2015ApJ...803L..12W,2021ApJ...907...41K, 2024ApJ...975..168W}. However, they exhibit lower variability in comparison to that of helmet streamers \citep[e.g.,][]{Wang07, Lee_2021L}. \citep{Crooker2012}. An analysis of near-Earth solar wind inferred to originate in pseudostreamers reveals the frequent presence of inverted radial magnetic fields without any change in the pitch angle of suprathermal electrons \citep{owens_solar_2013}, and as such may be considered SBs. These SBs have been attributed to interchange reconnection near the base of pseudostreamers between closed loops inside the pseudostreamer and the open magnetic field \citep{chapter5}.

In conclusion, pseudostreamers of various scales can be another potential source of SBs. However, they may not be a sufficient source of SBs. Further dedicated studies that combine coronal magnetic field modeling together with imaging and/or spectroscopic observations are needed to verify their role as SB precursors.

\subsection{Thermal Nonequilibrium in Helmet and  Pseudostreamers, and Fan-spine Topologies}
Thermal nonequilibrium (TNE) is a well-known pattern in plasma dynamics \citep{1991ApJ...378..372A,klimchuk_distinction_2019, antolin_thermal_2019} with significant implications for coronal heating that may also play a role in the dynamics of large-scale magnetic-field structures and, to some extent, in the modulation of SBs. TNE is a counterintuitive mechanism in which stable atmospheric conditions produce periodic or intermittent thermodynamic evolution of the coronal plasma. This highly nonlinear process can dominate the plasma dynamics of a coronal structure when heating is concentrated mainly at coronal loop footpoints and is quasi-steady or intermittent on timescales below the radiative cooling time of the ambient plasma. Such localized heating produces cycles of evaporation of the chromospheric plasma, which reaches coronal temperatures before catastrophically cooling and condensing part of the coronal plasma.
These cycles have typical periods of a few hours for steady heating (not so much for intermittent heating, though)  \citep[e.g.,][]{AntF_2022}, which is compatible with the timescales of SB patches \citep[e.g.,][]{ShiPV_2022}. As a result, loops undergoing TNE contain discrete regions at a temperature one to two orders of magnitude lower than the surrounding corona and densities raised by the same amount.

\begin{figure*}[!ht]
\centering
\includegraphics{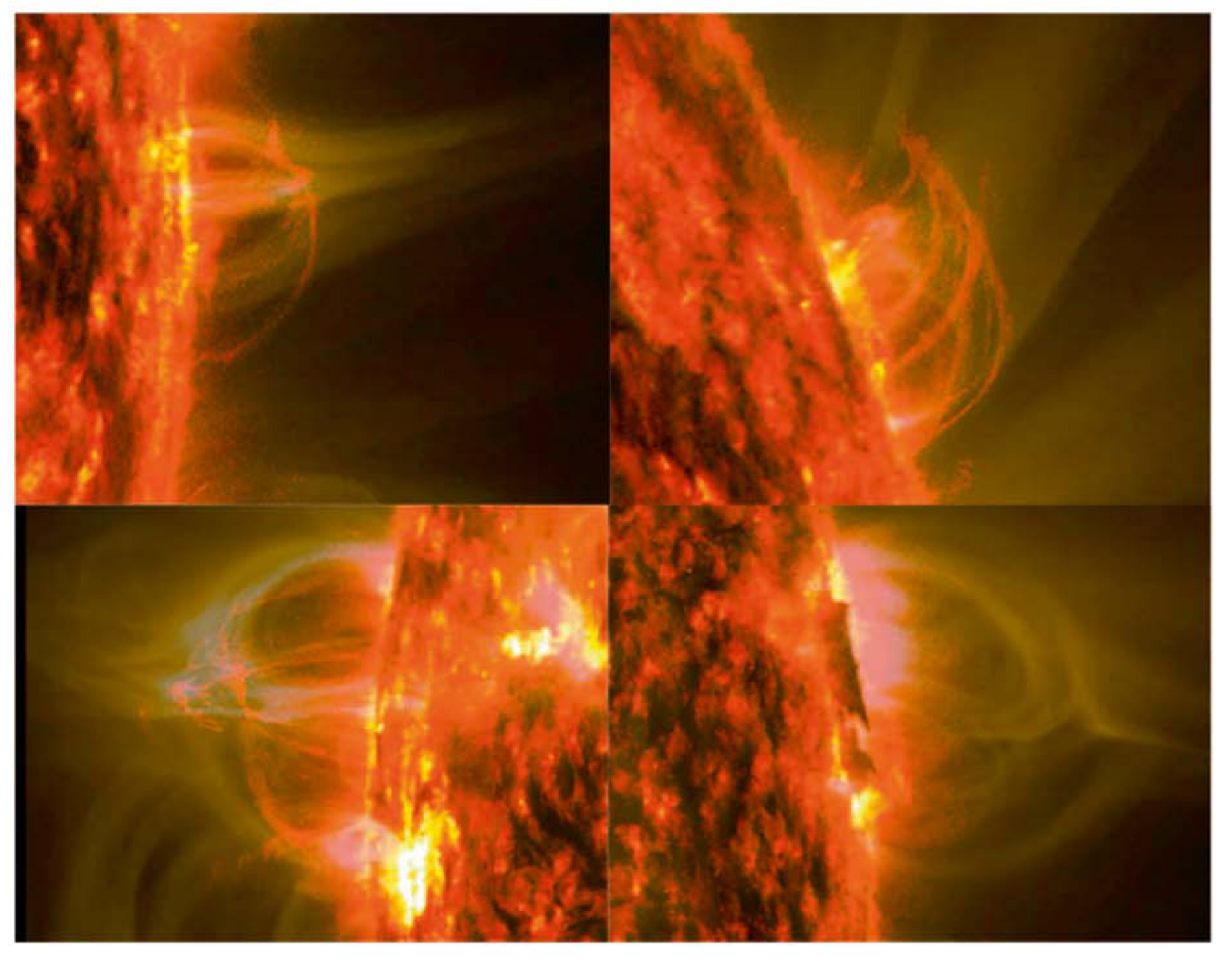}
\caption{Coronal rain observations in fan-spine topologies and a pseudostreamer. The observations (composite image taken in the SDO/AIA 171 and 304~{\AA} channels) were obtained on 16 April 2015 (top left), 5 May 2015 (top right), 13 February 2015 (bottom left), and 25 February 2016 (bottom right). Image reproduced with permission from \citet{mason_observations_2019}, copyright by AAS}
\label{mason_29}
\end{figure*}

Although TNE has been studied intensely in coronal loops and prominence magnetic fields, observational and numerical explorations of TNE in helmet streamers, pseudostreamers, and fan-spine topologies (also known as 3D nulls) are relatively recent. One phenomenon widely attributed to TNE is coronal rain, which is common in loops \citep{schrijver2001, muller_dynamics_2003, muller_dynamics_2004, antolin2010, antolin_thermal_2019, AntF_2022} as well as fan-spine configurations and pseudostreamers \citep{mason_observations_2019}. As shown in Fig.~\ref{mason_29}, coronal rain consists of discrete or elongated blobs of cool plasma forming and falling along coronal loops. Similar condensation events appear to be directly linked to reconnection between loops or between open and closed magnetic field \citep{li_coronal_2018, li_repeated_2019, li_relation_2020, chen_coronal_2022}. 

Of greater relevance to heliospheric phenomena is the effect of TNE on streamer blobs forming at the tips of helmet streamers. 2.5D MHD simulations have shown that TNE can significantly drive both plasma and magnetic-field dynamics of streamers, especially near their cusps where $\beta \sim 1$ \citep{schlenker_effect_2021}. At the open-closed boundary near the streamer top, these simulations show either an opening of the field lines into the solar wind or a reconnection pinch-off at the streamer apex that releases plasmoids into the solar wind. Hence, the interplay between TNE and the magnetic field could play a dominant role in the observed dynamics. This study also found that coronal rain develops in the closed loops of the streamer and falls toward the footpoints. Further investigation is needed to determine whether the interplay between TNE and interchange reconnection in these structures plays any role in modulating SB patches. 

\subsection{Structure and Dynamics of the S-Web}
\label{subsec:S-web}
\begin{figure*}[!ht]
\centering
\includegraphics[width=0.9\textwidth]{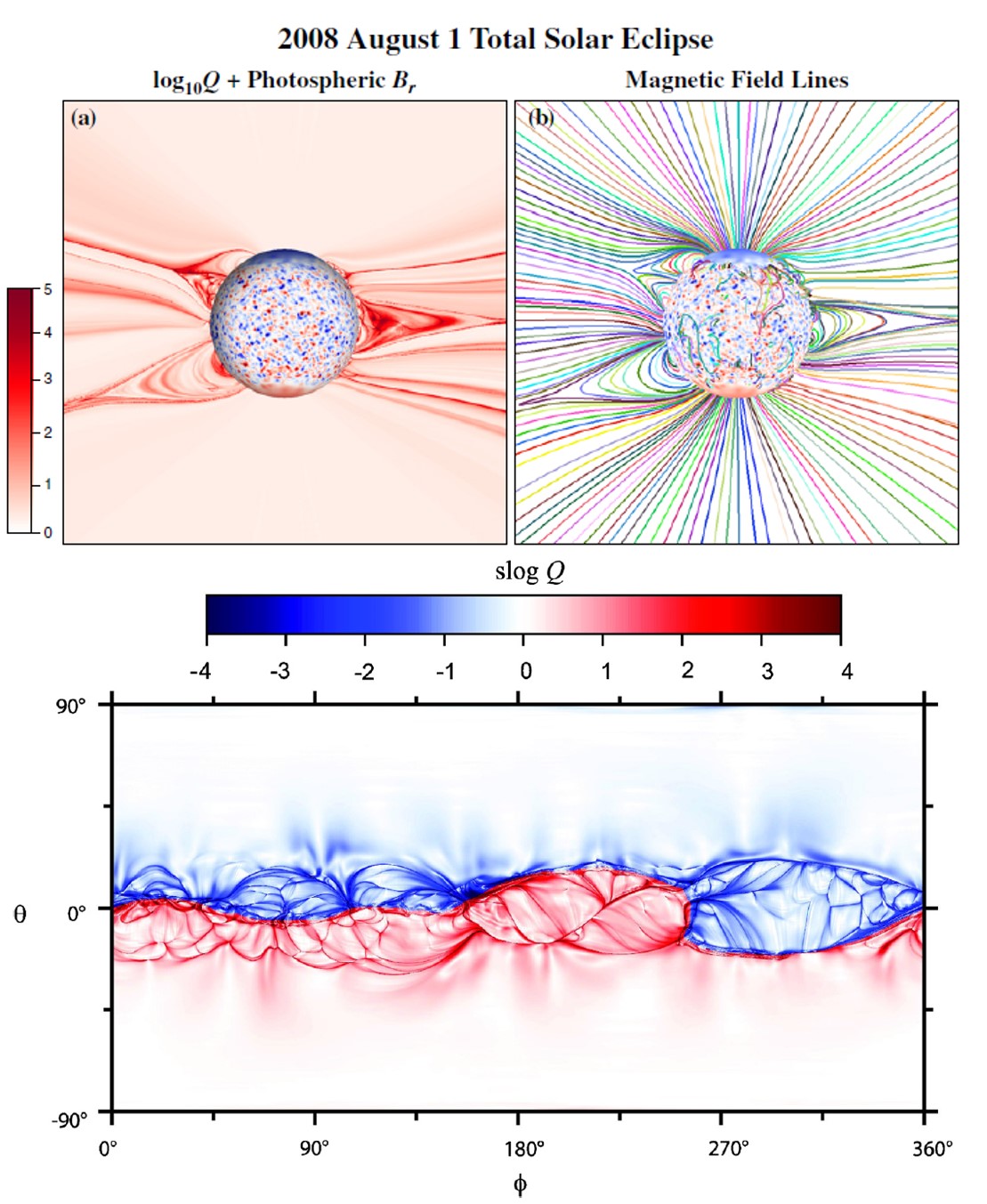}
\caption{Global coronal magnetic field and S-web structure. Observation-constrained 3D global MHD model of the large-scale magnetic field during the 1 August 2008 solar eclipse (top panel) and the S-web imprint (slog Q map) at $3 R_{\odot}$ (lower panel with adjacent color bar). High-Q regions correspond to the separatrices of the skeleton structure (e.g., the helmet streamer top at the interface between fields of opposite sign), while numerous weaker structures are the QSLs associated with narrow corridors of open flux connecting CHs. Image reproduced with permission from \citet[][top panel]{Antiochos11} and \citet[][bottom panel]{Titov11}, copyright by AAS}
\label{fig:LargescaleB_SWeb}
\end{figure*}

The global large-scale magnetic field is primarily structured by the above-described topological features: helmet streamers, pseudostreamers, and fan-spine configurations \citep{Platten14}. Static MHD models of the global coronal field are now reasonably successful in reproducing the white-light observations recorded during eclipses and by coronagraphs \citep[e.g.,][]{Rusin10, Mikic18,Yeates18,Lamy19}. These structures also provide a lowest order map of magnetic connectivity to the solar wind \citep{MacNeice18}.

The helmet-streamer PIL and the open fan separatrices of the pseudostreamers (and, to a lesser extent, the spines of the fan-spine structures) can be captured in maps of the connectivity \citep{Titov11} at a few solar radii away from the solar surface (see examples in Fig.~\ref{fig:LargescaleB_SWeb}). In addition to separatrices, numerous weaker structures in the corona from 1.5 to 6~R$_{\odot}$ are associated with quasi-separatrices \citep{Demoulin96a}, which are narrow regions of strong gradients in connectivity measured by the squashing factor Q \citep{Titov03}, or rather more typically the signed log version of it, slog-Q \citep[slog-Q map; ][]{Titov11}. These quasi-separatrices are due to the two tightly linked features of pseudostreamers and low-latitude CHs. The appearance of such CHs is constrained by magnetic topology, i.e., low-latitude CHs can only exist if they are connected to a polar CH by a narrow corridor of open field or if they are topologically connected to a pseudostreamer \citep{Antiochos07, Titov12, Scott18}. Although these corridors are possibly too narrow to be directly observable as dark lanes in EUV images of the low atmosphere, the flux systems on either side should display opposing directionality, yielding arcades similar to hair on either side of a part. Efforts should be made to observe definitive signatures of these corridors to connect the associated coronal topological features (dense web of separatrices and quasi-separatrices) that define the S-Web, and connect them more firmly to suggested evidence in the corona \citep[e.g.,][]{Chitta23}. The diverse narrow corridors connecting the low-latitude CHs all generate high-latitude arcs that appear in connectivity/slog-Q maps, such as in Fig.~\ref{fig:LargescaleB_SWeb} \citep[e.g.,][]{Higginson17,Scott18}. Because of the inherent complexity of the photospheric magnetic field, even at solar minimum, the S-web can be equally complex.  Since the separatrices and quasi-separatrix surfaces of the S-web are preferential sites for interchange magnetic reconnection \citep{Antiochos11, Linker11, Masson12b, Masson14, Aslanyan21, Wyper22, Pellegrin23}, the S-web may play a key role in generating the slow solar wind \citep{Antiochos12}, for which there is some observational evidence \citep[e.g.,][]{2022ApJ...933...95K, 2023ApJ...950...65B}.  

The S-web overall corresponds to the underlying topological features structuring the solar corona. At large spatial scales (tens of Mm), it is a slowly changing maze of topological structures (see Fig.~\ref{fig:LargescaleB_SWeb}, Fig.~2 in \cite{2023Sci...381..867C}, and Table~\ref{tab:table_largescale}). Though the time scales of the changes in the S-web are much longer than the durations of individual SBs, the S-web structure and its distribution can be particularly relevant to the statistical properties and the spatial distribution of SBs. The slow-changing S-web may be of importance as the host structure for smaller, more dynamical phenomena occurring within the width of the (quasi-)separatrices. It is important to note that, as of today, most S-web computations rely on PFSS models derived from synoptic maps in which smaller-scale magnetic features have been smoothed. Hence, the actual solar S-web shall actually be formed of even smaller and smaller QSL structures, which have, to this date, never been studied and/or tried to be determined because of the limitation of deriving a full-Sun PFSS solution using magnetograms at the actual best spatial resolution. To conclude, our understanding of the small-scale S-web, its properties, and its dynamics is simply nonexistent, even though its presence is strongly conjectured. It is the dynamics of this smaller-scale S-web that may be particularly relevant to the statistical properties and spatial distribution of SBs. 

S-web features are of importance because they are the required structures in some proposed formation mechanisms for SBs \citep{chapter5}. Indeed, intermittent transient mechanisms, such as interchange reconnection \citep{Higginson18}, torsional Alfv\'en waves \citep{Lynch14, karpen2017, Wyper22}, and reconnecting instabilities \citep{Gannouni23}, may develop relevant features in these large-scale structures and provide seeds for amplifying magnetic shears \citep[e.g.,][]{Toth2023}. Moreover, the velocity shears inevitably present at interfaces between different (quasi-)connectivity domains may be important for formation models based on Kelvin-Helmholtz and Rayleigh-Taylor instabilities \citep{Einaudi1999,2020ApJ...902...94R}. Therefore, on longer timescales, the modulation and dynamics of the large-scale structure of the coronal magnetic field can affect the generation and properties of SBs.

We should also envisage that the S-web is also (in addition to producing reconnection events) an extension of narrow corridors at the solar surface. Reconnection in these corridors produces outflows along the S-web, resulting in slow wind at high latitudes and across a range of longitudes.

To summarize, it is impossible to disregard, nor to confirm, that the  S-web structure and its distribution can be particularly relevant to the statistical properties and
the spatial distribution of SBs. Both the slow-changing large S-web structures and the yet-to-be-studied smaller-scale S-web QSLs may be of key importance
as the host structure for dynamical phenomena occurring within QSLs and channeling SBs from their origins.

\subsection{Structures and Transients in the Young Solar Wind: Striae and Flocculae}
 
\begin{figure}[!ht]
\centering
\includegraphics[width=0.28\textwidth]{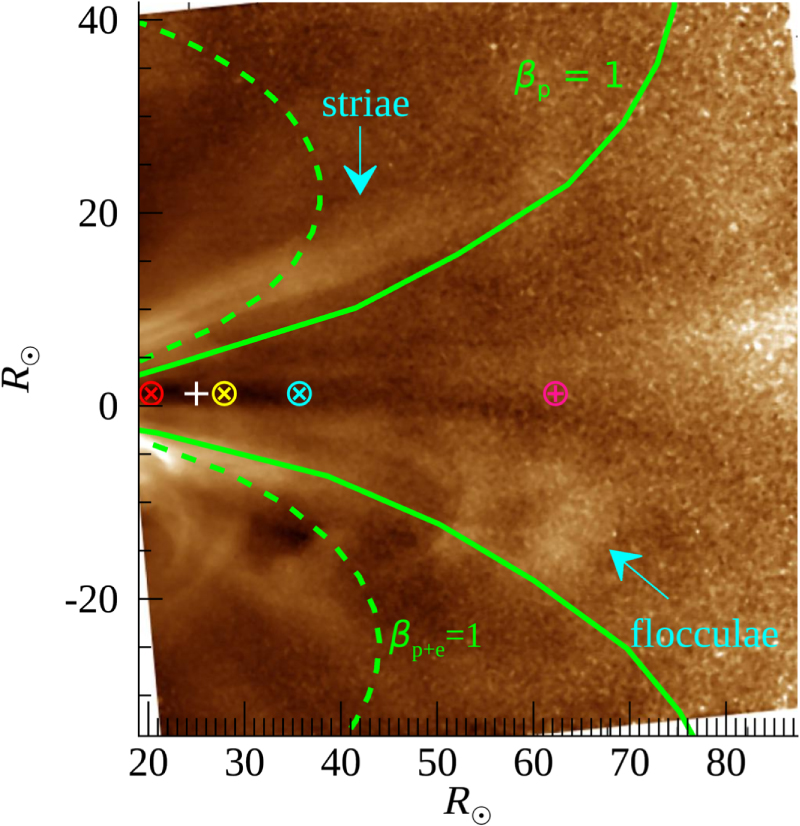}
\includegraphics[width=0.32\textwidth]{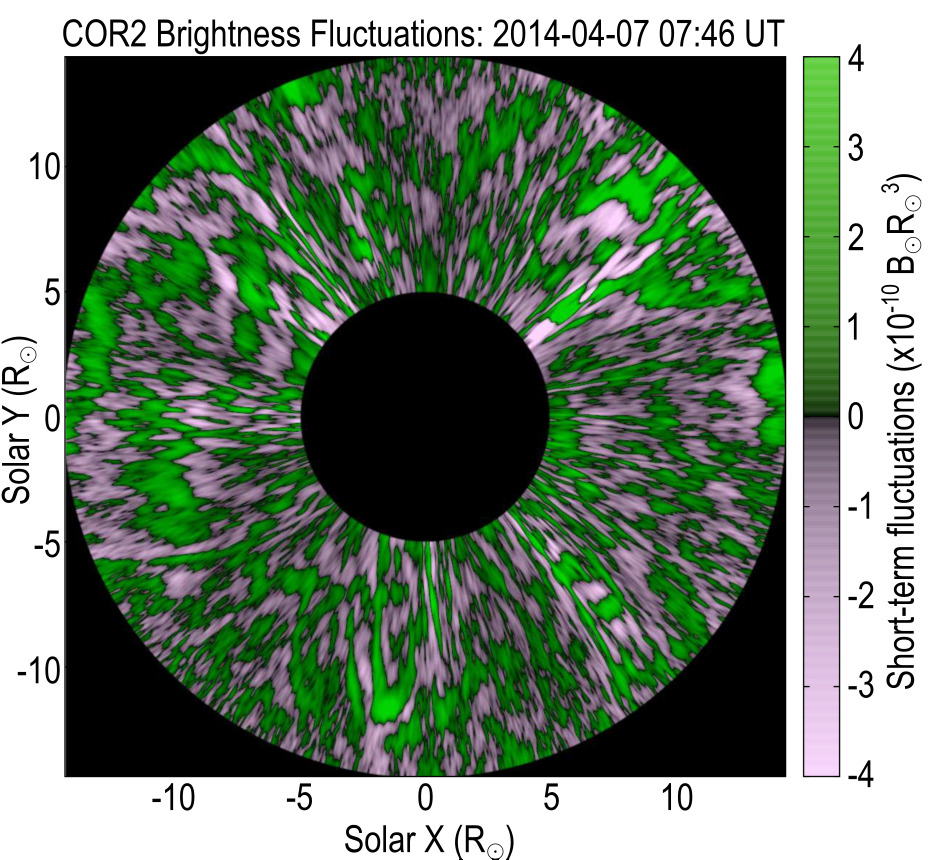}
\includegraphics[width=0.36\textwidth]{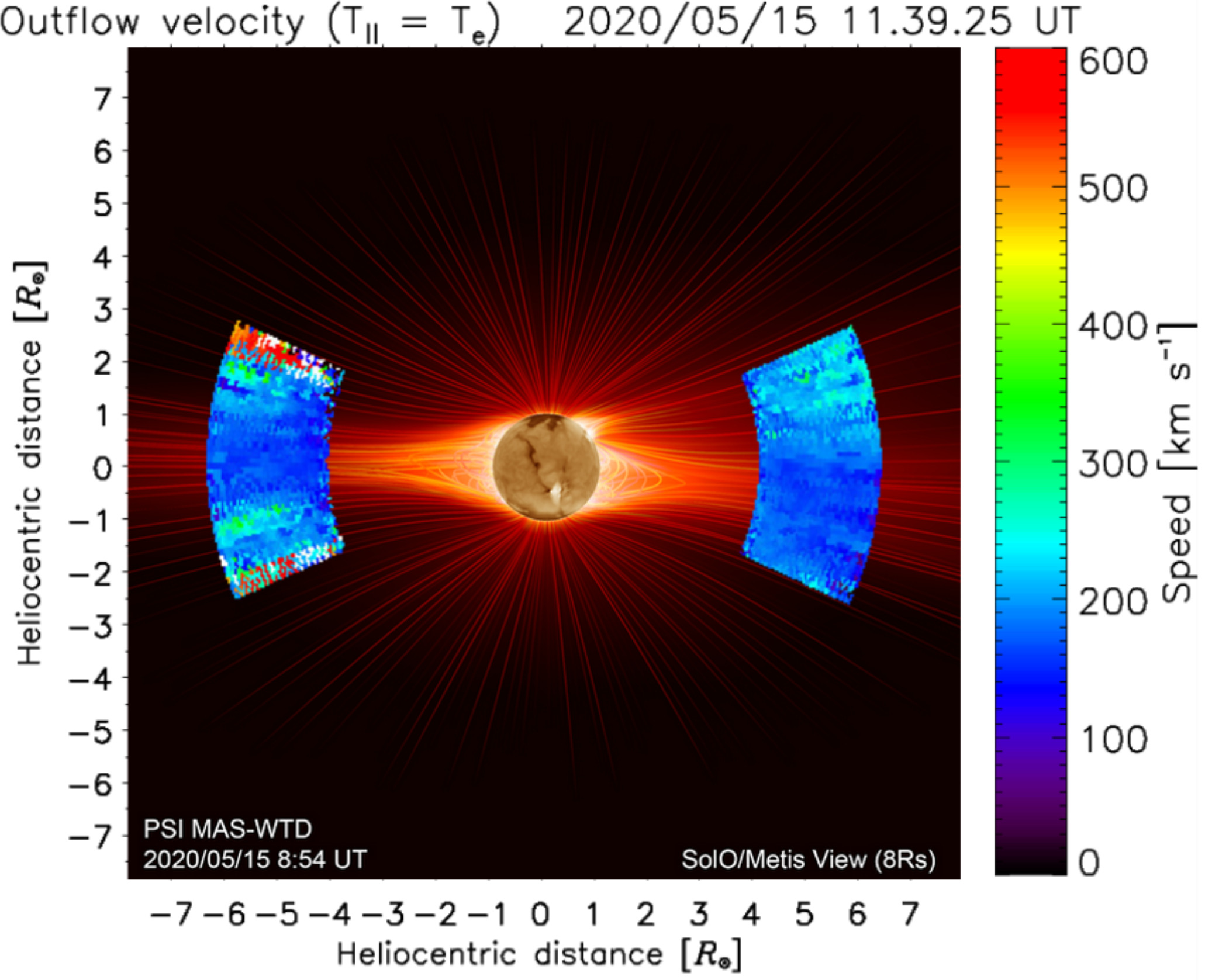}
\caption{
Coronal and heliosphere observations. Left panel: Dichotomy in the appearance of the outer corona/inner heliosphere in the STEREO/SECCHI/HI1  observations of ``striae" and  ``flocculae"  along with tracings of the plasma $\beta=1$ surfaces from a model \citep{2018ApJ...856L..39C}; Middle panel: Intermittent outflows observed in the outer corona by STEREO/SECCHI/COR2; Right panel: Outflow velocity maps in the outer corona from SO/Metis. Images reproduced with permission from \citet[][left panel]{2020ApJ...902...94R}, \citet[][middle panel]{2018ApJ...862...18D}, copyright by AAS and  \citet[][right panel]{2021A&A...656A..32R}, copyright by ESO}
\label{fig:outflow}
\end{figure}

\citet{2016ApJ...828...66D} analyzed highly processed STEREO/SECCHI/HI1 images during solar minimum conditions. They found that the coronal/inner heliospheric scene is characterized by two distinct types of structures: elongated, termed ``striae", and puff-like, termed ``flocculae" (see Fig.~\ref{fig:outflow}a). The striae experience a gradual dimming at apparent distances in the range $\approx$44--88~{\rs}, while the flocculae smoothly turn on in the range $\approx$29{--}75~{\rs}.  \citet{2016ApJ...828...66D} postulated that flocculae are formed in situ in the young solar wind and do not form in the corona, as transient outflows with similar morphologies, such as streamer blobs \citep[e.g.,][]{1999JGR...10424739S}. This transition in the appearance of the corona/inner heliosphere occurs more frequently and was attributed to the onset of large-scale turbulence in the young solar wind, giving rise to the flocculae \citep{2016ApJ...828...66D}. At the same time, the striae, which originate in the corona, start to fade away gradually. 

Numerical modeling attributes the striae-to-flocculae transition to the first plasma-$\beta$ equal unity surface in the open CH regions where solar-wind compressible dynamics may compete with magnetic effects (see Fig.~\ref{fig:outflow}, left panel) \citep{2018ApJ...856L..39C}. This surface is also close to the  Alfv\'en surface, where the solar wind becomes super-Alfv\'enic and once more loses its magnetic ``ties" with the corona \citep[e.g.,][]{2021PhRvL.127y5101K}. The formation of flocculae in the young solar wind is an element of near-Sun SB formation models \citep{chapter5} based on shear-driven instabilities such as the Kelvin-Helmholtz instability \citep[KHI; e.g.,][]{2020ApJ...902...94R}. Nonlinear Kelvin-Helmholtz-like instabilities and the resulting large-scale turbulence occur in regions of differential outflows of the nascent solar wind at the interfaces between solar wind streams with different speeds. As shown in Fig.~\ref{fig:outflow}, middle panel, STEREO/SECCHI/COR2 deep-exposure image sequences from  2 to 6~{\rs} reveal intermittent coronal outflows at multiple spatial scales \citep{2016ApJ...828...66D},  which could form the basis of the differential outflows. Likewise, SO/Metis outflow maps in the outer corona (Fig.~\ref{fig:outflow}, right panel) also show evidence of differential outflows \citep{2021A&A...656A..32R}. 
Numerical simulations of stream-stream interactions in the wind also exhibit KHI and the resulting evolution \citep{Parhi1999, Suess2000}. While the STEREO, SOHO, and SDO spacecraft are approximately 1~AU from the Sun, near-Sun vantage point observations by PSP/WISPR and SO/HI allow for sharper views of the flows and structures discussed in this paragraph. In addition, full-sky polarimetric observations by the recently launched Polarimeter to UNify the Corona and Heliosphere \citep[PUNCH,][]{punch} mission will provide a much-improved and comprehensive context of the texture in the corona and inner heliosphere.

While differential outflows and Kelvin-Helmholtz instabilities in the nascent solar wind have been observed, the latter only on the flanks of a CME  \citep[e.g.,][]{PaoSL_2024}, there is no clear association between these phenomena and the SBs in the solar wind observed by PSP. Therefore, further analysis combining remote sensing observations with those from in situ measurements must be performed to gain additional insights.

\subsection{Outflows from Open Flux Regions}
It was realized already in the Skylab era \citep{Zirker77} that high-speed (fast) solar wind with $v > 500$~\kms\ may be traced back to polar CHs. Despite extensive efforts to characterize the direct linkages, early studies could not progress beyond the fact that the fast wind originates in CHs and is transported along the open-field structure \citep[see review by][and references therein]{Sch_2006}. The sources of slow wind ($v < 500$~\kms) have been backtracked to the edges of ARs and CHs \citep[e.g.,][]{WanS_1990, Sch_2006, BroUH_2015, 2019Natur.576..237B, JanTM_2008, UpenC_2020}, as well as high-latitude S-web narrow areas (Sect.~\ref{subsec:S-web}), the latter and the former not being mutually exclusive. Rather, open flux at AR boundaries, in quiet Sun (QS) regions, and at the boundaries of large CHs could all be strongly dynamic with interchange reconnection and component reconnection \citep{Rappazzo12}, a prevalent phenomenon, and therefore, a source of fluctuations and potentially SBs.

\subsubsection{Outflows from Polar and Equatorial CHs}\label{subsec:ubiq_outfl}

Spectroscopic UV and EUV observations of CHs and QS from SUMER, EIS, and IRIS have revealed that CHs show persistent outflows across a range of temperatures \citep[e.g.][]{HasDL_1999,StuS_1999,StuS_2000,TuCM_2005,XiaM_2004,TriNS_2021}. Using SUMER observations recorded simultaneously in the chromosphere Si~{\sc ii}, transition region C~{\sc iv}, and at the base of the corona (Ne~{\sc viii}), \cite{HasDL_1999} showed that, while both QS and CH regions exhibited significant outflows along the network boundaries, diffuse blue-shifted emission (upflows) also appeared outside the network boundaries in CHs. In contrast, the QS patch (small-scale area of a few hundred square km) displayed red-shifted emission (downflows) outside the network boundaries. These observations suggest that the solar wind is very likely rooted in localized regions along the network boundaries at the coronal base.
  
A more comprehensive study of radiance, Doppler shifts, and spectral-line widths in lines with formation temperatures from 8000~K to 1.4~MK in CHs and QS found that the spectral lines with peak formation temperatures above 40\,000~K had reduced radiance, enhanced blue-shifted emission, and broader spectral-line widths in CHs than in the QS \citep{StuS_1999,StuS_2000,XiaM_2004}. However, at lower transition-region and chromospheric temperatures, the intensities were comparable \citep{KayTSP_2018, UpenT_2021}. The fact that CHs and QS regions are distinguished from one another at coronal temperatures but not in transition-region temperatures may be explained using loop statistics: while the number densities of small, low-lying loops in CHs and QS are similar, many more long, closed loops exist in QS than in CHs \citep{WieS_2004}. 

While \citet{HasDL_1999} directly observed sources of fast solar wind in the lower atmosphere above the network, a key question has remained: at what temperature or atmospheric height does the wind start? Another detailed study \citep{TuCM_2005}, which combined SUMER observations with the extrapolated coronal potential magnetic field, indicated that the solar wind starts at heights of 5{--}20~Mm above the photosphere. Based on the extrapolation, it was suggested that the closed loops in the transition region of CHs undergo interchange reconnection with open field lines, thereby leading to upflows. Such interchange reconnection may lead to Alfv\'enic fluctuations and precursors to SBs in the source of the fast solar wind.
\begin{figure}
    \centering
    \includegraphics[width=0.9\textwidth]{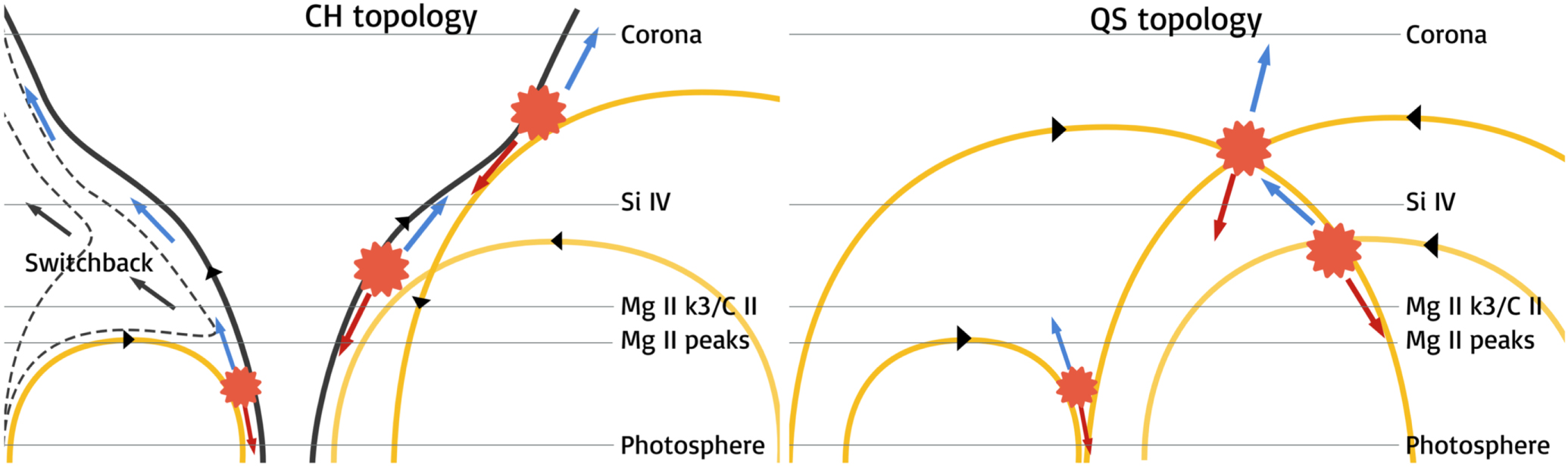}
\caption{Reconnection scenarios in CHs and the QS. Left panel: CH topology where interchange reconnection between open and closed field lines drives jets into the corona. Right Panel: In the QS, magnetic reconnection between closed loops forms other closed loops with swapped footpoints and brighter QS emission. Image reproduced with permission from \citet{UpeT_2022}, copyright by AAS}
    \label{fig:topo_qs_ch}
\end{figure}

\citet{TriNS_2021} analyzed the Si~{\sc iv}
line intensities, Doppler shifts, and non-thermal widths from IRIS observations of QS and CH regions, directly incorporating photospheric magnetic-field data. For identical magnetic flux densities, CH emission intensities were lower than in the QS. Both regions showed redshifted emission, implying downflows whose speeds increased with magnetic flux, though some CH areas displayed upflows with a similar scaling. Non-thermal velocities were comparable, indicating similar turbulence levels. These differences can be understood by combining the observations from \citet{TriNS_2021} with loop-statistics arguments \citep{WieS_2004}  and impulsive heating via magnetic reconnection: in the QS, reconnection likely occurs among short and long closed loops, whereas in CHs it occurs between short loops and open field \citep[see Fig.~\ref{fig:topo_qs_ch} and][]{UpenT_2021}, producing jets and possibly powering plumes and the solar wind. This scenario links transition-region signatures to solar-wind formation and unifies coronal heating in the QS and CHs. Transient CHs may also exist within the QS, though quantifying open flux there remains challenging due to limited magnetogram resolution; however, it represents a crucial step in assessing the global amount of interchange reconnection and its impact on heliospheric features, such as SBs. 
Studies including C~{\sc ii} and Mg~{\sc ii} h \& k lines \citep{UpenT_2021, UpeT_2022} and  2.5D MHD numerical simulations \citep{UpenT_2025} confirmed this framework, suggesting reconnection and impulsive heating occur at various heights, sometimes producing bidirectional chromospheric flows that may further drive the solar wind. Some analyses propose that interchange-reconnection-generated kinked field lines could appear as near-Sun SBs \citep[see e.g.,][]{Fisk2020, TriNS_2021,UpeT_2022, kumar_new_2023}. Magnetic reconnection is a multi-scale process and may occur between field lines threading an extended current sheet or at its boundary, with shear being converted into twist at different scales. MHD simulations tend to show that kinks rapidly straighten \citep{Landi05}, leaving instead transverse fluctuations in 2D and torsional Alfv\'enic fluctuations in 3D, propagating ahead of slower ejecta \citep[e.g.,][]{2009ApJ...691...61P,karpen2017,Uritsky2021,TouPF_2024} that lack the large deflections typical of SBs \citep[][]{Wyper22}. Thus, additional processes may be needed to form SBs from interchange-reconnection fluctuations \citep[e.g.,][]{2006GeoRL..3314101L,Toth2023,chapter5}.

\subsubsection{Outflows at Active-region Boundaries}
\begin{figure}
    \centering
      \includegraphics[width=0.8\textwidth]{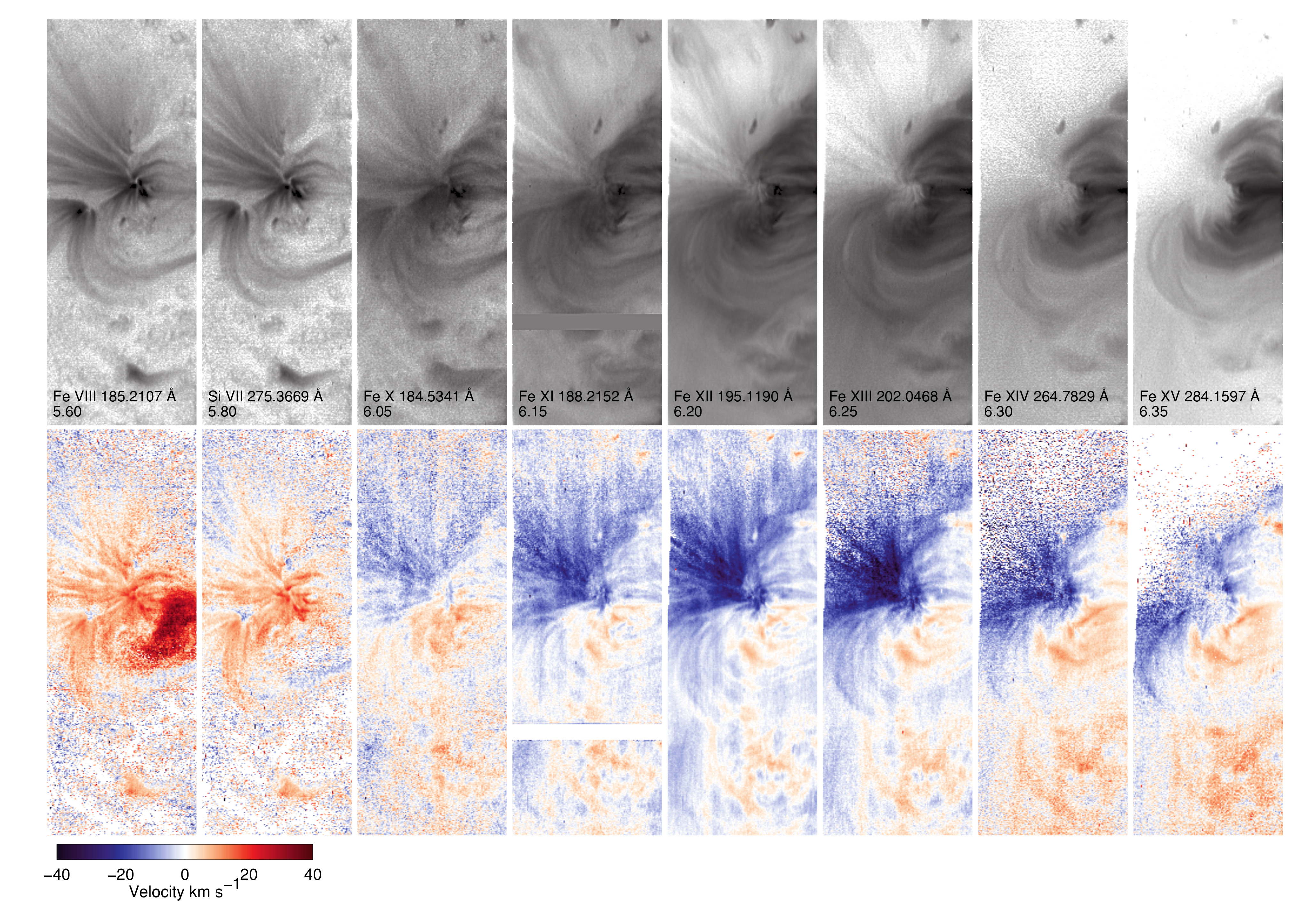}
    \includegraphics[width=0.8\textwidth]{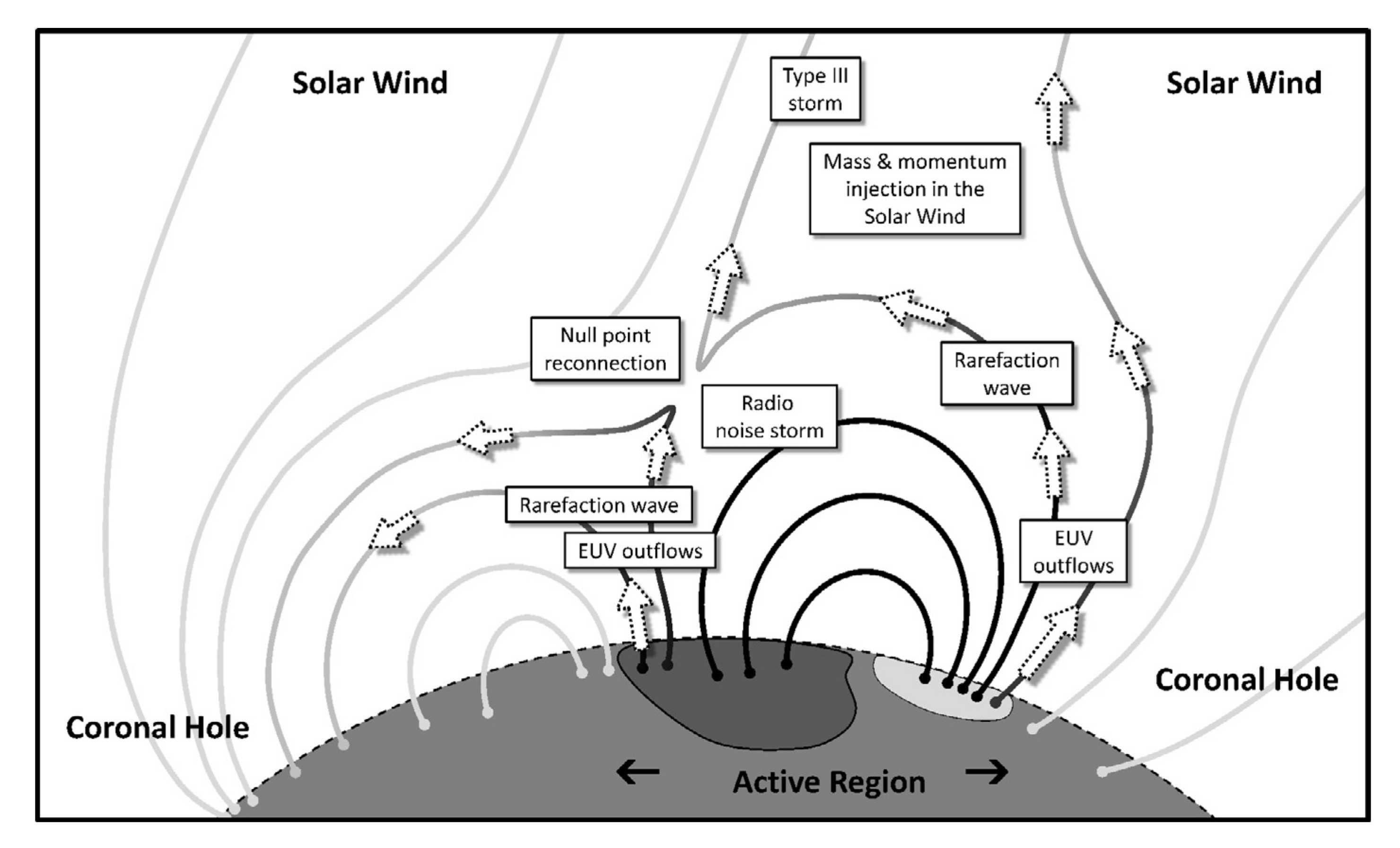}
    \caption{AR outflows. Top panel: Intensity (top row) and corresponding Doppler maps (bottom row) obtained in various spectral lines as labeled. Bottom panel: Possible scenario of the observed association between AR outflows and the formation of the slow solar wind. Images reproduced with permission from \citet[][top panel]{2011ApJ...727...58W} and \citet[][bottom panel]{2011A&A...526A.137D}, copyrights by AAS and ESO, respectively}
    \label{fig:ar_upflow}
\end{figure}

Interplanetary scintillation observations of slow wind streams during solar minimum show that in some instances their sources are associated with open magnetic field at the CH boundary rooted in one polarity of an AR \citep{1999JGR...10416993K}. Later soft X-ray observations by Hinode/XRT and spectroscopic observations from Hinode/EIS confirmed the existence of high-temperature ($>$1{--}3~MK) upflows (see top panel of Fig.~\ref{fig:ar_upflow}) from the edges of ARs with projected speeds between 50 and 200~{\kms} \citep[e.g.,][]{2007Sci...318.1585S, 2008A&A...481L..49D, 2008ApJ...676L.147H, TriMD_2009}. These upflows are possibly associated with magnetic reconnection occurring in quasi-separatrix layers (QSLs) or null-point locations \citep{2011A&A...526A.137D, 2011ApJ...727L..13B, 2017PASJ...69...47H}. By combining spectroscopic measurements with magnetic-field extrapolation and radio observations, \citet{2011A&A...526A.137D} put forward a unified scenario (see bottom panel of Fig.~\ref{fig:ar_upflow}) for the AR outflows, formation of the solar wind, and Type III radio storms \citep[see also][]{BraAD_2011}.

Another important piece of evidence on the association between solar wind and AR outflows was obtained by measuring the First Ionization Potential (FIP) bias. 
An increase by a factor of 2{--}4 of low-FIP elements (e.g., Mg, Fe, Si) compared to high-FIP elements (e.g., C, N, O) is a phenomenon known as the FIP effect.
Measuring the chemical composition in such outflow regions over five days using Hinode/EIS data revealed that the silicon abundance is always enhanced over that of sulfur by a factor of 3{--}4, consistent with the enhancement factor of low-FIP elements measured in situ in the slow solar wind \citep{2011ApJ...727L..13B}. However, in situ observations of slow wind composition also exhibit significant variability, which may make the FIP measurements alone an unreliable indicator of solar wind origin.

High spatial-resolution observations recorded by Hi-C combined with EIS spectroscopic measurements by \citet{2020ApJ...894..144B} indicate that the AR outflow emission may comprise two components. The first may come from locations of interchange reconnection between closed AR loops and adjacent open field, while the second may arise from dynamic activity in the upper transition region, as also reported by \citet{2021ApJ...918...33H}.
The first component would lead to an increase by a factor of 2{--}4 of low-FIP elements, whereas the second will have no excess of low-FIP elements. Such a scenario also explains the variable composition observed in the slow solar wind. Spectroscopic studies of the chromospheric and transition-region counterparts of high-temperature AR outflows also found upflows, suggesting a deeper origin for these outflows \citep{2020ApJ...903...68P,2021ApJ...918...33H}.  

  Hinode/EIS observations of low coronal outflows observed at some AR boundaries \citet{2023ApJ...950...65B} suggest that the narrow areas with superradial expansion at the edge of ARs predicted by the S-web theory (see Sect.~\ref{subsec:S-web}) are probable sources of slow solar wind streams characterized by short periods of moderate Alfv\'enicity and occurrence of SB events. The question remains whether the physical mechanism producing AR outflows can generate precursors of SBs. Given the ongoing interchange reconnection, this seems highly likely.

\begin{figure}[!ht]
\centering
\includegraphics[width=0.7\textwidth]{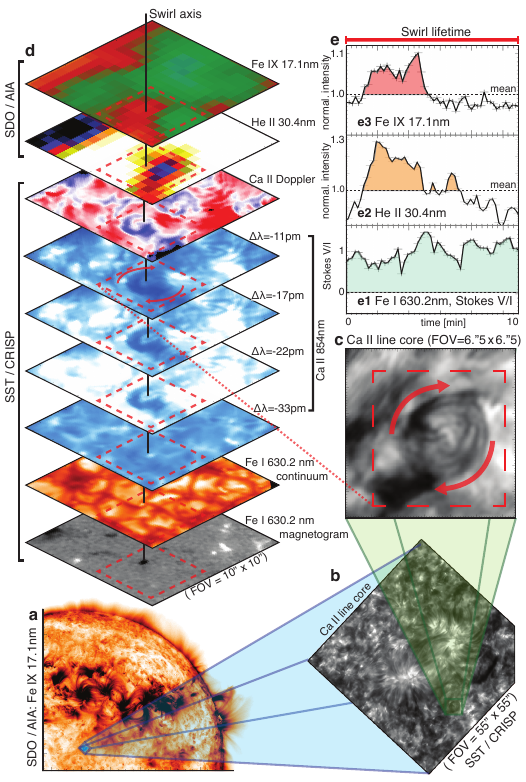}
\caption{Vortices/swirls in various layers of the solar atmosphere as recorded by SDO/AIA and SST/CRISP, as labeled. Image reproduced with permission from \citet{2012Natur.486..505W}, copyright by Springer}
\label{fig:vortex}
\end{figure}

\section{Dynamic Transient Events in the Low Solar Atmosphere}
\label{sec:transients}
\subsection{Small-scale Vortices} \label{subsec:vortex}
Solar observations \citep[e.g.,][]{2008ApJ...687L.131B, 2010ApJ...723L.139B, 2009AA...507L...9W, 2020ApJ...895...67S}, 3D numerical simulations \citep[e.g.,][]{2012AA...541A..68M, 2013ApJ...770...37K, 2016AA...596A..43C}, and relevant theory \citep[e.g.,][]{1975SoPh...42...79S, 1984A&A...140..453S} have revealed that the lower atmosphere should be awash with twisted and swirling small-scale flows \citep[see also][]{2023SSRv..219....1T}. 

Although the observed manifestations of vortex flow at different atmospheric heights may result from different drivers, the prevalent belief is that they are mainly magnetic in nature. Vorticity at granular scales mainly results from turbulent subsurface convection and its interaction with magnetic fields. Inside supergranules, subsurface convection appears on the surface in the form of bright granules with typical sizes of 1~Mm, whose boundaries form a network of dark intergranular lanes. There, the conservation of angular momentum carried by converging flows from neighboring cells of cooled plasma, which is redirected downwards and back into the convection zone, results in the formation of hydrodynamic intergranular vortex flows \citep[IVFs;][]{2014PASJ...66S..10W}. This formation process is commonly known as the ``bathtub effect" \citep{1994ASIC..433..471N}. However, strong magnetic flux concentrations often emerge within intergranular lanes and are observed as photospheric bright points (BPs). Thus, photospheric BPs co-spatial with an IVF are often forced into rotation. Since the magnetic field effectively couples different layers of the atmosphere, it transfers this rotation into the upper layers of the atmosphere, forming funnel-like spiraling magnetic structures known as atmospheric vortex flows \citep[AVFs;][]{Velli99, 2014PASJ...66S..10W}, also referred to as magnetic tornadoes -- though the latter terminology is ambiguous and has been applied, incorrectly, to quiescent prominence barbs as demonstrated by \cite{Panasenco14}. While the observable imprints of such AVFs in the chromosphere are identified as chromospheric swirls \citep{2009AA...507L...9W, Tziotziou2018AA}, their response has also been identified in the transition region and lower corona as rapidly rotating magnetic structures (see Fig.~\ref{fig:vortex}) \citep[][]{2012Natur.486..505W, Tziotziou2018AA}.

\begin{figure}
    \centering
      \includegraphics[width=0.9\textwidth]{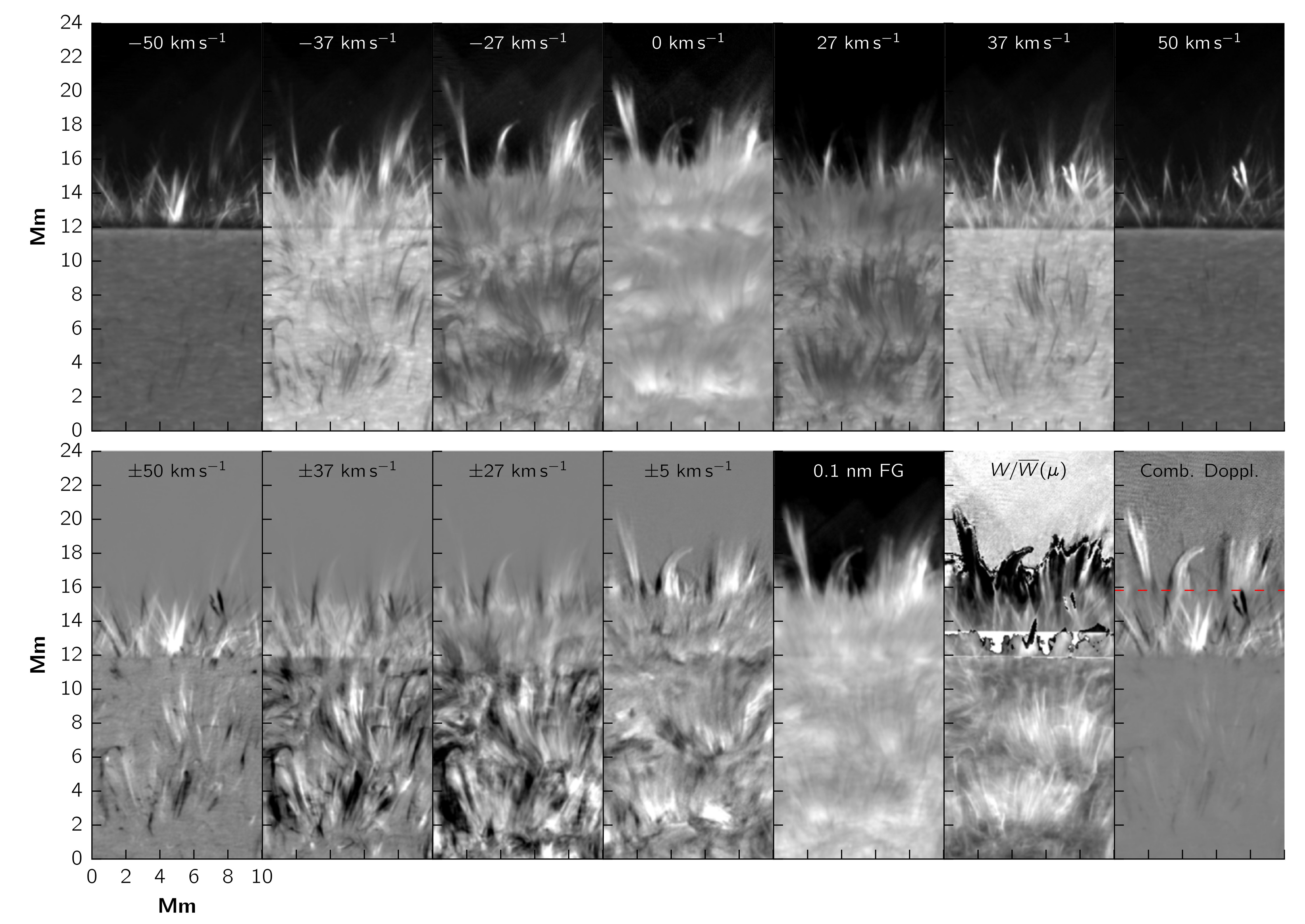}
    \caption{H$\alpha$ observation of spicules and fibrils or mottles. Top panels: H$\alpha$ intensities at $\pm$50~{\kms}, $\pm$37{~\kms}, $\pm$27{~\kms} and line center (0~\kms) as labeled. Bottom panels: from left to right, with labels shown at the top: Doppler velocity images at $\pm$50, $\pm$37~\kms, $\pm$27, and $\pm$5~\kms, filtergram with 0.1~nm Gaussian FWHM, H$\alpha$ line width divided by the mean width at each $\mu$ value, and combined Dopplergram. Image reproduced with permission from \citet{2016ApJ...824...65P}, copyright by AAS}
    \label{fig:spicules}
\end{figure}

Observations of such vortex motions in the photosphere and chromosphere have provided significant information about their radius ($\sim$0.5{--}2~Mm), average flow speed on disk (up to 2~\kms), line-of-sight speed ($<$8 \kms), lifetime ($\sim$10~min), space-time density (a few 10$^{-3}$~Mm$^{-2}$~min$^{-1}$), and total number at any given time (3$\times$10$^{-3}${--}10$^{-1}$~Mm$^{-2}$) \citep {2023SSRv..219....1T}. Observations and simulations have also revealed that small-scale vortex flows predominantly form in the quiet internetwork regions but not at the network boundaries \citep{2008ApJ...687L.131B,2014PASJ...66S..10W}, in contrast to jets and other potential contributors to SBs. These vortices essentially represent flows along and accompanying twisted magnetic flux, which can excite and transport various MHD waves, including different Alfv\'{e}nic modes such as kink waves, sausage modes, and torsional Alfv\'{e}n waves, as both observations \citep{Tziotziou2019AA, 2020A&A...643A.166T} and to a larger extent simulations \citep[e.g.,][]{2011ApJ...727...17F, 2011AnGeo..29.1029F, 2013ApJ...776L...4S, 2016GMS...216..431V} have suggested. Numerical estimates of the Poynting flux delivered at the upper chromosphere--TR boundary and up to the low corona range from $\sim$440~W~m$^{-2}$ \citep{2012Natur.486..505W} to a few thousand W~m$^{-2}$ \citep[e.g.,][]{2020ApJ...894L..17Y, 2021AA...649A.121B}, making vortex-related MHD waves a potential mechanism for energy and momentum transport into the upper atmosphere. 

With their braided, twisted magnetic-field configurations, vortices can naturally produce strong magnetic-field deflections, providing possible chromospheric seeds for SBs. These deflections could form through nonlinear wave steepening of the torsional waves that are prevalent in such vortical structures \citep[e.g.,][]{ 2021ApJ...911...75M}.  Moreover, they could also form from Rayleigh-Taylor-type instabilities in upward jets that lead to the formation of roll-ups and true full reversals of the local magnetic field and eventually to the formation of plasmoids, through reconnection.  

Concerning the `survival' of magnetic deflection associated with vortices, simulations indicate that such deflections formed at lower atmospheric heights or jet-like flows quickly unfold in the low corona and are unlikely to survive to the observed distance of SBs \citep[e.g.,][]{2021ApJ...914....8M}. Simulations of interchange reconnection also show that strongly twisted nonlinear Alfv\'en waves and kinked magnetic fields formed during reconnection rapidly untwist and straighten as they propagate away from the reconnection site \citep[e.g.,][]{Wyper22}. Detailed discussion of the survival of magnetic deflections generated by direct injection from convective motions can be found in Section~2.1 of \cite{chapter5}. Clearly, more research is needed to determine whether any SB seeds generated in the low solar atmosphere could aggregate and cascade to other scales without losing their magnetic deflection during their transit to the corona and into the solar wind.  Alternatively, some properties of SBs may develop or be further amplified by in situ evolution in the solar wind \citep[][]{2020ApJ...891L...2S, 2021ApJ...918...62M,2021ApJ...915...52S,chapter5}. 

Vortices have also been associated with jetting and outflows. The Lorentz force of the twisted magnetic flux could directly drive jet-like features \citep{2016ApJ...830..133K, 2017ApJ...848...38I}. Vortex motions may also generate magnetic substructures and, ultimately, shocks in merging flux tubes higher in the atmosphere that can drive thin, short-lived features interpreted as jets \citep{2018ApJ...857..125S}. Observations also suggest the generation of intergranular jets in the photosphere \citep{2011ApJ...736L..35Y}, which are associated with the formation and evolution of horizontal vortex tubes resulting from baroclinicity \citep{2023SSRv..219....1T}. Other simulations reveal that continuous accumulation of kinetic energy and magnetic and gas pressure around the temperature minimum layer in the lower atmosphere may lead to the formation of highly twisted ring-like structures \citep{2013ApJ...770...37K}. These structures could become unstable and produce spontaneous flow eruptions and associated Alfv\'en waves. 

As discussed above, however, small-scale vorticity is primarily generated within the solar internetwork. Therefore, it would be difficult to explain the proposed ``supergranular modulation" of SBs as a result of extensive vortex-generated SBs, if SB patches and the large-scale network are indeed linked \citep[e.g.,][]{Bale2021,Fargette2021}. If any individual vorticity-generated SBs finally survive and propagate in the solar wind, they could merely be a fraction or a subclass of the observed SB population. This conjecture has yet to be conclusively evaluated. 
\subsection{Spicules}
\label{spicules}

Spicules are elongated jet-like structures observed at or close to the solar limb, primarily in chromospheric lines \citep[see reviews by][]{ Beckers1968, Pasachoff2009, 2000SoPh..196...79S, 2012SSRv..169..181T}. The term spicule refers to different small-scale dynamic features (see Fig.~\ref{fig:spicules}), named differently depending on their appearance and locations, e.g., mottles {--} on-disk QS structures rooted in the network; fibrils {--} in AR plages \citep{Beckers1968, Bray1974}; straws \citep{Rutten2007}; macrospicules, which have some similarities to spicules, but it is not known whether
they have the same or a different driving mechanism \citep[][and refs. therein]{Yamauchi2004}; and rapid blue-shifted and red-shifted excursions \citep{Langangen2008, Rouppe2009}.  Apart from field-aligned flows, spicules exhibit oscillations, waves, swaying, and torsional motions, which implies the existence of magnetic twist and are suggestive of upward propagating Alfv\'{e}n waves \citep{DePontieu2007a, DePontieu2012, DePontieu2014, Suematsu2008, Zaqarashvili2009, Sekse2013,BasAN_2025}. 

The existence of two different populations with distinct observational characteristics, Type~I and Type~II spicules, has been suggested \citep{DePontieu2007, Zhang2012a}. Type~I spicules are short-lived structures (up to 20~min and typically 5{--}10~min) with mean velocities of the order of 20{--}30~{\kms}, heights of 5\,000--10\,000~km and widths around 1\,000--1\,500~km \citep{2012SSRv..169..181T}. They appear as small-scale jets, often following parabolic paths (suggesting propagation in magnetic loops) followed by downward motion \citep{Hansteen2006, DePontieu2007a}. Type-II spicules were first detected in  Ca~{\sc ii}~H~3968~{\AA} emission by Hinode/SOT/NFI. They are reported to have much shorter lifetimes (10{--}150~s), higher speeds ($\sim$100~\kms), and smaller widths of 150{--}700 km \citep{DePontieu2007, Pereira2012}, and mostly exhibit upward motion followed by rapid fading with a less-obvious falling phase \citep{2015ApJ...806..170S}. 

Type~II spicules have been linked to the so-called rapid blue and red excursions observed on the solar disk \citep{Rouppe2009}. Spicule-like structures are also observed in UV and EUV (e.g., by IRIS and SDO/AIA) that may result from heating of the chromospheric plasma to UV temperatures \citep{2014Sci...346A.315T}, thereby explaining their disappearance from chromospheric spectral lines \citep[see also][]{Klim_spi, TriK_2013}. Their EUV appearance has also been linked to the transition-region sheath surrounding chromospheric spicules \citep{Judge2008}. 

Type~I spicules are commonly believed to be driven by shocks generated by the leakage of magnetoacoustic oscillations from the photosphere to the chromosphere \citep[e.g.,][]{Sterling1990, DePontieu2004, Hansteen2006}. In contrast, Type~II spicules are thought to result from magnetic reconnection \citep{Sykora2009}. It has also been suggested that Alfv\'en waves might generate spicules \citep[e.g.,][]{Hollweg_1982, 1999ApJ...514..493K, 2017ApJ...848...38I}. Using the SOT/BFI observations, \cite{2010ApJ...719..469J} estimated that  2~$\times$~10$^7$ spicules exist on the Sun at any given time. This estimate is 20 times lower than that of \citet{1972ARA&A..10...73B}.

Though QS spicules form in abundance at the chromospheric network, the spicule mass largely falls back towards the Sun, in particular for type I spicules. Nevertheless, transverse/rotational fluctuations associated with both spicule types might propagate into the corona, providing a background of Alfv\'enic fluctuations and potential SB seeds. However, the survival of these fluctuations throughout the dense chromosphere and transition region is questionable, and dedicated modelling efforts are needed to explore this possibility. 

\subsection{Surges} 
\label{subsec:surges}

\begin{figure}
    \centering
      \includegraphics[width=0.8\textwidth]{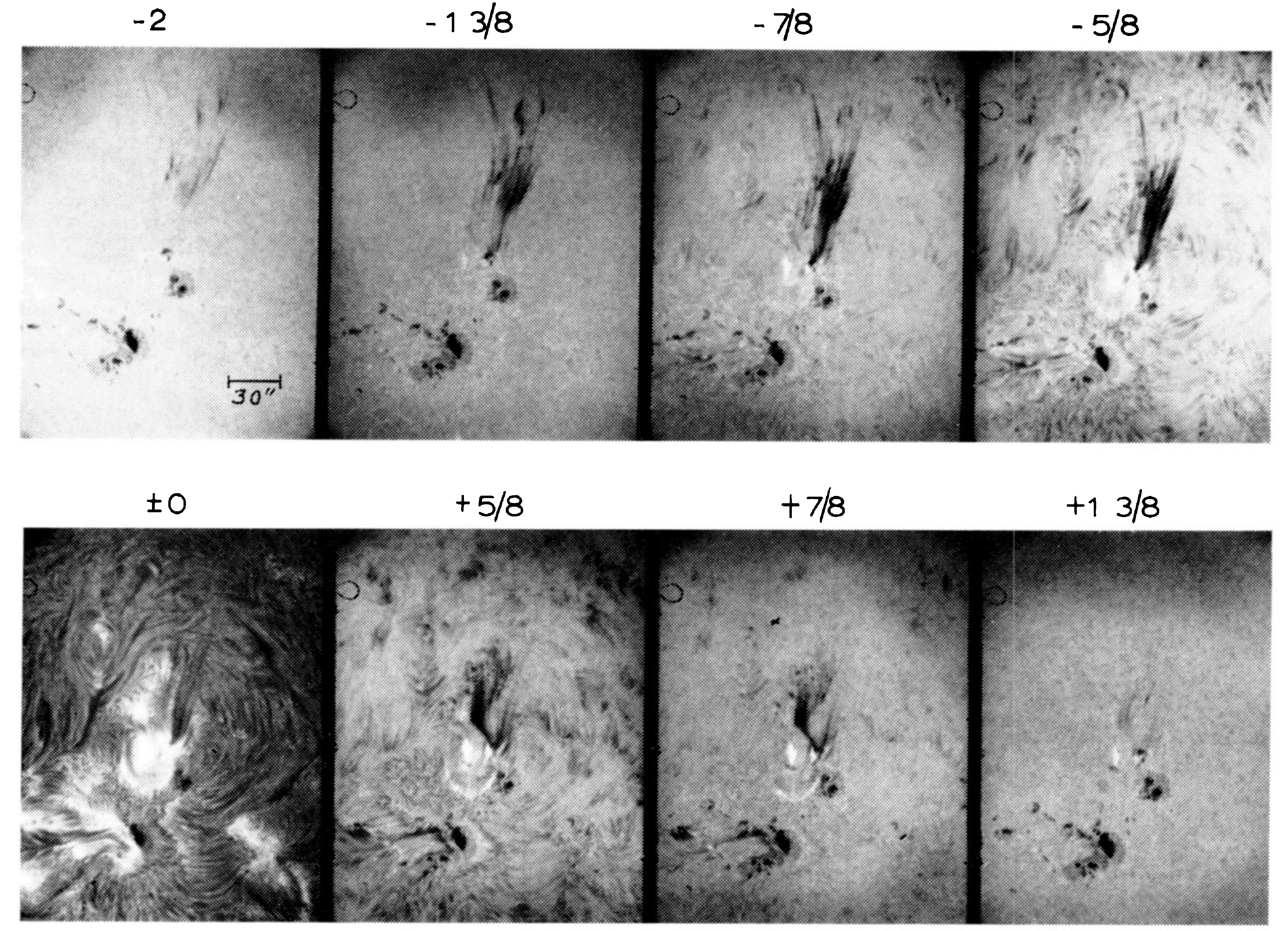}
    \caption{H$\alpha$ observations of a surge.  The eight-frame sequence is
obtained by stepping the filter bandpass across the H$\alpha$ profile through $-$2, $-1\frac{3}{8}$, $-\frac{7}{8}$, $-\frac{5}{8}$, $\pm0$, $+\frac{7}{8}$, $+\frac{5}{8}$, and $+1\frac{3}{8}$~\AA\ in 20~s. The sequence demonstrates the complex structure of surges. Image reproduced with permission from   \citet{1973SoPh...28...95R}, copyright by Springer}
    \label{fig:surge}
\end{figure}

The first observational description of surges dates back nine decades \citep{1937CMWCI.568....1M}, long before the start of the space age. Therefore, most of the early descriptions of these features are based on H$\alpha$ observations. Surges represent collimated flows of cool chromospheric plasma (see Fig.~\ref{fig:surge}), and their trajectories may be straight or slightly curved depending on the morphology of the magnetic field  \citep{1973SoPh...28...95R,Mac_surge}. They are ejected at speeds ranging from 20 to 50~{\kms}, reaching a height of up to $\approx$200~Mm. Within about 10{--}30~minutes, the surged plasma typically drains along the ascent path \citep{Bru_1974} and disappears. High-resolution observations of surges suggest that they consist of fine substructures \citep[thread-like features, e.g.,][]{2008AdSpR..42..803L,Li_surge}, where each substructure is connected to brightenings in its source region \citep[][]{1973SoPh...28...95R}.  Higher-cadence observations reveal their recurrent nature and other associated phenomena such as helical motions and oscillations \cite[e.g.,][]{2007A&A...469..331J, JibC_surge}. Surges are often associated with microflares seen as bursts of EUV and UV radiation at their footpoints \citep[e.g.,][]{Siverio_surge_uv}, which are believed to result from sudden energy release presumably caused by magnetic reconnection \citep[e.g.,][]{2007A&A...469..331J}. 

Multiwavelength observations have shown that some surges are associated with hotter EUV and X-ray jets \citep{1981SoPh...69..135S,1999ApJ...513L..75C,1992PASJ...44..265S,1996ApJ...464.1016C}, while others may expel predominantly cool material. Both flux emergence and flux cancellation have been suggested as the source of surges \citep[e.g.,][]{1973SoPh...28...95R,1993ASPC...46..507K,1996SoPh..169..357G}. Though knowledge of the occurrence rate of surges is uncertain, it is conjectured that surges and their associated EUV/X-ray jets may account for some SBs, a subject that requires further scrutiny. 

\subsection{Mini-eruptions: Jets and Mini-CMEs}

Narrow, transient, collimated brightenings in the solar atmosphere are commonly found to be jets (see Figs.~\ref{jets_multi}, \ref{jets_madj}, and \ref{mulay2016}). Although they may share a common origin with surges, namely reconnection, it is the layers of the solar atmosphere and temperature ranges that distinguish them: jets are produced higher up and observed in EUV and X-rays rather than in H$\alpha$, and are often accompanied by radio bursts.

Jets \citep[for reviews see,][]{2016SSRv..201....1R, 2021RSPSA.47700217S} are more easily observed in CHs and at the periphery of ARs due to the presence of open magnetic flux (hence a dark background), but jets also occur in closed-field-dominated regions \citep[e.g.,][]{Tri_jet,HouT_2021}. \citet{Innes_etall:2009} found omnipresent small-scale eruptions in the QS in 171~{\AA} observations by STEREO/SECCHI/EUVI. These eruptions were named mini-CMEs because of their close resemblance to large-scale CMEs. Some coronal jets have been interpreted to cause observable effects in situ in the heliosphere, supporting the hypothesis that they may be precursors of SBs \citep{2012ApJ...750...50N,2020ApJS..246...45H,2020ApJ...896L..18S,2021ApJ...920L..31N}.

\begin{figure*}
\centering
\includegraphics[width=0.8\textwidth]{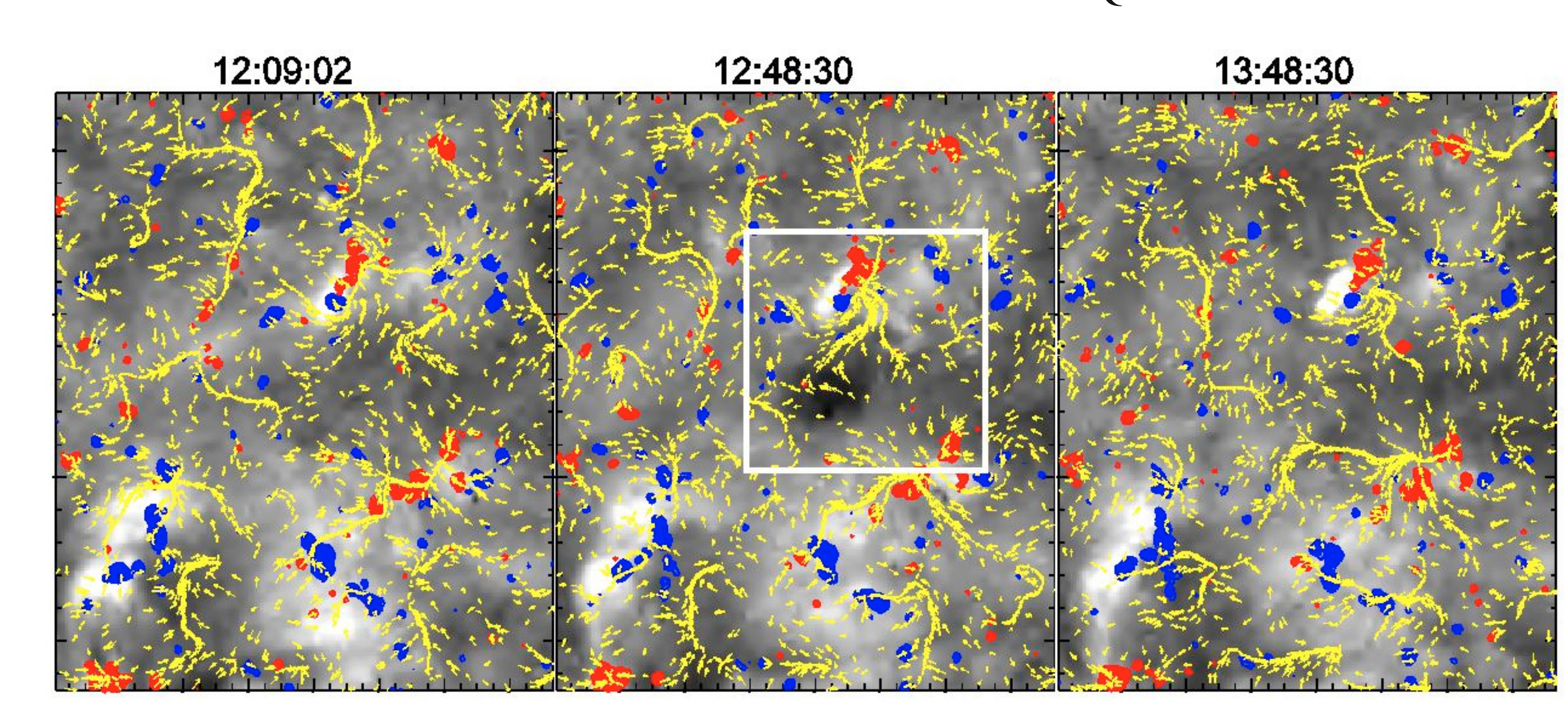}
\caption{Mini-CMEs above the supergranular junction in the center of the white-line square. STEREO/SECCHI/EUVI 171~\AA\ images (greyscale background). The magnetic field concentrations of positive (negative) polarities saturated at $\pm$40~G are shown in blue (red) colors. The photospheric flows are shown with yellow arrows. The FOV size is 150\arcsec $\times$ 180\arcsec, centered at (285\arcsec, 78\arcsec). The white box marks the location of the event. Image reproduced with permission from \citet{Innes_etall:2009}, copyright  by ESO} \label{innes_2009_fig5}
\end{figure*}
Mini-eruptions, both jets and mini-CMEs,  typically occur at the junction of supergranulation network lanes, where supergranular flows can build up the shearing of the magnetic field (Fig.~\ref{innes_2009_fig5}) \citep[e.g.,][]{2016A&A...596A..15A}. X-ray and EUV small-scale flare brightenings are observed at the base of both jets and mini-CMEs.  They have been referred to in various ways, including ``bright loop" \citep[e.g.,][]{1995Natur.375...42Y}, ``hot loop" \citep[e.g.,][]{1995Natur.375...42Y}, ``small flare" \citep[e.g.,][]{1996PASJ...48..123S}, ``jet bright point''  \citep[e.g.,][]{2015Natur.523..437S} and ``microflare'' \citep[e.g.,][]{2011A&A...526A..19M,2012A&A...545A..67M}. We will use the term ``microflare" to refer to these brightenings, regardless of their global context.  

Magnetic flux emergence and cancellation can both be involved in the generation of jets and mini-CMEs. Their omnipresence, as observed in the photosphere over the whole Sun, together with their frequent occurrence, indicates a causal relation to the occurrence rate of magnetic SBs. 
Magnetic flux emergence was originally considered the main scenario of jet formation \citep[e.g.,][]{1992PASJ...44L.173S}. Although this sometimes occurs \citep[e.g.,][]{2024ApJ...962L..38D}, several studies report that magnetic flux convergence and cancellation starting several hours before the jet onset energize and trigger jets \cite[e.g.,][]{2017ApJ...844..131P, 2018A&A...619A..55M, 2018ApJ...853..189P, 2021ApJ...909..133M}. In contrast, other observational studies find little evidence for cancellation associated with jet energy buildup or initiation \citep{kumar2018,kumar2019a,kumar2019b}, consistent with numerical simulations of energy buildup and release through rotational and shearing footpoint motions \citep{wyper2016,2018ApJ...852...98W,wyper2018b,wyper2019}. {Clusters of small-scale loops in the QS, CHs, and in the vicinity of ARs known as coronal bright points \citep[CBPs;][and references therein] {1973ApJ...185L..47V,Golub1980,2019LRSP...16....2M} typically produce jets well after emergence and decay through gradual dispersion of magnetic flux \citep{kumar2019a}. Microflares, which appear as compact, bright regions, always accompany these mini-eruptions and are found above the CBP PILs, analogous to flare arcades.

\subsection*{Jets}
\begin{figure}
    \centering
    \includegraphics[width=1.0\textwidth]{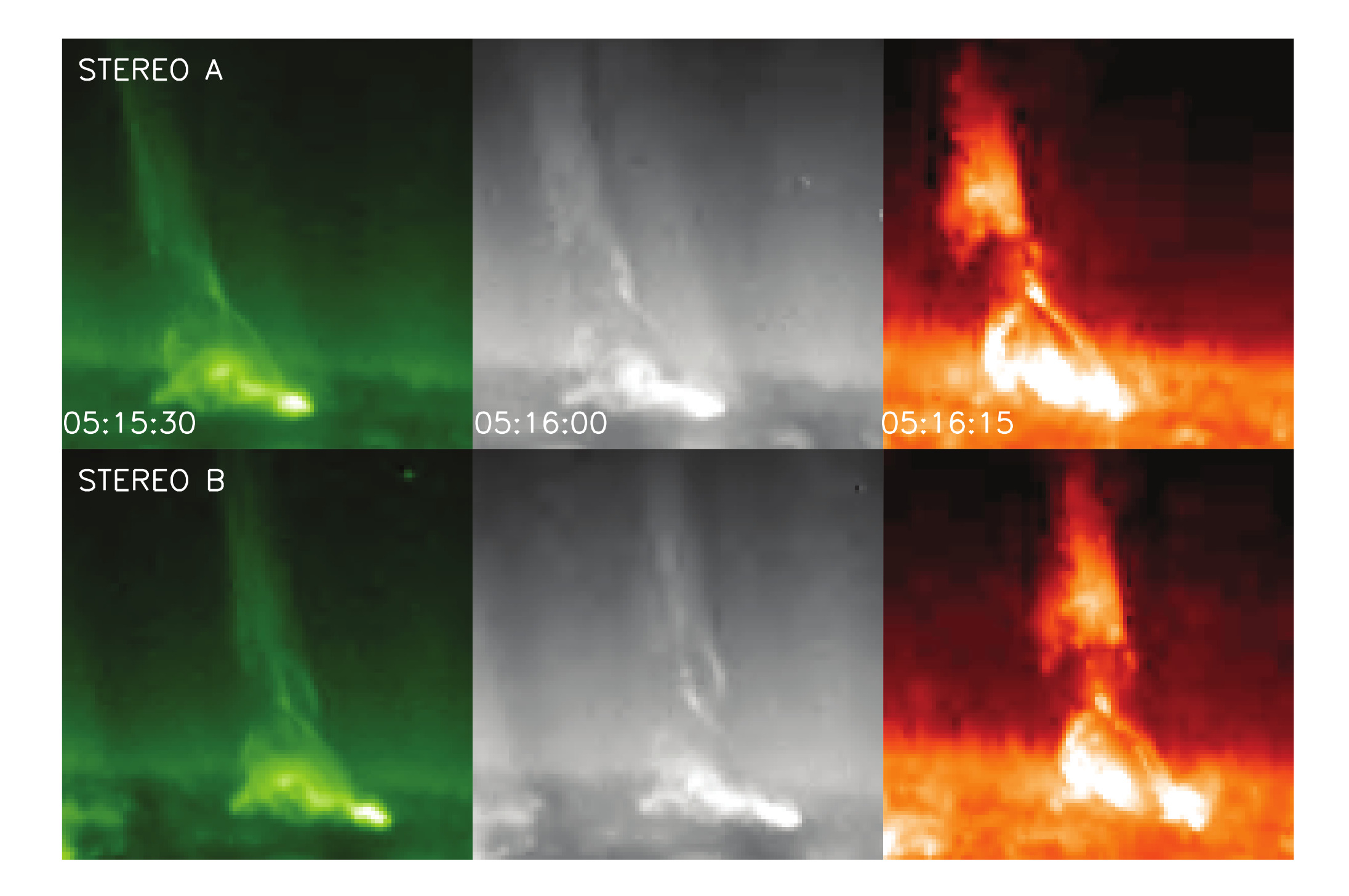}
\caption{Jets in EUV imaging data.  The helical structure of a jet is revealed in the EUVI/SECCHI/STEREO A and B in 195~{\AA} (left), 171~{\AA} (middle), and 304~{\AA} (right). Image reproduced with permission from \citet[][]{2008ApJ...680L..73P}, copyrights by Springer}
\label{jets_multi}
\end{figure}
\begin{figure}
    \centering
\includegraphics[width=0.9\textwidth]{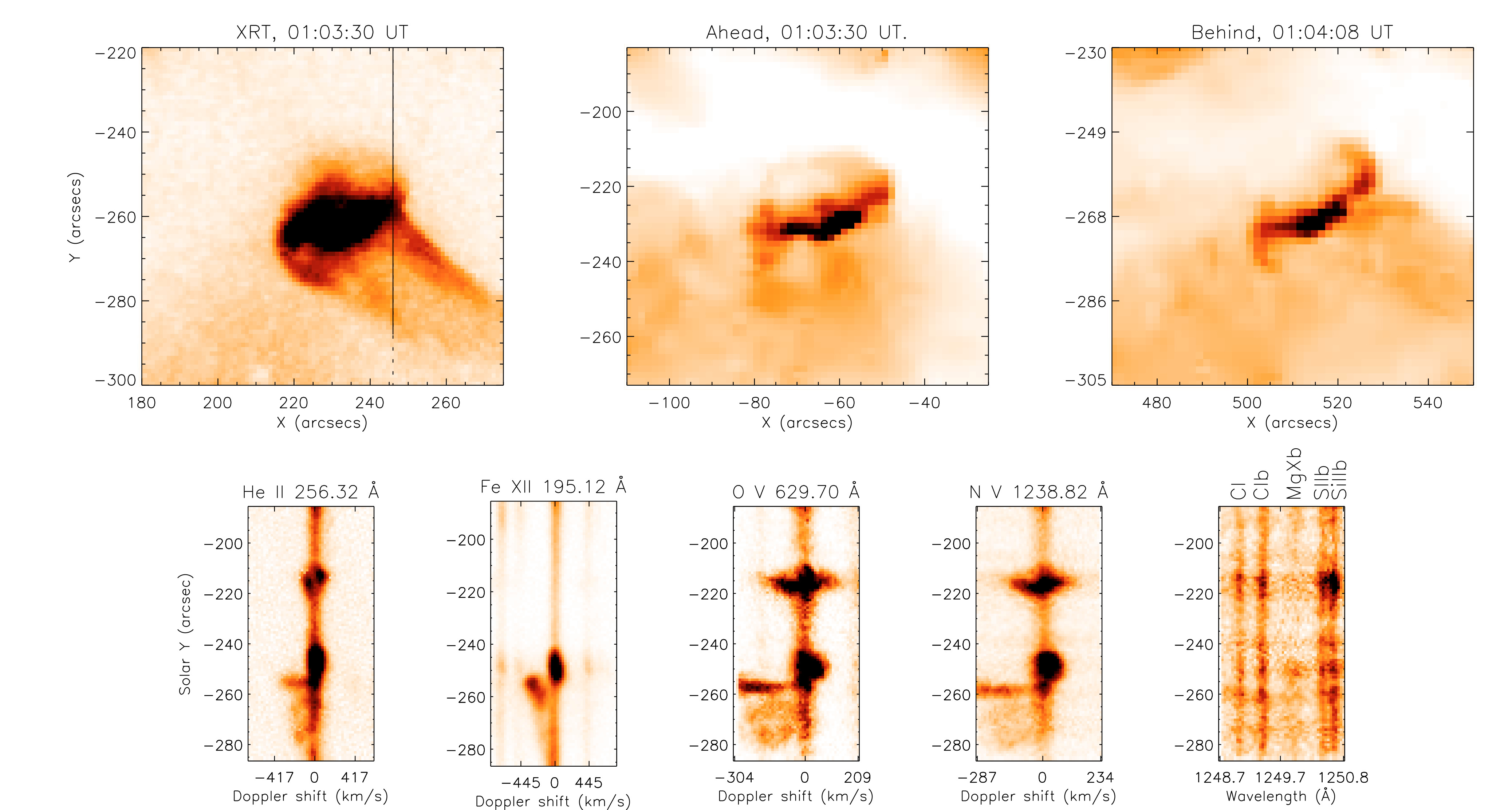}
\caption{X-ray and EUV imaging and spectroscopic observation of a CH jet. {Top row, left to right:} A jet as seen in X-ray emission by Hinode/XRT (left), and in EUV with EUVI/SECCHI/STEREO~A~\&~B at 193~{\AA} (middle and right). The overplotted vertical line is the slit position for Hinode/EIS. {Bottom row:} Jet emission as recorded along the EIS slit in several spectral lines, showing the blue-shifted emission from the jet and strong brightening from the reconnection site. Image reproduced with permission from \citet{2011A&A...526A..19M}, copyright by ESO}
\label{jets_madj}
\end{figure}

Jets are the most common small-scale transient activities in both polar and equatorial CHs.    Given the excellent reviews on observed properties and physics of jets \citep[e.g.,][]{2016SSRv..201....1R, 2000ApJ...542.1100S, 2021RSPSA.47700217S, 2021GMS...258..221S, 2022FrASS...920183S}, here we mainly summarize jet properties directly pertinent to possible solar precursors of SBs. 

In-depth studies of coronal jets started with the launch of the {\sl Yohkoh} satellite in the 1990s with X-ray observations of jets from CBPs and ARs (for details on AR jets, see Sect.~\ref{subsubsec:arjets}). Subsequently, jets were identified in EUV images from several solar missions, including SOHO/EIT, TRACE, STEREO/SECCHI/EUVI, SDO/AIA, IRIS, Hi-C, and SO/EUI. Jets are observed in the chromosphere, transition region, and corona as well as in the extended corona in white-light coronagraph images.
Note that the macrospicules discovered by \cite{Bohlin75} are now also considered to fall within the jet category. The terms used to describe potentially overlapping phenomena are summarized in Table~\ref{tab:maria}.
Jets form on a wide range of spatial scales and are always associated with small-scale flare-like activity, as mentioned above. It is now well-established that jets are driven by magnetic reconnection. Additional evidence for the dominant role of reconnection in jets comes from observations of plasmoids or plasma blobs (presumably magnetic islands) in bidirectional flows and bright arcades composed of reconnected field lines \citep[e.g.,][and references therein]{kumar2019a,kumar2019b,2021RSPSA.47700217S,2024A&A...687A.190H}.

\begin{figure*}[!ht]
\centering
\includegraphics[width=0.85\textwidth]{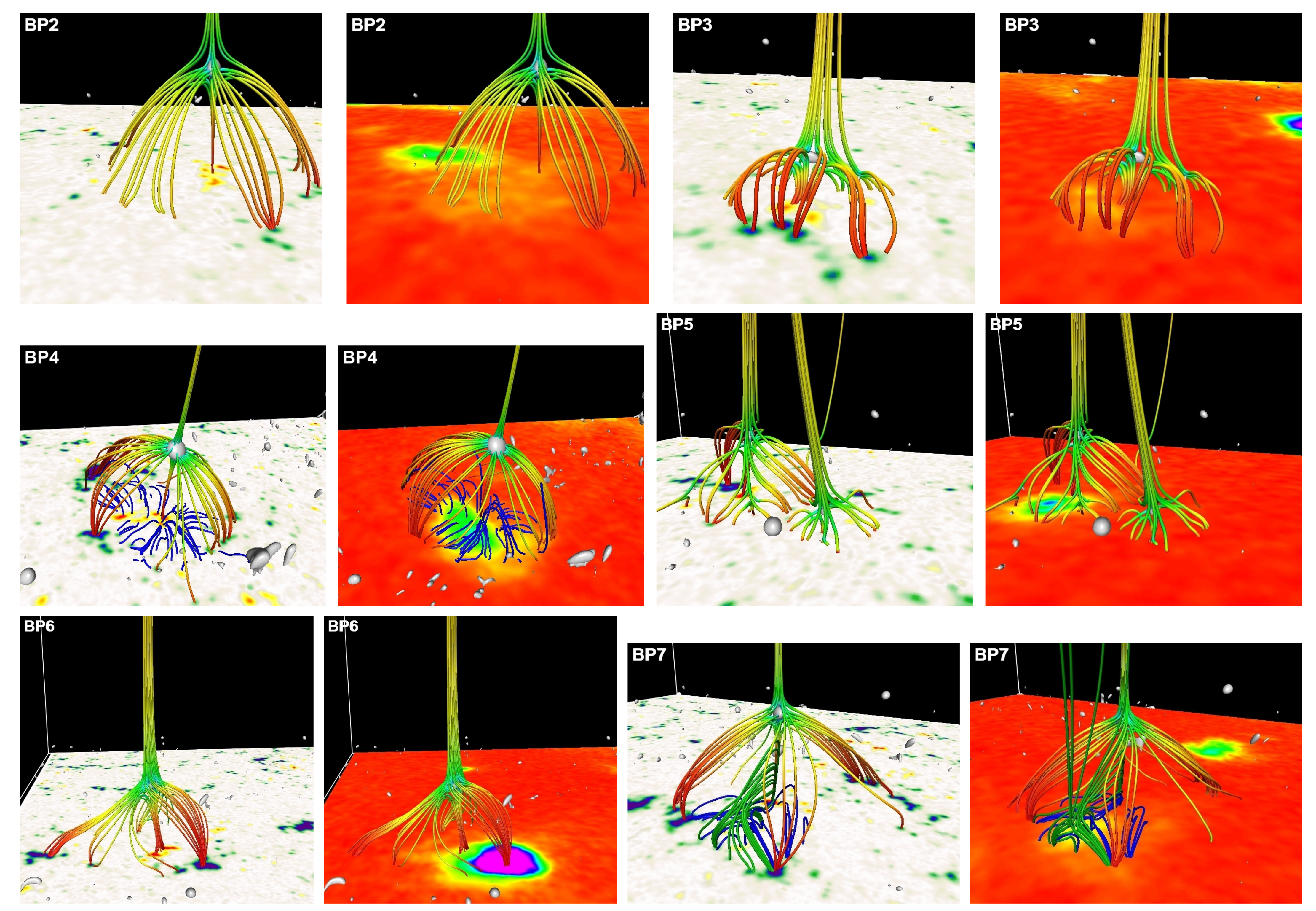}
\caption{3D null-point topologies with fan separatrix surfaces and spines in a CH for six different CBPs as labeled. In each example, the extrapolated field is shown on a magnetogram obtained from Hinode/SOT (left panels) and X-ray images recorded with Hinode/XRT (right panels). Here, the X-ray emission is mainly located at the footpoints of the closed loops (color shading), while the white blobs denote the locations of null points. Image reproduced with permission from  \citet{2017A&A...606A..46G} copyright by ESO}
\label{gaalsgard_2017_fig4}
\end{figure*}

Before multi-instrument observations became commonly available, jets were categorized according to the wavelength range in which they were observed (X-ray, EUV, etc.) or the characteristic temperature of the solar region in which they were observed (chromospheric, transition region, and corona). Most jets are observed to originate in fan-spine magnetic configurations, exhibiting a bright dome structure at the base surmounted by a cusp and a bright spire (Figs.~\ref{gaalsgard_2017_fig4} and \ref{duan_2024_fig2}). 
In such fan-spine topologies \citep{Lau1990}, the fan encloses closed flux, the inner spine extends from the null to the solar surface, and the outer spine extends outward into open field or closes remotely \citep{antiochos1996}. 
Continual flux emergence through the photosphere creates a ``magnetic carpet" of mixed polarities \citep{Schrijver1997}, forming small-scale fan-spine topologies in the corona, yielding numerous sites favorable for magnetic reconnection. Such coronal magnetic-field topologies generally appear as CBPs seen in EUV and soft X-ray images. The often anemone-like appearance of CBPs indicates that their underlying magnetic topology is fan-spine, with the bright CBP loops residing beneath the dome \citep{2017A&A...606A..46G}. The well-established connection between jets and CBPs further strengthens the theoretical and modeling evidence for fan-spine topologies as key sources of small-scale solar activity. In fact, small-scale bright points and faint loops are the main magnetic-field structures building the solar atmosphere outside ARs \citep{2023A&A...678A..32M}. Those occurring in CHs are of greatest interest in the context of SB sources because of their open outer spines (see Sect.~\ref{sec:transients}).

\citet{2010ApJ...720..757M} divided anemone jets evenly into standard and blowout jets based on their morphology. According to this division, blowout jets have an additional jet spire and are generally more energetic than standard jets \citep{2013ApJ...776...16P}.
Later studies presented evidence that both blowout and standard jets (but not all) result from mini-filament eruptions \citep[e.g.][]{2015Natur.523..437S,2020A&A...643A..19M,2022ApJ...927..127S}. 
These two terms are not universally used, however, since they can reflect observational limitations rather than physical properties. Furthermore, regarding the universality of the mini-filament eruption mechanism, recent studies have demonstrated that diverse phenomena—including two-sided-loop jets \citep[e.g.,][]{2019ApJ...883..104S}, jets in fan-spine topologies \citep[e.g.,][]{2023ApJ...942L..22D}, and macro-spicules—are all fundamentally driven by mini-filament eruptions \citep[e.g.,][]{2019ApJ...885L..11S}.

The plasma temperature structure in individual jets is complex. Some jets contain both cool and hot plasmas, although they may appear either simultaneously or sequentially \citep[e.g.,][]{1999ApJ...513L..75C, 2007A&A...469..331J, Innes_etall:2010,2011A&A...526A..19M, schmieder2022, kayshap2024}. In addition, the cool component might be seen in absorption in EUV lines. The presence of mini-filaments --- cool, dense plasma aligned with the PIL --- inside the dome of many jet sources also leads to the coexistence of hot and cool plasmas, especially when erupting mini-filaments become mixed with the jet outflows.

\begin{figure*}[!ht]
\centering
\includegraphics[width=0.8\textwidth]{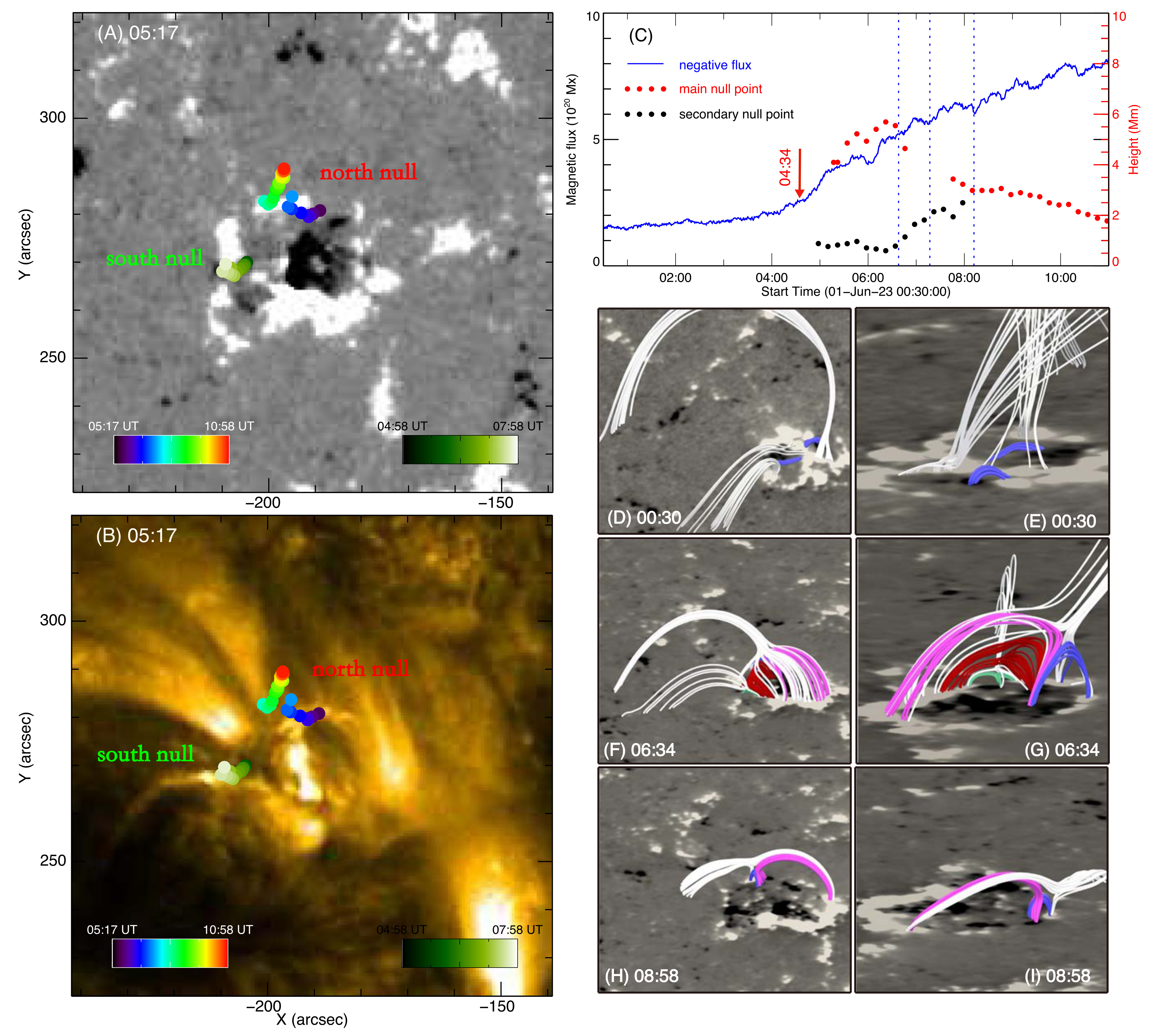}
\caption{Coronal magnetic topology evolution associated with a jet. Left column: HMI magnetogram with null point locations (panel A) and the corresponding AIA~171~{\AA} image. The evolution of the negative magnetic flux and null point heights, as labelled, is shown in panel C (top panel).  Panels (D), (F), and (H) depict the evolution of the two fan-spine structures overplotted on HMI magnetograms. Panels (E), (G), and (I) present close-up views of the same structures shown in panels D, F \& H but from different perspectives. Image reproduced with permission from \citet{2024ApJ...962L..38D}, copyright by AAS}
\label{duan_2024_fig2}
\end{figure*}
The coronal jets that are easily detectable in EUV images are possibly only the tip of the iceberg of the full coronal jet population. Often, QS jets are hard to distinguish when observed in projection on the solar disk, due to the strong background emission and/or low emissivity of the jet. When the density or filling factor (and thus the EUV intensity) contrast between the outflow material and the surrounding corona is not sufficiently high, jets may be poorly visible or even invisible in imaging observations. Such jets were first detected using a combination of imaging (SDO/AIA) and spectroscopy (Hinode/EIS) within CHs and at QS/CH boundaries by \citet{young_dark_2015}. The study revealed significant outflow signatures from CBPs in the Doppler images, with no associated jet-like structures visible in AIA EUV images, naming them `dark' jets. Those jets' properties appear similar to regular coronal jets (e.g., speeds of 50 to 100~\kms).  The study suggests that their low intensity indicates their mass flux is about two orders of magnitude lower than that of other jets. Note also that 3D numerical models of CBPs show that null-point reconnection produces weak jet-like upflows with a very faint EUV counterpart resembling the `dark jets' \citep[see Fig.~\ref{fig:dark_jetsboundaries}, e.g.,][]{ pariat2015,nobrega-siverio_deciphering_2023}.

Possibly the tiniest jets observed thus far are the picoflare jets \citep{2023Sci...381..867C} detected by SO/EUI. These CH jets are a few hundred kilometers across, last for about 20 to 100~s, and may reach speeds of about 100~{\kms}.  Like other jets, they are also triggered by reconnection and possibly contribute to the solar wind \citep[see also][]{TriNS_2021, UpenT_2021, UpenT_2025}.

The IRIS spectrometer has revealed the presence of high-velocity jets at transition-region temperatures in the network, where typically spicules are observed  \citep{2014Sci...346A.315T}. Best seen in the IRIS  1330~{\AA} slit-jaw images, they have typical widths of $\sim$300~km, speeds ranging from 80--250~\kms, lifetimes in the range of 20--80~s, and lengths from 4 to 10~Mm, with some reaching up to 15~Mm. \cite{KolPT_2024} showed that jets observed by IRIS have enhanced nonthermal velocities of 50{--}60~km~s$^{-1}$. They were consequently investigated as possible transition-region counterparts of jetlets (see Sect.~\ref{subsec:jetlets}) and picoflare jets at the base of plumes \citep{2019SoPh..294...92Q}. Several of the IRIS jets exhibit the characteristic fan-spine morphology, along with their reportedly high velocity and nonthermal width, suggesting that they also result from magnetic reconnection and can be classified as a category of jet phenomena. This makes them possible candidates for SB precursors. 

\subsection*{Mini-CMEs}

Similar to CMEs, most mini-CMEs (also referred to as mini-eruptions) display the classical three-part structure: a bright loop moving ahead of a dark cavity and a bright core that corresponds to an eruptive prominence/filament. A statistical analysis of 79 coronal jets observed by STEREO/SECCHI/EUVI found that five events corresponded to scaled-down versions of CMEs \citep{2009SoPh..259...87N}. As shown in Fig.~\ref{innes_2009_fig5}, mini-CMEs originate at the boundaries of supergranular cells \citep{Innes_etall:2009}, which is particularly pertinent in the context of SB sources. Most mini-CMEs manifest scaled-down CME features: ejection of cool plasma, which is believed to be erupting mini-filaments; microflaring activities in the source region that appear as compact brightenings; and regions with depleted coronal emission, known as dimmings.
 
\begin{figure}[ht!]
    \centering
    \includegraphics[width=0.9\textwidth]{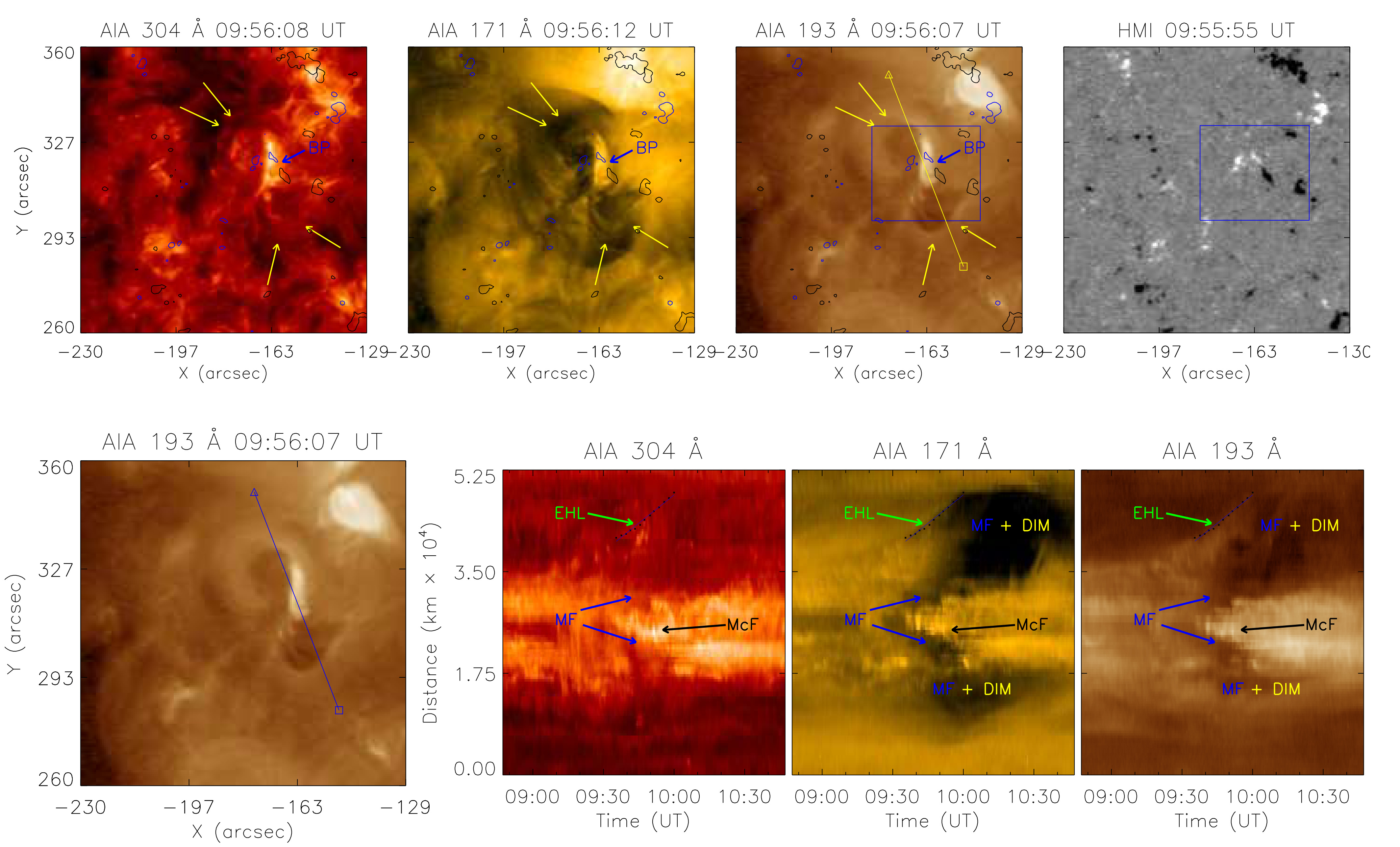}
    \caption{Mini-CME from a CBP. Top row, from left to right: AIA 304~\AA, 171~\AA, and 193~\AA\ images of a mini-CME and corresponding HMI line-of-sight magnetogram saturated at $\pm$50~G. The yellow arrows point at the erupting MF. The blue and black contours outline the positive and negative fluxes at 50~G. Bottom row, from left to right: AIA 94~\AA\ image of the CBP during an eruption showing the microflare. The solid line indicates the slice that has been extracted to produce the time-distance panels in the AIA 304~\AA, 171~\AA, and 193~\AA\ channels. The bottom of the slice is marked with a triangle, and the top with a square. The solid lines in the time-distance panels are a linear fit of visually selected points that outline the bright or dark front of the eruptive features. The following abbreviations are marked on the panels: BP -- bright point, EHL -- erupting hot loops, MF -- mini-filament, DIM -- dimming, McF -- microflare. Image reproduced with permission from \citet{2018A&A...619A..55M}, copyright by ESO}
\end{figure}

Mini-CME events appear to be associated with CBPs at their bases \citep{Innes_etall:2009, Podladchikova_etall:2010}.  To investigate this link, \citet{2018A&A...619A..55M} analyzed 42 CBPs in the QS and found that the majority (76\%) produced at least one eruption during their lifetime. As all the events were studied in projection on the disk, it is unknown how many of them produced jets (collimated flows that could escape into the heliosphere) or evolved as failed eruptions. \citet{2022A&A...660A..45M} studied a microflare that was interpreted as evidence for magnetic reconnection in a CBP related to a mini-CME/jet in IRIS imaging and spectroscopic data. The study demonstrated that microflares seen in images are located at the base of the jets (also denoted as ``explosive events''), which originally were only observed spectroscopically as non-Gaussian spectral line profiles with Doppler shifts representing outflow speeds of 50--200~\kms\ \citep[e.g.,][]{1983ApJ...272..329B, Huang_etall_2014}. \citet{Panesar_etall:2020} studied ten QS eruptions that evolved into jets, all of which originated in CBPs.

\subsection*{Jets in ARs} \label{subsubsec:arjets}

Jets in ARs can have a different and more complex morphological appearance compared to those in CHs and QS. This is primarily due to the complexity of AR magnetic fields, which causes an equally complex appearance of jet emission in imaging and spectroscopic data. An example of an AR jet observed with SDO/AIA is shown in Fig.~\ref{mulay2016}. Counterparts of some AR jets may have been observed in the heliosphere \citep{2020ApJ...896L..18S}.  While some non-AR jets also may reach the heliosphere \citep[e.g.,][]{1998ApJ...508..899W,2015ApJ...806...11M,2018JPhCS1100a2024S}, the heliospheric counterparts of AR jets may be easier to track than CH or QS jets due to their higher average energy content.

\begin{figure}
\centering
\includegraphics[width=\textwidth,angle=0]{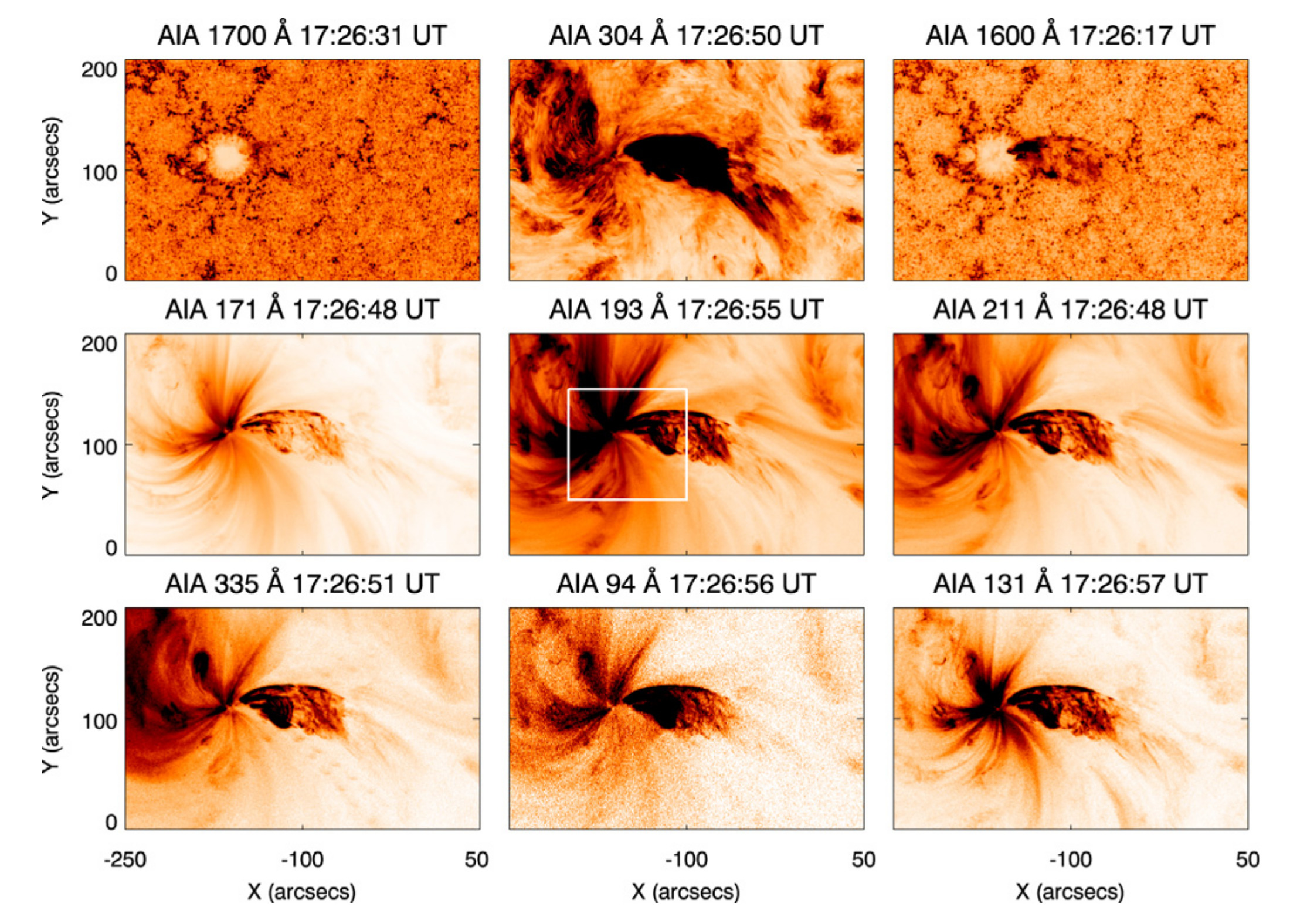}
\caption{AR jet appearance in several UV and EUV channels (as noted above each image) of SDO/AIA. Image reproduced with permission from \citet{2016A&A...589A..79M}, copyright by ESO}
\label{mulay2016}  
\end{figure}

It is also common for AR jets to occur repeatedly from nearly the same location over a short period of time; in some cases, these can be termed as homologous \citep[e.g.,][]{2002ApJ...575..542W,2008A&A...491..279C,2015ApJ...815...71C,2016ApJ...822L..23P,2016SoPh..291..859Z,2019ApJ...873..110P,2018ApJ...861..105S,2018ApJ...852...10L,2020ApJ...891..149P}. Note that homology is observed in non-AR jets as well \citep{2024ApJ...962L..38D, 2015ApJ...815...71C}.
Although many AR jets have some morphological and energetic differences from typical QS and CH jets \citep[e.g.,][]{2016ApJ...821..100S,2017ApJ...844...28S,2019ApJ...873..110P,2020ApJ...891..149P,2022ApJ...935..172P}, AR jets likely operate in the same way as their QS and CH counterparts \citep{2024ApJ...960..109S}. Fundamentally, they are the result of the eruption of stressed filament channels in a fan-spine configuration.

\subsubsection{Physical Properties of Jets: Velocity, Electron Temperature, Density, and Frequency}
\label{subsubsec:jetproperties}
Out of 100 jets identified in {\sl Yohkoh}/SXT data, at least 66\% were associated with ARs \citep{1996PASJ...48..123S}. Therefore, early measurements of jet properties were disproportionately influenced by sample bias toward AR jets. The lengths of the jet spire ranged from a few$\times 10^4${--}$4 \times 10^5$\,km, with an average of $1.5 \times 10^5$\,km; widths ranged from 5000 to $10^5$\,km, averaging to $1.7\times10^4$\,km.  Their projected speeds ranged from 10{--}1000\,km\,s$^{-1}$, with an average of $\sim$200\,km\,s$^{-1}$, and lifetimes ranged over 100{--}16\,000\,s following a power-law distribution. For 16 jets observed with SXT, \citet{2000ApJ...542.1100S} found thermal energies of $10^{19}$--$10^{22}$\,J; by comparison, the energy for two CH jets was estimated to overlap the lower end of this range with values of $\sim$10$^{19}$--10$^{20}$\,J \citep{2013ApJ...776...16P}.

Studies of both EUV and SXR jets and their associated transient brightenings (microflares) have found no significant differences in their physical parameters between polar and equatorial CHs, and no longitudinal dependence \citep{2010AnGeo..28..687N,2013ApJ...775...22S}. The mean properties of X-ray jets in CHs are: length~$\approx$ 5.0 $\times$ 10$^4$~km, lifetime $\approx$ 10~min, speed $\approx$ 170~{\kms}, width $\approx$~8000~km, and thermal energy $\approx$ $7 \times 10^{18}$~J. The mean and estimated properties of the microflares are lifetime $\approx$ 5~min, area $\approx$ 3.0~$\times$~10$^7$~km$^2$, and thermal energy $\approx$ 8~$\times$~10$^{18}$~J \citep{2007PASJ...59S.771S}. 

CH jet spires have temperatures of $\sim$1.5--2.0\ MK determined from intensity ratios in different EUV and soft X-ray filters \citep{2013ApJ...776...16P,2015A&A...579A..96P}, while AR jet spires contain plasmas of both $\sim$1\,MK and several MK \citep{2016A&A...589A..79M,2017A&A...598A..11M,2017A&A...606A...4M}. Jet footpoints can be much hotter, in excess of $\sim$10\,MK, while also containing plasma at transition-region and chromospheric temperatures \citep[e.g.,][]{2000ApJ...542.1100S,2011A&A...526A..19M,2018SoPh..293..160M}. In short, jets are multithermal phenomena both at their source and as they propagate outward. 

\citet{huang_statistical_2023} visually identified 27\,832 jets in polar and equatorial CHs over 24 days using multi-scale Gaussian normalization \citep{2014SoPh..289.2945M} processed AIA~193~{\AA} images (T $\sim$ 1~MK), i.e., $\approx$1160 jets per 24~h, or a rate of 3.7~$\times~10^{-11}$ km$^{-2}$ hr$^{-1}$. This rate is slightly higher than that of X-ray jets (T $\sim$ 3~MK), $\sim$1.25~$\times$~10$^{-11}$ km$^{-2}$ hr$^{-1}$ \citep{2013ApJ...775...22S}. This is not surprising, however, as not all jets reach X-ray temperatures. The study estimates that PSP may sample 0.12--0.73 jets per day, which is far below the SB count rate. 

We should also note that one jet can trigger a subsequent one, as shown by  \cite{2021ApJ...912L..15T}, who demonstrated the existence of homologous jets \citep[see also][]{2016A&A...589A..79M}. This can easily account for the observed clustering or high frequency of these events, which could also explain SB occurrence in patches \citep[assuming a causal link between the two phenomena, e.g.,][]{2023Natur.618..252B}.

\citet{Innes_etall:2009} estimated the occurrence rate of mini-CMEs at about 1400 per day over the whole Sun. A follow-up study, which applied automatic detection of mini-CMEs to the same EUVI data set as \citet{Innes_etall:2009} in addition to AIA data, found almost twice the rate obtained from the visual selection \citep{Alipour_etall:2012}. As shown in Fig.~\ref{innes_2009_fig5}, mini-CMEs originate at the boundaries of supergranular cells \citep{Innes_etall:2009}, which is particularly pertinent in the context of SB sources. 

Comparing the occurrence rates of jets and mini-CMEs with those of SBs is one way to search for a statistical link between these two classes of phenomena. We detail recent attempts that combine remote-sensing and in situ observations in Sect.~\ref{sec:connect-RS-IS}.

\subsubsection{Helical Structure of Coronal Jets}
\label{subsubsec:jettwist}
\begin{figure}
\centering
\includegraphics[width=0.5\textwidth]{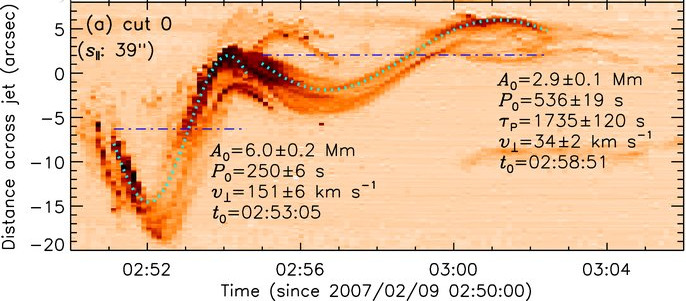}
\includegraphics[width=0.4\textwidth]{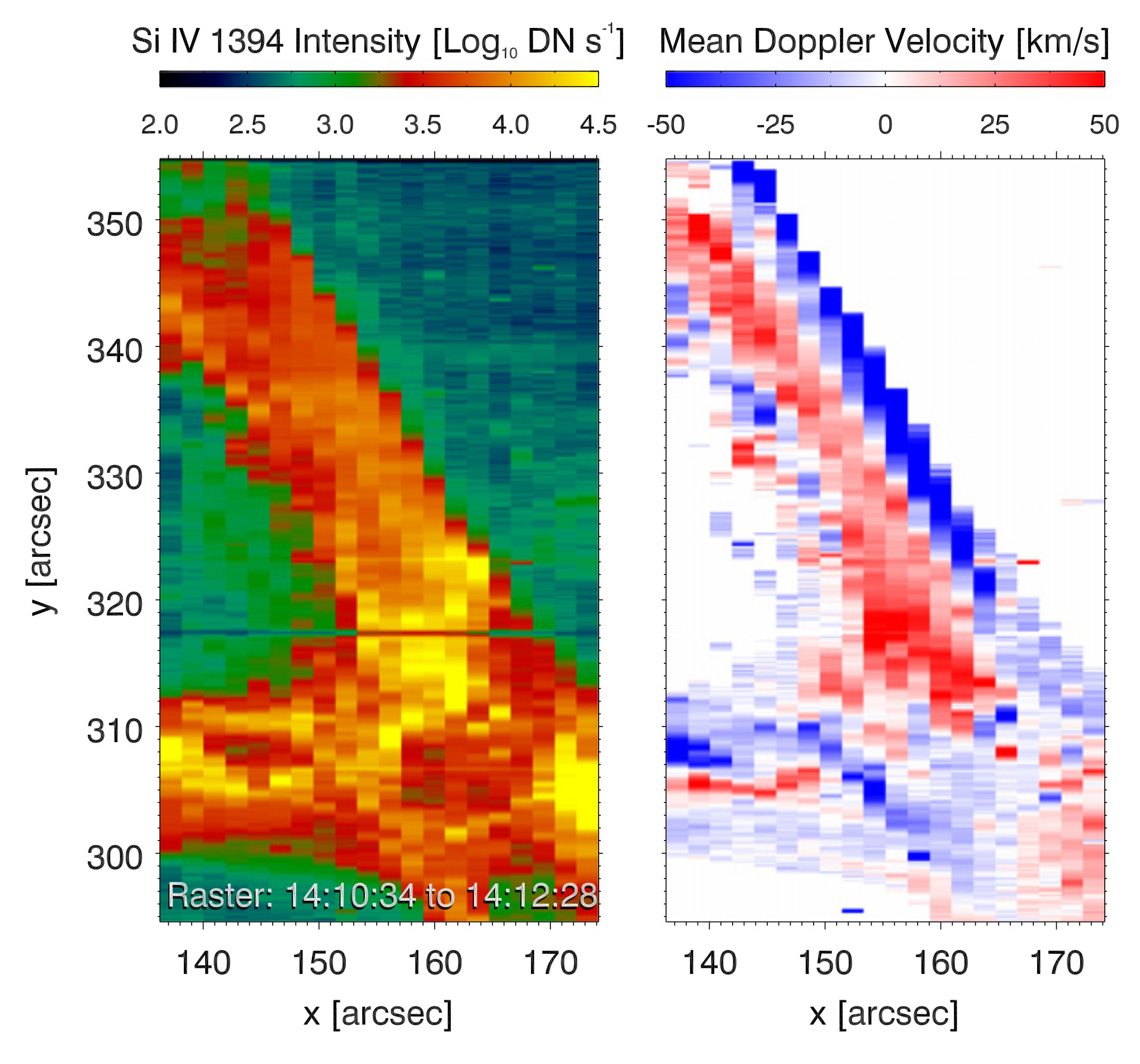}
\caption{Observational evidence of jet twist. {Left panel:} Hinode/SOT observations showing evidence of helical structure in a jet based on sinusoidal time-space intensity profiles across the jet axis. {Right panel:} IRIS observations showing evidence of helical structure in a jet based on opposite Doppler shifts across its axis. Images reproduced with permission from \citet[][left panel]{2009ApJ...707L..37L}  and \citet[][right panel]{2015ApJ...801...83C}, copyright by AAS} \label{fig:twist}
\end{figure}

Twist indicates the amount of magnetic free energy available in a source region, which is a prerequisite for driving a jet. Jets frequently exhibit such helical structure (example shown in Fig.~\ref{jets_multi}) associated with apparent untwisting as they propagate outwards. Such untwisting motions have been interpreted as the transfer of magnetic twist and subsequent relaxation resulting from reconnection between pre-existing \citep[e.g.,][] {1986SoPh..103..299S,2013ApJ...769L..21A, lee15,2008ApJ...673L.211M} or dynamically formed flux ropes  \citep[e.g.,][]{2009ApJ...691...61P, 2017Natur.544..452W, 2018ApJ...852...98W} and the ambient open magnetic field. Nonlinear force-free (NLFF) magnetic-field extrapolations indicate the presence of a pre-eruption flux rope as the seed of a helical jet \citep{2013A&A...559A...1S}. \citet{2011ApJ...735L..43S} reported an example of an unwinding polar jet observed by AIA. The unwinding of this jet was evidenced by the observations of multiple, rotating threads within the body of the jet. The twist in this jet was estimated from 1.17{--}2.55 turns. This untwisting drives a nonlinear torsional wave travelling at the local Alfv\'en speed \citep[e.g.,][]{2009ApJ...691...61P,lee15,karpen2017}.

Recent Metis observations by \citet{2025ApJ...982..142R} reported evidence of helical structures as far as 3 solar radii, which resulted from the eruption of a jet-like narrow CME in a pseudostreamer. The observed helical structures were attributed to interchange reconnection between the narrow CME and the pseudostreamer, giving rise to torsional Alfv\'en waves.

Helical jet structures and related untwisting are observed almost exclusively with  EUV or SXR imaging and spectroscopic observations in the corona below 0.5~{\rs} \citep[e.g.,][]{kumar2018}. Earlier studies using {\sl Yohkoh}/SXT observations showed that about 10\% of the SXR jets have a helical structure \citep{2007PASJ...59S.745S}. Unambiguous evidence of the presence of helical structure and untwisting in a polar coronal jet was based on stereoscopic observations by STEREO/SECCHI/EUVI when the two spacecraft were separated by 11$^{\circ}$ \citep{2008ApJ...680L..73P}. A statistical study of 79 CH jets recorded by stereoscopic STEREO/SECCHI/EUVI observations reported that 31 exhibited helical morphologies  \citep{2009SoPh..259...87N}, for jets launched from both polar and equatorial CHs \citep[e.g.,][]{2010AnGeo..28..687N}. Similarly, recent high-resolution Hi-C and SO/EUI observations show evidence of helical structures and untwisting in coronal jets \citep{2019ApJ...887L...8P, 2023ApJ...944...19L}

Twisting motion has also been observed in chromospheric jets. For example, \citet{2009ApJ...707L..37L} reported clear evidence of twist in a chromospheric jet observed by Hinode/SOT, demonstrated by the presence of oscillating patterns in time-distance intensity plots across the jet axis (Fig.~\ref{fig:twist}, left panel). \citet{2018MNRAS.476.1286J} reported a mini-filament eruption accompanied by an untwisting jet, observed by SDO/AIA. The chirality of the mini-filament was consistent with the sense of rotation of the jet, which implies a transfer of twist from the magnetic field supporting the mini-filament to the jet. 

Evidence of helical structure in jets has also been detected in Doppler maps obtained from spectroscopic observations, where oppositely directed Doppler shifts across the jet's axis suggest plasma rotation around the axis. As shown in Fig.~\ref{fig:twist}, right panel, several spectroscopic investigations of jets with Hinode/EIS or IRIS have found evidence for untwisting \citep{2014PASJ...66S..12Y,2015ApJ...801...83C, 2018A&A...616A..99K, 2019ApJ...887..154L, schmieder2022}.
 
Observing helical structures in jets poses greater challenges than merely observing the jets themselves, as it requires adequately resolving the jet's evolving fine structure. Therefore, the spatial and temporal resolution of a given instrument limits the visibility of helical jet structures and, hence, the estimated number of twists/turns. Comparison of the multi-viewpoint STEREO observations of \citet{2008ApJ...680L..73P}  with synthetic images from MHD simulations of \citet{2009ApJ...691...61P} suggests that the observed jet had approximately 1.1 turns. Using SDO/AIA observations, various studies have reported different numbers of turns in different jets: for CH jets, 1.17{--}2.55  and 1.0 turns \citep{2011ApJ...735L..43S,2012RAA....12..573C}, while for an AR jet a typical value is 0.9 turns \citep{2013RAA....13..253H,2018ApJ...852...10L, 2019FrASS...6...44L}. Of 14 large polar-CH jets imaged in the 304~\AA\ channel of SDO/AIA, all showed evidence of untwisting \citep{2015ApJ...806...11M}. By following conspicuous features in time,  twists in the range 0.75{--}2.25 (average value of 2.25) and rotation speeds in the range 60{--}220~{\kms} (average value of 110 \kms) were deduced. 
An important action item is to investigate whether helical morphologies are frequently found in the smaller, more frequent jetting detected by SDO/AIA and SO/EUI \citep[e.g.,][]{2023ApJ...945...28R,2023Sci...381..867C}. 

Another question to address through theory and modeling is how these twists, which are observed at low altitudes in the atmosphere, may survive to become SBs in the young solar wind \citep[see][]{chapter5}. Using 3D MHD parametric simulations, \cite{TouPF_2024} showed that solar jets can propagate into the super-Alfv\'{e}nic wind. Magnetic untwisting waves associated with the jets show SB-like signatures. However, large deflections (above 90 degrees) were not observed in simulations. The authors hypothesized that a two-step scenario, involving additional in situ processes, may be required to produce them.
\subsubsection{Plumes and Jetlets} \label{subsec:jetlets} 

Solar coronal plumes are hazy, ray-like structures best observed in polar CHs \citep[e.g.,][]{2001ApJ...546..569D}. Plumes are rooted at supergranulation boundaries (i.e., the network) and are seen to reach heights up to 30~{\rs} in eclipse observations. They have also been observed in equatorial CHs, but their extension is hard to follow due to the strong background emission of the surrounding QS or nearby ARs. They are typically observed as bright regions in white-light emission far into the heliosphere, and in EUV/XUV emission in the lower atmosphere at their footpoints \citep[for review see][]{2015LRSP...12....7P}. Plumes are twice as dense (N$_e$ = 1.2$\times$10$^9$~cm$^{-3}$) as the surrounding CH,  with a typical temperature of about 0.8~MK \citep[e.g.,][]{2024ApJ...974..163M}. Because this is $\sim$~0.1{--}0.3~MK hotter than the background corona \citep{2003A&A...398..743D}, plumes are best visible in the SDO/AIA 171~\AA, STEREO/SECCHI 171~\AA, SO/EUI/FSI 174~\AA, and SOHO/EIT~173~\AA\ channels. Their average lifetime is about 20~h, but they can last 2{--}3 days and as long as two weeks, depending on their size. CBPs often exist at the base of plumes in EUV and X-ray observations. The smaller CBPs produce jetlets, while the larger CBPs are the main sources of jets observed in CHs. A plume's average lifetime is similar to that of CBPs found at their base. 

\begin{figure}
    \centering
      \includegraphics[width=1.0\textwidth]{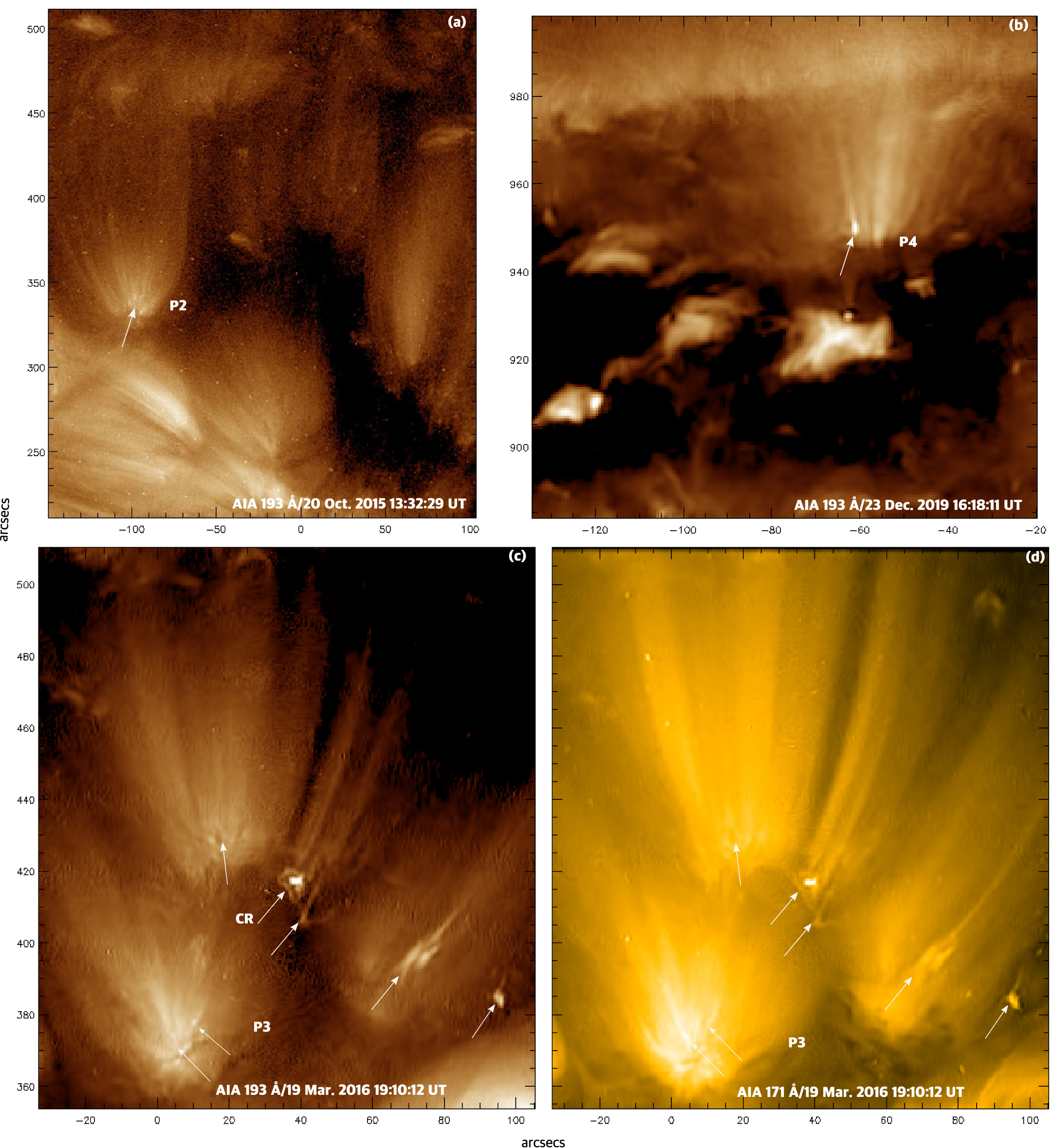}
    \caption{SDO/AIA 193~{\AA} and observations of plumes and jetlets. The arrows point to brightenings associated with jetlets. Plumes in an equatorial (a) and a polar (b) CH. Plume P3 and the neighboring plumes are located in a plage and network region. CR refers to a circular ribbon. Image reproduced with permission from  \citet{2022ApJ...933...21K}, copyright by AAS}
    \label{fig:enter-label}
\end{figure}

While much is known about plumes based on multiwavelength observations, their formation mechanism remains a topic of debate. A causal link has been suggested by some observations of jets preceding plume appearance \citep{2008ApJ...682L.137R, 2014ApJ...793...86P, 2024ApJ...974..163M}, but other studies have not confirmed a one-to-one connection. \citet{2014ApJ...787..118R} suggested that plumes are formed and sustained by reconnection-driven small-scale jets (named jetlets) and tiny bright points that are found in the plume footpoints. \cite{2018ApJ...853..189P, 2019ApJ...887L...8P} speculated that some jetlets are likely driven by the same type of mini-flux rope eruption that drives many coronal jets. The jetlets, however, are too small to be resolved in detail in the present observations and, thus, confirm this scenario. Therefore, whether or not jetlets are driven by the same physical mechanism as coronal jets remains open. 

 Jetlets also exhibit quasiperiodic intensity fluctuations at temperatures above 1~MK with a period of 3{--}5 minutes, indicating that p-mode waves might trigger episodic interchange reconnection at chromospheric or transition-region heights in small-scale fan-spine topologies at the base of the plumes \citep{2022ApJ...933...21K}. Plumes are composed of several bright filamentary structures, called plumelets \citep{Uritsky2021}, which have been traced back to jetlets, rooted in brightenings at the plume base in SDO/AIA images \citep{2022ApJ...933...21K}. Indeed, the appearance of microstreams observed beyond 1~AU in Ulysses data has been attributed to such jetting activities within plumes \citep{2012ApJ...750...50N}.

Analysis of in situ and remote-sensing data during PSP's Encounter 10, when the spacecraft was magnetically connected to a small area within a CH for a few days, showed that the velocity fluctuations at PSP also exhibited 3{--}5 minute periods, suggesting a link with the underlying episodic jetlet activity within the CH \citep{kumar_new_2023}. The appearance of SB patches and their quasiperiodic nature designates plumes and jetlets as one of the likely solar sources \citep[see][for details]{chapter3}. \citet{2023ApJ...945...28R} analyzed off-limb coronal images from SDO/AIA to determine whether SBs truly originate at the base of the corona. Their study found evidence for jetlets emanating from all regions of the Sun. Jetlets thus emerged as potential candidates for solar precursors of SBs. However, it remains uncertain whether jetlets alone are numerous enough to account for most or all of the high-amplitude fluctuations observed by PSP.

\begin{figure*}
    \centering
    \includegraphics[width=1.0\textwidth]{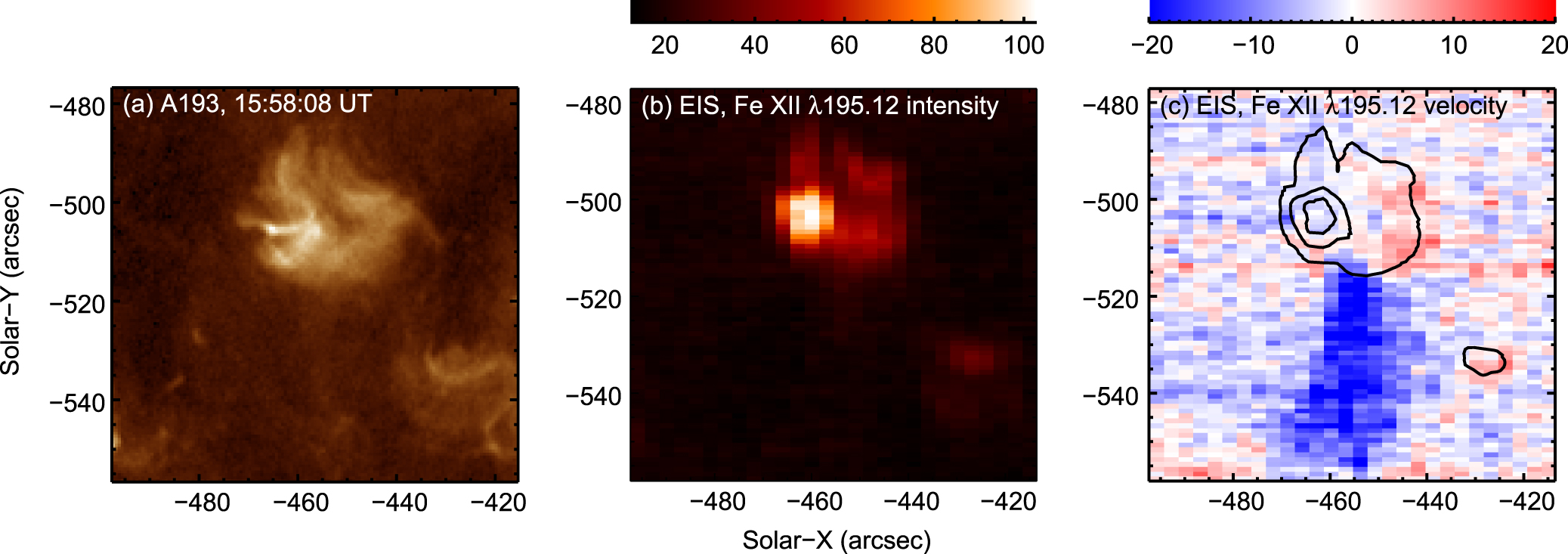}
    \includegraphics[width=0.42\textwidth]{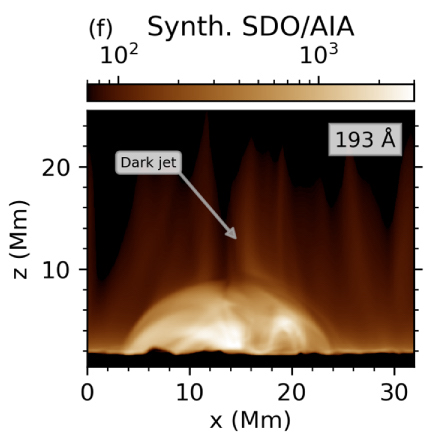}
    \includegraphics[width=0.34\textwidth]{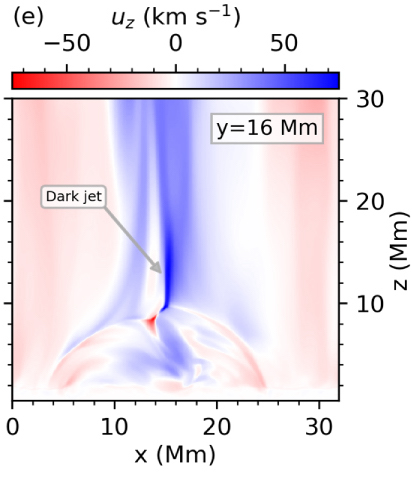}
    \caption{Top panel: EUV imaging (left and middle panels)  observations and  Doppler velocities in the Fe~{\sc xii}~195.12~{\AA} line (right panel).  Bottom panel:  
Synthetic imaging EUV observation (left panel) and LOS velocity of a dark jet derived from simulations performed with a 3D radiative-MHD model using the Bifrost code. Images reproduced with permission from \citep[][top panels]{young_dark_2015} and \citet[][bottom panels]{nobrega-siverio_deciphering_2023}, copyrights by AAS}
    \label{fig:dark_jetsboundaries}
\end{figure*}

\begin{table*}[!ht]
\centering
  \begin{threeparttable}[b]
\caption{Energetics of observed small-scale eruptive transient solar eruptive phenomena. }
\begin{tabular}{llll}
\hline 
Phenomenon & Region & Energy (Joule)   & References \\
\hline
Vortices & chrom& ${10}^{17}-{10}^{19}$&\citet{2023SSRv..219....1T} \\
Spicules & chrom &${10}^{16}-{10}^{18}$&\citet{tsiro04}  \\
&&& \citet{2011Sci...331...55D}\\
Jets \& mini-CMEs &  corona &${10}^{16}-{10}^{21}$    
& \citet{2015ApJ...806...11M}\\
&&& \citet{2016SSRv..201....1R} \\
Nanojets & corona& ${10}^{17}$ & 
\citet{2021NatAs...5...54A} \\
Point-like & corona &${10}^{17}-{10}^{18}$            & {\cite{2018A&A...615A..47S}}\\
EUV Brightenings&&&\\
Microjets&corona&${10}^{16}$& \citet{2021ApJ...918L..20H}\\
Jetlets & corona&  ${10}^{15}-{10}^{17}$ &   \citet{2022ApJ...933...21K}\\
Picoflare jets & corona&$\gtrsim{10}^{13}$\tnote{*} & \citet{2023Sci...381..867C}\\
\hline
\label{tab:table_en}
\end{tabular}
\vspace{-0.3cm}
\begin{tablenotes}
\item[*]{We adopt a value of $\gtrsim{10}^{14}$~J; for details see Sect.~\ref{sec:terminology}.}
\end{tablenotes}
\end{threeparttable}
\label{table2}
\end{table*}

\subsubsection{Energetics of Small-scale Eruptions}
\label{sec:energy}
We summarize in Table~\ref{tab:table_en} the energetics of various small-scale transients in the solar atmosphere. Such phenomena are frequently assigned different names in the literature, although they often reflect a common behavior and phenomenology. For example, the entries corresponding to ``jets", ``jetlets", and ``picoflare jets" all refer to collimated transient outflows, i.e., jetting activity, observed at different spatiotemporal scales, perhaps with different instruments. We nevertheless opted to maintain the terminology proposed by the corresponding authors. Given that the study of energetics involves several restrictive and instrument-dependent assumptions, the quoted energies should be taken as order-of-magnitude estimates that range between 10$^{13}${--}10$^{21}$~J. We emphasize that our compilation is not exhaustive and merely provides an overview of the energetics of these phenomena.

As shown in Table~\ref{tab:table_en}, the range of energies involved is quite large. Moreover, the ever-increasing spatial and temporal resolution of next-generation instruments enables the detection of an increasing number of smaller-scale events and a further extension of the energies towards lower values. {It is important to note that the quoted energies almost exclusively refer to the kinetic, thermal, and radiative energies of the plasma. Exceptions include the chromospheric swirl energy estimates,  which are based only on Poynting fluxes \citep {2023SSRv..219....1T, 2022A&A...663A..94D}, and the estimated energy associated solely with torsional Alfv\'en waves in two polar CH jets \citep{2015ApJ...806...11M}. Note that the Alfv\'en-wave energy flux is potentially relevant to SBs with a solar origin. 

3D MHD simulations of a resistive-kink jet estimated that around 70\% of the energy released in coronal jets is transported outward via the Poynting flux of the associated  Alfv\'en waves \citep{2009ApJ...691...61P}. Therefore, the estimates provided in Table~\ref{tab:table_en} could well represent lower limits on the energy budgets involved in jet phenomena. An important action item is to perform more detailed calculations of the Alfv\'en-wave energy content of jetting events using spectroscopic observations and modeling/theory. In this respect, spectroscopic observations of nonthermal velocities could be exploited in the search for jet-related Alfv\'en waves near their sources \citep[e.g.,][]{1990ApJ...348L..77H}, in addition to near-Sun in situ observations by PSP.

\subsection{Jet Propagation in the Heliosphere}
\label{sec:helio}

\begin{figure}[!ht]
\centering
\includegraphics[width=1.0\textwidth]{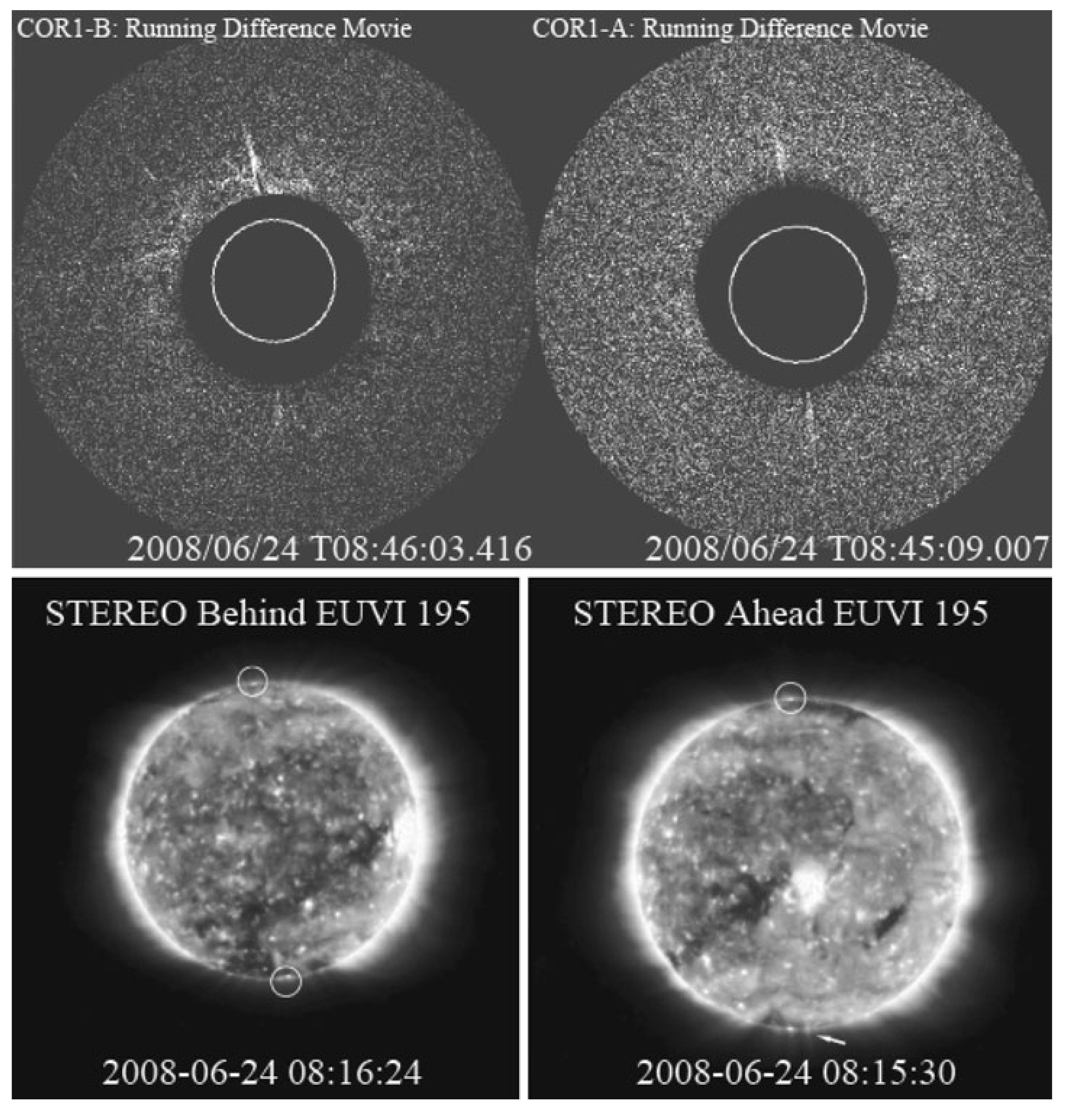}
\caption{Solar jet propagation at large distances. CH jets are seen in coronagraph (SECCHI/COR1 from 1.5 to 4~\rs \, top panels) and EUV (SECCHI/EUVI, bottom panels) images. The CBPs associated with these jets are marked with circles in the bottom panels. The arrow points to a jet whose source region (a CBP) is invisible, as it possibly lies behind the limb. Image reproduced with permission from \citep{2010SoPh..264..365P}, copyright by Springer}
\label{paraschiv_10_fig1}
\end{figure}
In this section, we discuss coronal jets that are also observed to extend into the inner heliosphere by white-light (WL) coronagraphs and heliospheric imagers. In SOHO/EIT observations of 27 polar jets emanating from polar CHs during solar minimum, all events had WL counterparts in the LASCO/C2 coronagraph observations spanning 2{--}6~{\rs} \citep{1998ApJ...508..899W}. The observed WL jets had angular widths of $\approx$2--4$^{\circ}$ and propagated non-radially along the divergent polar magnetic fields.

\citet{2009SoPh..259...87N} investigated 79 polar CH jets using STEREO/EUVI observations and successfully traced their counterparts in the STEREO/COR1 field of view (1.5--5~\rs). \citet{2010SoPh..264..365P} performed a large statistical survey featuring more than 10\,000 WL jets observed by STEREO/SECCHI/COR1/ during solar minimum conditions. For a subset of 1090 (observed only by STEREO-A), 1143 (observed only by STEREO-B), and 501~WL jets (observed simultaneously by STEREO-A and STEREO-B) with lifetimes exceeding 20 minutes, they found that about 75\%, 73\%, and 78\% of these events, respectively, had counterparts in STEREO/EUVI data (see Figure~\ref{paraschiv_10_fig1}). \citet{2014ApJ...784..166Y} traced three large SXR jets from the low to the outer corona using data from Hinode/XRT, SOHO/LASCO/C2 (2.5{--}6~\rs), and STEREO-A/COR2 (2.5{--}15~\rs), respectively, and recovered the signatures of these jets in the inner heliosphere using Solar Mass Ejection Imager (SMEI) 3D reconstructions. Extensions of EUV jets up to 2~{\rs} above the limb have been detected in eclipse observations \citep{2018ApJ...860..142H}. Because the coronagraphic spatiotemporal resolution limits the observed extensions of jets into the upper corona, smaller-scale activity may well have a stronger impact on the inner heliosphere than presently understood.
At least part of the narrow-CME population detected by coronagraphs, i.e., CMEs with widths less than about 15{--}20$^{\circ}$, likely corresponds to coronal jets \citep[e.g.,][]{2001ApJ...550.1093G,2003AdSpR..32.2631Y,Wang2006}. 

The above results demonstrate a rather tight correlation between EUV/SXR jets observed in the inner (below 0.5~{\rs}) corona and WL jets observed in the outer corona. Therefore, an important action item will be to trace and characterize coronal jets observed in EUV and SXR through the upper corona into coronagraph fields-of-view with STEREO/HI/SECCHI, PSP/WISPR, SO/HI, SO/Metis, PROBA-3/ASPIICS, and PUNCH observations. Coronal jets have also been linked with $^3$He-rich solar energetic particle (SEP) events detected by in situ instruments in the inner heliosphere \citep[e.g.,][]{Wang2006, Nitta2008, Bucik2018}. However, some of these events appear to be associated with flares rather than jets or CMEs \citep{Mason2003}. More research is needed to determine whether the presence of anomalous composition in the heliosphere is a reliable and unique indicator of jetting or other frequent reconnection activity at the Sun. Composition measurements of SBs would also be very useful in this regard \citep[see e.g.,][]{2024ApJ...974..198R}. Note, however, that there are no instruments on board PSP measuring composition, and from SO, very limited data are available as yet.

\subsection{Connecting Remote-sensing Observations with In Situ Data on SBs and SB Patches} 
\label{sec:connect-RS-IS}
Several studies of jetting activity using remote-sensing observations compare the properties of the studied events (occurrence, time scales, spatial scales) with quantitative and/or qualitative aspects of SBs. \citet{lee_solar_2022} used the visible-wavelength spectropolarimeters of the BBSO/GST to study spicules along the chromospheric network at a CH boundary. They found that the angular scale of these structures may overlap with those reported for SBs in PSP observations \citep{Fargette2021}. The follow-up study of \citet{2024ApJ...963...79L} included the analyses of SB time scales (such as waiting time reported in \citet{DudokdeWit2020}) to compare them with these solar spatial scales (inter-distances between spicules). The authors concluded that both distributions are compatible. {\cite{2026A&A...705A.188P} employed a simple method relying on magnetic flux conservation in flux tubes that connect SBs with their solar sources and inferred the cross-sectional radius of potential triggers in the low solar atmosphere in the range $\approx$ 10--26\,000 km.

As mentioned in Sect.~\ref{subsubsec:jetproperties}, \citet{huang_statistical_2023} identified about 1160 coronal jets per day in polar and equatorial CHs using 24 days of SDO/AIA images, and estimated that PSP would encounter such ejections no more than once per day near its perihelia. The authors conclude that this rate appears to be compatible with the rate of small-scale magnetic flux ropes observed in the PSP data but not with the rate of SBs. However, this study did not consider small-scale events such as jetlets and spicules. 

\citet{2023ApJ...945...28R} used SDO/AIA and GOES Solar Ultraviolet Imager (SUVI) EUV images together with BBSO/GST magnetograms to study the occurrence of magnetic flux cancellation driving such small-scale jetting activity. They estimate around $5\times10^5$ jetlets occur per day on the whole Sun, which is significantly higher than the frequency of larger coronal jets. 
As mentioned in Sect.~\ref{spicules}, the occurrence of spicules derived from Hinode/SOT/BFI observations, which presumably refers to the more energetic high-velocity Type-II spicules, is 10$^7$ events on the whole Sun at any given time.  
It remains to be determined whether perturbations generated by jetlets and spicules can evolve into the SBs observed by PSP. Those occurrence rates (jets, jetlets, spicules) should be further examined and compared with the occurrence rate of SBs, which can vary over time depending on the definition of SB and the employed detection methods \citep{chapter3}.

Such observations also critically require co-temporal in situ observations to determine whether the main properties of SBs would be found in such structures \citep{chapter3}. 
Indeed, the above-mentioned studies provide interesting comparisons between the remote sensing of jetting activity and the observational and theoretical properties of SBs. However, they do not use co-temporal solar wind in situ time series, which will ultimately be critical to identifying the potential solar sources of SBs. 

So far, very few attempts have been made to combine co-temporal in situ and remote-sensing observations. \citet{de_pablos_searching_2022} tracked the solar source of the solar wind streams observed during PSP's first encounter with the Sun. They performed cross-correlations of the time series of the radial magnetic field, radial proton velocity, mass flux, proton temperature, and proton number density with the EUV light curves from different features selected in SDO/AIA images. They found high correlations for some of the parameters explored. However, the use of nonsynchronous observations may prevent the drawing of a causal link between solar and solar wind variability. 

\citet{kumar_new_2023} explored possible periodicities found in SDO/AIA time series of jetlets in plumes and spikes in the radial proton velocity in PSP data from Encounter 10 and found them to be comparable (on the order of a few mins to up to 20 mins). However, it is important to note that the time series were detrended with filters having a cutoff frequency very close to the detected periods before the wavelet analyses were performed. Such detrending may cause spurious detections \citep[see e.g.,][]{2016ApJ...825..110A}. Therefore, further comparative studies of this kind are required. Similarly, \citet{2024ApJ...968L..28H} analyzed possible periodicities found in the appearance of coronal brightenings observed with SDO and solar wind velocity spikes (associated with SBs) observed with SO. They found similar periodicities (ranging from 7 to 9 hours) in the velocity spikes time series, coronal brightening counts time series, and total photospheric magnetic flux time series. The correlations between time series are, however, based on visual inspection only.

The methodology used to identify the solar sources of the SBs' solar wind streams (connectivity) is crucial in linking SBs to transient solar events. \citet{2025A&A...694A.181B} performed such a study, taking into account a range of uncertainties, including the variation of the PFSS source surface height and the dependence on the solar wind propagation velocity, taking into account different propagation scenarios. The authors performed statistical tests on hourly time series of solar jets and SBs for two co-rotation periods in PSP's encounters 10 and 15, for which the connectivity and availability of on-disk sources were satisfactory. They found no correlation between SBs and jet counts. However, they established a matching level of activity between jets and SBs, depending on the size of the CH connected to the PSP. The event selection methods and detectability in remote sensing observations, along with the evolutionary properties of the events, are additional constraints on establishing a reliable correlation.

\citet{2024NatAs...8.1246H} linked ten solar jets observed with SDO to ten SB clusters observed with PSP. They used a similar approach to \citet{2025A&A...694A.181B} to determine the source of solar wind, testing different parameters (source surface height, propagation velocity) to validate the connectivity and take into account the propagation models. The final uncertainty on the time shift between the two time series is about an hour, which is smaller than that obtained by \citet{2025A&A...694A.181B}. It is important to note, however, that \citet{2024NatAs...8.1246H} chose data from the 4th encounter of the PSP, which was excluded from the study of \citet{2025A&A...694A.181B}. The conclusions of \citet{2024NatAs...8.1246H} are, therefore, in stark contrast with those of \citet{2025A&A...694A.181B}. The different conclusions reached by the two studies are also due to methodological differences, as well as differences in the datasets. Thus, the causal relationship between SBs and small-scale transients in the low solar atmosphere remains highly debatable and requires further in-depth investigations.

\section{Comparison of SB and Transient Solar Event Energetics}
{An order-of-magnitude comparison between the energy carried by SBs observed during PSP Encounter~1 and the energetics of small-scale transient solar phenomena (Table~\ref{tab:table_en}) may help assess the plausibility of a solar origin for SBs. The comparison should be taken with caution, given that it is essentially a dimensional analysis based on single-point measurements from the PSP. Because there is little in situ evidence of energy injection into SBs \citep{Sioulas25}, the energy calculated for PSP SBs should be a lower limit for the energy of the progenitor phenomena, as coronal heating and solar wind acceleration mostly extract energy from the fluctuations.

In estimating SB energy budgets, we adopt two observationally supported assumptions: (1) SBs are Alfv\'enic \citep[e.g.,][]{2019Natur.576..237B,DudokdeWit2020,2020ApJ...904L..30B,2021A&A...650A...3L}, and (2) they are thin, elongated (cylindrical) structures with width $d$, length $L_{SB}$, and aspect ratio $\kappa = L/d$ \citep[][]{2020ApJS..246...45H,2021A&A...650A...1L}. The SB energy is then
\begin{equation}
E_{SB}=\epsilon_{w}V_{SB},
\label{eq:sbe}
\end{equation}
where $\epsilon_{w}$ and $V_{SB}$ denote the Alfv\'en wave energy density and SB volume. Following \citet{1971ApJ...168..509B},
\begin{equation}
\epsilon_{w}=\tfrac{1}{2}\rho{\delta v}^{2} +\tfrac{{\delta B}^{2}}{8\pi},
\label{eq:eden}
\end{equation}
with $\rho$ the mass density and $\delta v$, $\delta B$ the velocity and magnetic field perturbations. 
In writing Equation \ref{eq:eden}, no distinction between outward/inward propagating Alfv\'en waves is made as, for example, in \citet{2011ApJ...743..197C}.The SB volume is
\begin{equation}
V_{SB}=\pi r_{SB}^{2}L_{SB}, \quad r_{SB}=d/2,\; L_{SB}=\kappa d.
\label{eq:sbv}
\end{equation}
We use PSP Encounter~1 values for $\delta v$, $\delta B$, and proton density of 10--100~\kms, 10--100~nT, and 500~cm$^{-3}$, respectively \citep{2019Natur.576..237B,DudokdeWit2020,2021A&A...650A...5F,2021A&A...650A...1L}, and SB width and aspect ratio of 20\,000--89\,000~km and 11-59 \citep{2021A&A...650A...1L}. By randomly sampling these parameters $10^{6}$ times and solving Equations~\ref{eq:sbe}--\ref{eq:sbv}, we obtain $E_{SB}$  value spanning $7\times10^{11}${--}$1.2\times10^{26}$~J, with mean (median) of $10^{15}$ ($5\times10^{14}$)~J. These values are consistent with several entries in Table~\ref{tab:table_en}, especially considering that jet energies may be underestimated (Sect.~\ref{sec:energy}). Moreover, the \citet{2015ApJ...806...11M} estimates for jet-driven torsional Alfv\'en waves ($10^{20}$~J at 0.03~\rs, $10^{18}$~J at 1.3~\rs) are comparable to the inferred SB energetics. 

Although these back-of-the-envelope estimates are simplified, they suggest broad energetic consistency between SBs and small-scale transient solar phenomena, motivating more refined calculations of Alfv\'en-wave energy release during jetting.

\section{Small-scale Transient Event Observations: Terminology vs Phenomena}  \label{sec:terminology}
Here, we clarify the nomenclature of small-scale solar transient phenomena considered as possible precursors of magnetic SBs in the solar wind. A crucial question is whether different names have been assigned to phenomena driven by the same physical mechanism but observed along different lines of sight, with different instruments (imagers or spectrometers), wavelength ranges, or spatial and spectral resolutions.   
Transient brightenings (TBs) observed at chromospheric, transition-region, and coronal temperatures are particularly prone to redundant labeling; however, progress is being made toward a unified classification based on their underlying physics. After accounting for the characteristics of each instrument, \citet{2003A&A...409..755H} concluded that the TBs known as blinkers \citep[CDS observations of transient brightenings; e.g.,][]{1999A&A...351.1115H}, network and cell brightenings \citep[e.g.,][]{2000A&A...362..371H}, and EUV brightenings \citep{1998A&A...336.1039B} represent the same phenomenon. Multi-height observations, covering the chromosphere through the corona, have established that blinkers (i.e., TBs) are the EUV manifestations of transient small-scale energy-release events originating at coronal, transition-region, and/or chromospheric heights \citep{2012A&A...538A..50S}. A study of IRIS spectra from the site of a mini-CME associated with a microflare found that the spectral signature of ongoing reconnection was identical to that of the so-called explosive events \citep{2022A&A...660A..45M}. Likewise, when a small-scale EUV brightening was simultaneously recorded in SO/EUI images and IRIS spectra, its spectral characteristics matched those of explosive events \citep{2023A&A...676A..64N}. 

The morphological appearance, spectral properties, and dynamic evolution of small-scale eruptive transients indicate that they belong to a continuum of events differing only in energetics and spatial scales.  Jets, mini-CMEs, microjets, jetlets,  and picoflare jets are therefore plausibly the same phenomenon. At the lower end of the energy range, picoflare jet energies have been estimated at $\gtrsim10^{13}$~J, assuming an electron density one order of magnitude lower than expected even for such small jets \citep{2021RSPSA.47700217S}. Consequently, their kinetic energy should more realistically be $\gtrsim10^{14}$~J.

\begin{sidewaystable}[htp]
\centering
\caption{Transient (eruptive) small-scale events$\rm ^1$.}
\label{tab:maria}
\begin{tabular}{lllll}
\hline 
Name & Instrumentation  & Location& Magnetic field  & Corresponding events\\
\hline
X-ray and EUV jets & {\sl Yohkoh}, SOHO, TRACE, & SGN & bipolar &  Surges, EEs, MSs, MFs \\
 &  Hinode, SDO &  &  &  \\
mini-CMEs & STEREO/SECCHI, AIA & SGN& bipolar/multi-polar &  X-ray/EUV jets, \\
 &&   & & macrospicules, TBs\\
Microjets & EUI  & SGN &bipolar/multi-polar &\\ 
 Jetlets&AIA, EUI, IRIS& SGN& bipolar/multi-polar& plumes\\
 Network jets&IRIS, SST&SGN&bipolar/multi-polar&picojets, type II spicules, network jets\\
 Picojets& EUI  & SGN& bipolar/multi-polar& network jets, type II spicules, microjets  \\
 MSs & EIT, AIA, TRACE, EUI &SGN& bipolar/multi-polar& \\
 Surges & BBSO, SST, SOHO, IRIS & SGN& bipolar/multi-polar &\\
TBs&HRTS, TRACE, XRT, EIT, AIA, EUI& SGN& bipolar/multi-polar& jets at all scales\\
 &CDS, IRIS, EUI, SUMER&&& blinkers, EEs\\
 EEs&HRTS, SUMER, IRIS&SGN&bipolar/multi-polar &jets, surges, TB\\
 Blinkers&CDS&SGN&bipolar/multi-polar&EEs, jets, MSs\\
\hline
\end{tabular}
\footnotetext[ ]{The following abbreviations were used: SGN -- Supergranulation Network, TBs -- Transient Brightenings, EEs -- Explosive events, MSs -- Macrospicules, MFs -- Mini-filaments }
\end{sidewaystable}
\section{Summary, Conclusions, and Open Issues }
\label{sec:summary}

Table~\ref{tab:maria} summarizes the main types of known small-scale transients, the instruments used to detect them, their locations, and their interrelations with jets and other transient phenomena. 
The key point is that all events in Table~\ref{tab:maria}, except for vortices, occur within the supergranulation network and are associated with embedded bipoles. When high-resolution photospheric magnetic-field data are available, the sources of TBs are consistently found to involve fan-spine configurations or more complex variants thereof, including separatrix layers or  QSLs that appear in high-resolution magnetic-field maps as part of the S-web. Mini-CMEs, certain jets, erupting 
mini-filaments, and microflares are signatures of filament-channel eruptions, essentially scaled-down versions of the larger events that produce flares and
CMEs. As noted in Sect. \ref{subsec:psfstne}, the presence of null points, separatrices, and QSLs in such configurations naturally leads to current-sheet formation and magnetic reconnection.  

In summary, all transient events discussed above, except for vortices and Type I spicules, plausibly represent plasma ejections driven by magnetic reconnection with a kinetic energy that extends over 5{--}7 orders of magnitude, from 10$^{15}$ to 10$^{22}$~J (see Sect.~\ref{sec:energy}). 
Having compiled this range of possible solar contributors to SBs, either direct or indirect, we next consider the criteria used to distinguish between the 
more and less probable solar sources. The primary criteria are:
\begin{itemize}
\item{Are these events sufficiently frequent to account for SBs?}
\item{Do these events produce plasma and magnetic-field structures that can propagate through the intervening atmosphere to SB heights?}
\item{Do these events have sufficient energy to propagate to SB heights?}
\item{Can these events explain the observed clustering denoted SB patches?}
\end{itemize}

Taking into consideration the above-mentioned criteria, in Table~\ref{tab:table_final}, we provide a final summary and evaluation of potential solar sources for SBs. In the left column, we list the feature names and discuss their pros (middle column) and cons (right column) regarding their suitability  of SB sources. Given the current state of understanding, all these potential sources appear likely. We further note that the small-scale deflections that appear associated with these phenomena may not survive to the observed distances of SBs. However, as found in the few numerical studies published thus far, deflections can eventually amplify to become larger \citep[see][for a review of the formation mechanisms of SBs]{chapter5}.

Current observations are largely unable to provide unambiguous answers to these questions. The primary observational gap has been the lack of uninterrupted imaging of the corona from 1.5 to 4~\rs\ from the solar center, between the outer limits of UV/EUV/X-ray imagers and the inner limits of currently operational coronagraphs. With the successful launch \& operation of Metis on board SO and ASPIICS on board Proba-3, we have some coverage of this range that may provide the missing link between the dynamics from 1.5 to 4~\rs. However, we note that differences in FOV, resolution, temporal coverage, and viewing angle often make it difficult to combine observations of individual events to derive a comprehensive picture. The absence of coronal magnetic field observations also prevents definitive measurements of disturbances that could discriminate among models of SB progenitors. For smaller transients (e.g., jetlets), photospheric and/or chromospheric magnetograms with the highest possible spatial resolution are essential to determine the underlying physical mechanisms of energy buildup and release. Combined spectroscopic and imaging coverage of transients over a wide temperature range has led to important insights, but such coverage has been scarce. However, this would change in the near future with the upcoming Multi-Slit Solar Explorer \citep[MUSE;][]{muse} and the EUV High-throughput Spectroscopic Telescope (EUVST) on board the Solar-C mission \citep{Solar-C}. The advent of multiple spacecraft at different distances and viewing angles from the Sun has enabled the 3D structure of some transients to be deciphered, and more such opportunities will occur in the future. Coronagraphs and heliographic imagers with improved sensitivity, cadence, and resolution, as well as FOV inner boundaries as close to the solar surface as possible, would provide crucial links between EUV images of small-scale eruptive activity and in situ data by tracing their transport from the lower solar atmosphere into the heliosphere. Thus far, most (if not all) SB measurements have been taken close to the ecliptic, so their latitudinal distribution remains unknown. This is particularly crucial for determining whether the S-web plays a role in their formation.

Returning to all events described in Sect.~\ref{sec:transients}, we examine each progenitor candidate in turn and list the arguments for and against its likelihood of generating SBs in Table~\ref{tab:table_final}. Note that the large-scale features described in Sect.~\ref{sec:large-Bfield}---helmet streamers, pseudostreamers, and the S-web---are not included. These features are instrumental in guiding, transporting, and/or modifying transient disturbances from the Sun to the heliospheric locations where SBs are observed, but may not themselves be responsible for generating SBs. However, if such locations are preferential to SB formation, a correlation should be found between SB distribution and clustering of the sources within these regions.

\begin{sidewaystable}[htp]
\centering
\caption{Evaluation of potential solar sources}
\label{tab:table_final}
\begin{tabular}{lll}
\hline 
Phenomenon & Pros & Cons\\
\hline
Vortices & produce torsional waves & occur in granule/supergranule centers, \\
 &  &  not network; too low in atmosphere \\
\\
Streamers & escape into wind &too infrequent; not network; in HCS only \\
\\
Pseudostreamers  & open into wind &too infrequent\\
 & likely associated with the complex S-web & more statistics is needed \\
\\
Spicules & numerous; generated in network & low in atmosphere \\
& produce torsional waves & escape to solar wind unlikely \\
\\
Surges & none & too infrequent \\
\\
CH jets & produce torsional Alfv\'en waves & too infrequent \\
\\
AR jets & produce torsional Alfv\'en waves, energetic & too infrequent,  \\
 & could produce subset of SBs & mostly closed loops, not network-associated  \\
\\
Jetlets (and pico-flares) & frequent, network-associated,  & global distribution,  direct evidence in wind not established\\
 & source of plumes, quasi-periodic \\
\\
QS jets and mini-CMEs & wide global distribution & need open flux, escape into wind not established \\
\hline
\end{tabular}
\end{sidewaystable}

Numerical models of SB formation and propagation should also play a major role in eliminating unsuccessful theories and establishing the range of conditions under which the remaining theories can explain observed SB properties. Broadly speaking, three categories of origin theories exist: in situ, in the solar atmosphere, and hybrid. This paper is only relevant to the latter two categories, which include mechanisms that inject flows and/or waves into the solar wind and hybrid models in which events starting at the Sun are converted to SBs in the outer corona/inner heliosphere. We refer the reader to \citet{chapter5} for a comprehensive review of SB formation theories.  

Finally, we note that while this review may give the impression that the potential sources of SBs are primarily near the solar surface, it does not exclude other possibilities, such as those originating from, e.g., the S-web, which is also touched upon in the manuscript. The main reason the review is more focused on  transients occurring near the solar surface is that the most readily recognized candidates in observations for supplying SBs to the solar wind are coronal jets (and other similar transient features). Moreover, the in situ observations indicate that SBs, when backmapped to the Sun, are linked to CHs. Hence, the phenomena that may trigger Alfv\'enic fluctuations with an occurrence rate matching SB rates are transient dynamical phenomena, such as jets and mini-CMEs. However, this does not rule out other possibilities, some of which are covered in this chapter. Our approach here is to present the candidates based on what is available in the literature as of now.  Ultimately, however, the source of SBs is not yet known, and future investigations will determine whether they are due to jets, or due to some other phenomenon, or due to some combination of these.

\bmhead{Acknowledgments} All authors acknowledge the workshop "Magnetic SBs in the Young Solar Wind" organized by the International Space Science Institute, Bern, Switzerland,  18{--}22 September 2023.  The authors thank the two referees very much for their careful reading, helpful comments, and corrections.  D.T. and C.F. acknowledge funding from the CEFIPRA Research Project No. 6904-2. M.M. acknowledges DFG grants WI 3211/8-1 and 3211/8-2, project number 452856778, and was supported by the Brain Pool program funded by the Ministry of Science and ICT through the National Research Foundation of Korea (RS-2024-00408396). M.M. thanks the International Space Science Institute, Bern, Switzerland, for the Visiting Scientist grant. J.T.K. acknowledges funding under multiple grants from NASA's Heliophysics Division.  M.V. acknowledges support via the FIELDS instrument suite of Parker Solar Probe. ACS thanks R. L. Moore and N. K. Panesar for stimulating discussions. A.C.S. received funding from the Heliophysics Division of NASA's Science Mission Directorate through the Heliophysics Supporting Research (HSR, grant No.~20-HSR20\_2-0124) Program, through the Heliophysics System Observatory Connect (HSOC, grant No.~80NSSC20K1285) Program, and the NASA/MSFC Hinode Project. S.P. acknowledges support by the ERC Synergy Grant `Whole Sun' (GAN: 810218). C.F. acknowledges funding from CNES and the Agence Nationale de la Recherche (ANR) for the CROSSWIND project under the grant ANR-24-CE31-2993. 
P.F.W. acknowledges support from STFC (UK) consortium grant ST/W00108X/1 and a Leverhulme Trust Research Project grant. E.P. and C. F. acknowledge support by the Action Th\'ematique Soleil-Terre (ATST) of CNRS/INSU PN Astro as well as from the APR program of CNES. C.F. \& E.P. acknowledge the support of the French Agence Nationale de la Recherche (ANR) for the JET2SB project under grant ANR-25-CE31-7416. The Parker Solar Probe was designed, built, and is now operated by the John Hopkins Physics Laboratory as part of NASA's Living with a Star (LWS) program (contract NNN06AA01C).

\section*{Declarations} Competing Interests: The authors declare no competing interests.
\bibliography{bibliography}

\begin{appendices}

\section*{Acronyms}
\begin{itemize}
\item[ ]{ACE}: Advanced Composition Explorer
\item[]{ASPIICS}: Association of Spacecraft
for Polarimetric and Imaging Investigation of the Corona of
the Sun
\item[ ]{AIA}: Atmospheric Imaging Assembly
\item[ ]{AR}: Active Region
\item[ ]{BBSO}: Big Bear Solar Observatory
\item[ ]{BFI}: Broadband Filter Imager (Hinode/SOT)
\item[ ]{CBP}: Coronal Bright Point
\item[ ]{CDS}: Coronal Diagnostics Spectrometer (on board SOHO)
\item[ ]{CME}: Coronal Mass Ejection
\item[ ]{CH}: Coronal Hole
\item[ ]{ECH}: Equatorial Coronal Hole 
\item[ ]{EIS}: EUV Imaging Spectrometer (on board Hinode)
\item[ ]{EUI}: Extreme-Ultraviolet Imager (on board SO)
\item[ ]{EUV}: Extreme-Ultra-Violet
\item[ ]{EUVI}: Extreme-ultraviolet Imager (on board STEREO/SECCHI)
\item[]{EUVST}: EUV High-throughput Spectroscopic Telescope
\item[ ]{FIP}: First Ionisation Potential
\item[ ]{FWHM}: Full Width at Half Maximum
\item[ ]{GST}: Goode Solar Telescope (at BBSO)
\item[ ]{HCS} Heliospheric Current Sheet
\item[ ]{HI}: Heliospheric Imager (on board SO)
\item[ ]{HMI}: Helioseismic and Magnetic Imager (on board SDO)
\item[ ]{IRIS}: Interface Region Imaging Spectrograph
\item[ ]{MDI}: Michelson Doppler Imager
\item[]{MUSE}: Multi-Slit Solar Explorer
\item[ ]{NFI}: Narrowband Filter Imager (Hinode/SOT)
\item[ ]{PDS}: Periodic Density Structure
\item[ ]{PHI}: Polarimetric and Helioseismic Imager
\item[ ]{PS}: Pseudostreamer
\item[ ]{PSP}: Parker Solar Probe
\item[ ]{PIL}: Polarity Inversion Line
\item[]{PUNCH}: Polarimeter to UNify the Corona and Heliosphere
\item[ ]{QS}: Quiet Sun
\item[ ]{SB}: Switchback
\item[ ]{SDO}: Solar Dynamics Observatory
\item[ ]{SECCHI}: Sun-Earth Connection Coronal and Heliospheric Investigation
\item[ ]{SJ}: Slit Jaw
\item[ ]{SOHO}: Solar and Heliospheric Observatory
\item[ ]{SOT}: Solar Optical Telescope (on board Hinode)
\item[ ]{SO}: Solar Orbiter
\item[ ]{STEREO}: Solar Terrestrial Relations Observatory 
\item[ ]{SUVI}: Solar Ultraviolet Imager (on 
board GOES-R)
\item[ ]{SXT}: Soft X-ray Telescope
\item[ ]{SJ}: Slit Jaw
\item[ ]{TNE}: Thermal nonequilibrium
\item[ ]{TRACE}: Transition Region  And Coronal Explorer
\item[ ]{XBP}: X-ray Bright Point
\item[ ]{XRT}: X-ray Telescope (on board Hinode)
\item[ ]{WISPR}: Wide-Field Imager for Parker Solar Probe
\end{itemize}

\end{appendices}

\end{document}